\documentclass[reprint,groupedaddress,superscriptaddress,aps,prd,preprintnumbers,onecolumn,12pt,fleqn,nofootinbib,eqsecnum,floatfix,showpacs,preprintnumbers,a4paper]{revtex4-2}

\usepackage{xcolor,color}
\usepackage{bm,amsmath,amssymb,dsfont}
\usepackage{graphicx}
\usepackage{afterpage}
\usepackage{enumerate}
\usepackage{relsize}
\usepackage{soul}
\usepackage{cancel}

\usepackage{tikz}

\definecolor{lcolor}{rgb}{0.5,0,0}
\definecolor{citcolor}{rgb}{0,0.3,0.0}
\usepackage[breaklinks,colorlinks,urlcolor=blue,citecolor=citcolor,linkcolor=lcolor]{hyperref}

\newcommand{\hor}{\mkern1.25mu}

\long\def\comment#1{ }

\newcommand{\ihat}{\bm{\hat{{\i}}}}
\newcommand{\jhat}{\bm{\hat{{\j}}}}

\newcommand{\beq}{\begin{equation}}
\newcommand{\eeq}{\end{equation}}
\newcommand{\bal}{\begin{align}}
\newcommand{\eal}{\end{align}}
\newcommand{\tr}{\mathrm{tr}}

\newcommand{\cf}{C_\mathrm{F}}
\newcommand{\nc}{N_\mathrm{c}}
\newcommand{\qs}{Q_\mathrm{s}}
\newcommand{\as}{\alpha_\mathrm{s}}

\newcommand{\abar}{\bar{\alpha}_\mathrm{s}}
\newcommand{\nn}{\nonumber\\}
\newcommand{\dif}{\mathrm{d}}
\newcommand{\bk}{\bm{k}}

\newcommand{\bx}{\bm{x}}
\newcommand{\bbx}{\bar{\bm{x}}}
\newcommand{\by}{\bm{y}}
\newcommand{\bby}{\bar{\bm{y}}}
\newcommand{\bu}{\bm{u}}
\newcommand{\bv}{\bm{v}}
\newcommand{\bz}{\bm{z}}
\newcommand{\bw}{\bm{w}}
\newcommand{\br}{\bm{r}}

\newcommand{\xP}{x_{\mathbb P}}
\newcommand{\YP}{Y_{\mathbb P}}
\newcommand{\mcal}{\mathcal}

\newcommand{\nabt}{\boldsymbol{\nabla}_T}
\newcommand{\lqcd}{{\Lambda_{\mathrm{QCD}}}}

\begin{document}

\title{When JIMWLK evolution really matters: the example of incoherent diffraction}


\author{T.~Lappi}
\email{tuomas.v.v.lappi@jyu.fi}
\affiliation{Department of Physics, University of Jyv\"{a}skyl\"{a}, P.O.~Box 35, 40014 Jyv\"{a}skyl\"{a}, Finland}
\affiliation{Helsinki Institute of Physics, P.O.~Box 64, 00014 Helsinki, Finland}
\author{D.N.~Triantafyllopoulos}
\email{trianta@ectstar.eu}
\affiliation{European Centre for Theoretical Studies in Nuclear Physics and Related Areas (ECT*)\\
and Fondazione Bruno Kessler, Strada delle Tabarelle 286, 38123 Villazzano (TN), Italy}

\date{\today}

\begin{abstract}

We consider high energy scattering in the effective theory of the Color Glass Condensate. The most convenient degrees of freedom are Wilson lines encoding multiple gluon exchanges, whose evolution with energy follows the JIMWLK equation. Instead of using the latter, very often one resorts to a Gaussian Approximation (GA), which is known to be remarkably accurate in describing a wide class of multi-gluon correlators whose expansion in the weak scattering limit starts with an exchange of only two gluons. Here we demonstrate, both analytically and numerically, that such an approximation is not valid for correlators which start with an exchange of four gluons. As a main example, we focus on incoherent diffraction in photon-nucleus collisions and we show that the discrepancy between the JIMWLK and the GA results is driven by weak scattering and further persists in the regime where unitarity corrections begin to become important. The JIMWLK calculation leads to cross sections which are systematically larger in all kinematic regimes of interest. 

\end{abstract}

\maketitle

\section{Introduction and Motivation}
\label{sec:intro}

An exciting part of the physics program at high energy scattering experiments both at the Large Hadron Collider (LHC) and the future Electron-Ion Collider (EIC) is to probe the high energy gluon saturation regime of QCD. In particular ultraperipheral collisions (UPCs) at LHC provide a relatively clean process where a simple dilute probe, the photon, scatters of a nucleus with a higher center of mass energy than is available at the EIC, whereas the latter will have access to a broader range of final states.  A convenient framework to understand these collision processes is provided by the Color Glass Condensate (CGC) effective theory for QCD. Here the physical picture of the scattering is one of partonic components of the dilute probe scattering off a ``shockwave'' made by the classical gluon field of the target. The degree of freedom used to describe the saturated gluon field in this picture is not a gluon number distribution, but the Wilson line, an eikonal scattering amplitude for a dilute projectile in the shockwave. The CGC picture provides a unifying description where inclusive, diffractive and exclusive observables can be described in the same framework in terms of the Wilson lines. 

For photon-mediated scattering it is convenient to work in the ``dipole picture''. Here, at leading order in the QCD coupling $\as$, the  photon splits into a $q\bar{q}$ dipole, with the interaction of the quark and the antiquark with the target given by their respective Wilson lines. The photon, and hence the $q\bar{q}$ pair before the scattering are color neutral. If also the final state is a color singlet, such as in exclusive scattering, or for the total cross section since it is related by the optical theorem to the elastic amplitude, the cross section is obtained from color singlet ``dipole'' operators of the Wilson lines, hence the name ``dipole picture''. For such a photon initial state the amplitude is a product of two Wilson lines, and consequently the cross section involves a product of four, in different configurations depending on the final state. Thus one in general needs expectation values of four fundamental Wilson line operators at leading order, and higher point operators at higher orders involving additional gluons crossing the shockwave.  In the CGC picture the energy dependence of these multi-point functions of Wilson lines is given by the Jalilian-Marian, Iancu, McLerran, Weigert, Leonidov, Kovner (JIMWLK) equation~\cite{Jalilian-Marian:1997qno,Jalilian-Marian:1997jhx,Weigert:2000gi,Iancu:2000hn,Iancu:2001ad}, which resums larges logarithms generated by the emission of soft gluons. The JIMWLK equation is an evolution equation for the probability distribution of Wilson lines as a function of rapidity. It can equivalently be recast as an infinite hierarchy of coupled evolution equations for expectation values of Wilson line operators, the Balitsky hierarchy \cite{Balitsky:1995ub,Balitsky:2001gj}. 

Working with an infinite hierarchy of coupled equations is a complication that one usually tries to avoid in phenomenological work, if possible, with the help of different approximations. Typically this is done by truncating the hierarchy using various forms of a mean field approximation in order to obtain given closed sets of equations. The most common approach is to use the Balitsky-Kovchegov (BK) equation~\cite{Balitsky:1995ub,Kovchegov:1999yj,Kovchegov:1999ua}, which can be obtained from JIMWLK by taking  the large $\nc$ limit, with $\nc$ the number of colors, in the evolution equation for the dipole operator.
If one wants to go beyond leading $\nc$, another truncation scheme is needed. The most common choice here is the Gaussian Approximation (GA, also referred to as Gaussian truncation)~\cite{Iancu:2002aq,Kovchegov:2008mk,Iancu:2011ns,Iancu:2011nj,Dumitru:2011vk}. The GA  can be viewed as a generalization of the MV model~\cite{McLerran:1993ni,McLerran:1993ka,McLerran:1994vd}, where in the large $A$ limit, with $A$ the atomic mass number of a nucleus, color charges are taken to be Gaussian local random variables.  In the GA the Gaussian property is preserved, but the color charge correlator is allowed to be nonlocal, enabling a different transverse momentum/coordinate dependence given by the JIMWLK evolution equation. The GA provides an economical way of obtaining higher Wilson line correlators. In practice one typically calculates the high energy evolution of the dipole correlator using the  BK equation  and then obtains higher point correlators in terms of the dipole, similarly as in the MV model~\cite{Blaizot:2004wv,Weigert:2005us,Dominguez:2008aa,Dominguez:2011wm}. As a result, the  GA has been used in many practical applications~\cite{Lappi:2012nh,Kovchegov:2008mk,Lappi:2020srm,Mantysaari:2019hkq} to replace a full JIMWLK calculation. Of course it is possible to directly use a numerical solution of the JIMWLK equation in phenomenological applications. This has been done e.g. to calculate transverse momentum dependent distributions (TMDs)~\cite{Marquet:2016cgx} (see Ref.~\cite{Cali:2021tsh} for a systematical study), or directly in calculations of cross sections~\cite{Lappi:2011ju,Schlichting:2014ipa,Schenke:2022mjv,Mantysaari:2024qmt,Mantysaari:2019csc}.

In the few cases where the GA has been directly compared to JIMWLK, they have been found to be in good agreement \cite{Dumitru:2011vk,Lappi:2015vta}. These studies have dealt with observables which at weak scattering start with a two-gluon exchange. Under these circumstances a proof that the GA is accurate (even though not exact) has already been given \cite{Iancu:2011ns,Iancu:2011nj}. It is the purpose of this paper to elucidate  what happens when the observable of interest does not obey the necessary conditions and to explicitly show that there are cases where the GA cannot be reliably used. The above discussion and the existing calculations give us clear hints for where to look. At large $\nc$ the Balitsky hierarchy truncates naturally, and one could expect a mean field approximation to work. Correlators that correspond to two-gluon exchange when the scattering is weak should obey the BFKL equation \cite{Kuraev:1977fs,Balitsky:1978ic} in this limit. This together with the unitarity limit for Wilson lines imposes quite strong constraints on what a given correlator can be, not leaving much room for deviations between the full JIMWLK result and the GA. This leaves us with correlators that, in the dilute limit, correspond to multiple gluon exchanges. We also expect potential differences between JIMWLK evolution and the GA to be relatively larger in observables suppressed at large $\nc$. A prime example is provided by incoherent diffraction, for example incoherent vector meson or dijet production. When we look at large enough values of the momentum transfer squared $|t|$, the incoherent cross section is expected to be dominated by color charge fluctuations, which correspond to the variance of the dipole operator of Wilson lines \cite{Demirci:2022wuy}. Here the leading $\nc$ contribution, corresponding to coherent diffraction, is specifically excluded from the definition of the experimental observable. As a diffractive observable, the amplitude starts with a two-gluon exchange and thus a minimum of four gluons are exchanged at the level of the cross section. Incoherent diffraction thus provides a concrete phenomenologically relevant example of a case where deviations from the GA could be expected to manifest themselves.

In this paper we show that indeed the full JIMWLK equation deviates significantly from the GA for some specific correlators, in particular for the case of incoherent diffraction. First we introduce the JIMWLK equation in Sec.~\ref{sec:JIM}. We present the GA, briefly discuss the necessary conditions for its validity and comment on its status in Sec.~\ref{sec:GA}. Then we demonstrate analytically, working in the dilute limit, how one rapidity step of JIMWLK evolution is enough to violate the GA when the aforementioned conditions are not met, in Sec.~\ref{sec:break}. We confirm these expectations by explicit numerical solutions of the JIMWLK equation in its Langevin form in Sec.~\ref{sec:num}. Subsequently in Sec.~\ref{sec:sigma} we address quantitatively the ratio of incoherent to coherent cross sections in the regime where the former is dominated by color charge fluctuations incorporated in JIMWLK evolution, before concluding in Sec.~\ref{sec:conc}.

\section{JIMWLK and Balitsky equations}
\label{sec:JIM}

The Color Glass Condensate (CGC) is a modern effective theory that describes the small-$x$ components of the wavefunction of an ultra-relativistic nucleus (or hadron). It is based on the concept that gluons carrying a small fraction $x$ longitudinal momentum of the nucleus behave like a stochastic distribution of classical color fields, generated by fast moving color sources with larger momentum fractions. Due to Lorentz time dilation at high energies, these sources appear static or ``frozen'' and in a suitable gauge, the resulting color field has only one non-zero component. Adopting the light-cone coordinates $x^{\mu} = (x^+,x^-,\bx)$, where $x^{\pm} = (t \pm x^3)/\sqrt{2}$ and $\bx=(x^1,x^2)$, the gauge field for a nucleus moving along the negative $x^3$ direction takes the form $A_a^{\mu}(x) = \delta^{\mu-} \alpha_a(x^+,\bx)$. The CGC is fully described by a weight functional $W_Y[\alpha]$, which defines the probability of a given field configuration $\alpha_a$. This functional allows one to compute gauge field correlators that encode the detailed structure and evolution of the nucleus with increasing rapidity $Y=\ln(1/x)$, starting from an initial rapidity $Y_0 = \ln(1/x_0)$. In the high energy regime $\abar (Y-Y_0) \gtrsim 1$, where $\abar = \as \nc/\pi$, this evolution is governed by a renormalization group equation known as the JIMWLK equation \cite{Jalilian-Marian:1997qno,Jalilian-Marian:1997jhx,Weigert:2000gi,Iancu:2000hn,Iancu:2001ad}. 
In its Hamiltonian form, it is expressed as
\begin{align}
	\label{H_on_W}
	\frac{\dif W_Y[\alpha]}{\dif Y} = H\, W_Y[\alpha],
\end{align}
where $H$ is the JIMWLK Hamiltonian which at leading logarithmic level reads  
\begin{align}
	\label{H_JIM}
	H = \frac{1}{8 \pi^3}
	\int_{\bu\bv\bz}\!
	\mcal{K}^i_{\bu\bz}\mcal{K}^i_{\bv\bz}
	\big(L_{\bu}^a  - U_{\bz}^{ab} R_{\bu}^b \big)
	\big(L_{\bv}^a  - U_{\bz}^{ac} R_{\bu}^c \big).
\end{align}
Here we introduced the compact notation $\int_{\bu} \equiv \int \dif^2 \bu$ and the Weizs\"{a}cker-Williams emission kernel
\begin{align}
	\mcal{K}^i_{\bu\bz} = \frac{(\bu-\bz)^i}{(\bu-\bz)^2}
\end{align}
and defined the Wilson line
\begin{align}
	\label{Wilson_U}
	U_{\bm{x}}
	\equiv \mathrm{P}
	\exp \left[ i g\int \dif x^+ \alpha_a(x^+,\bm{x})\, T^a \right]	
\end{align}
with the generator $T^a$ belonging to the adjoint representation and P standing for path ordering in $x^+$. The integration over $x^+$ extends along the whole real axis, but it is useful to remember that in practice it is localized around $x^+=0$ due to the Lorentz contraction of the target, often called the shockwave.  $L$ and $R$ in Eq.~\eqref{H_JIM} are ``left'' and ``right'' Lie derivatives, that is, they generate local color rotations to the left or to the right of Wilson lines according to
\begin{align}
	\label{LandR}
	L_{\bm{u}}^a V_{\bm{x}} = 
	i g \delta_{\bm{x}\bm{u}} t^a
	V_{\bm{x}},
	\qquad
	L_{\bm{u}}^a V_{\bm{x}}^{\dagger} = 
	-i g \delta_{\bm{x}\bm{u}}
	V_{\bm{x}}^{\dagger} t^a,
	\nn
	R_{\bm{u}}^a V_{\bm{x}} = 
	i g \delta_{\bm{x}\bm{u}}	
	V_{\bm{x}} t^a,
	\qquad
	R_{\bm{u}}^a V_{\bm{x}}^{\dagger} = 
	-i g \delta_{\bm{x}\bm{u}} t^a
	V_{\bm{x}}^{\dagger}.
\end{align}
With $t^a$ and $V$, as in the above, we will be denoting the respective generators and Wilson lines in the fundamental representation, although it should be clear that the relations in Eq.~\eqref{LandR} are valid in any representation. From these relations one immediately understands that  in reality $L$ and $R$ are functional derivatives with respect to the field $\alpha_a(x^+,\bx)$ at the end-points of the Wilson lines: the left one acts at the largest value of $x^+$ whereas the right one at the smallest value of $x^+$. Thus the JIMWLK Hamiltonian can be viewed as a second-order functional differential operator. It goes without saying that in order to fully define our problem, we must specify an initial condition at the initial rapidity $Y_0$. At least for a sufficiently large nucleus, $A \gg 1$, this is typically given by the McLerran-Venugopalan (MV) model \cite{McLerran:1993ni,McLerran:1993ka} to be defined later.

\comment{A convenient form for our purposes is \cite{Hatta:2005as}
\begin{align}
	\label{H_JIM}
	H = \frac{1}{8 \pi^3}
	\int_{\bu\bv\bz}\!
	\mcal{M}_{\bu\bv\bz}
	\big(L_{\bu}^a L_{\bv}^a
	+ R_{\bu}^a R_{\bv}^a
	- 
	2 U_{\bz}^{ab} R_{\bu}^a L_{\bv}^b \big),
\end{align}
defined the dipole kernel \cite{Mueller:1993rr}
\begin{align}
	\label{dip_ker}
		\mcal{M}_{\bu\bv\bz} = \frac{(\bu-\bv)^2}
		{(\bu-\bz)^2 (\bz - \bv)^2}
\end{align}}

Physical observables are represented by (color singlet) gauge-invariant operators constructed from the color fields. Their expectation values are obtained through functional averaging with the CGC weight function, namely
\begin{align}
	\label{Oave}
	\langle O \rangle_Y
	= 
	\int \mcal{D} [\alpha] W_Y[\alpha] O.
\end{align}
This demonstrates that, even though $W_Y[\alpha]$ is obtained from a quantum calculation, the averaging process itself is classical due to the separation of longitudinal scales. By differentiating Eq.~\eqref{Oave} with respect to $Y$ and performing two integrations by parts, we obtain the evolution equation
\begin{align}
	\label{dOdY_JIM}
	\frac{\dif \langle O \rangle_Y}{\dif Y} = 
	\langle H O \rangle_Y. 
\end{align}	
We will shortly see that this is no longer a functional equation, but rather an integro-differential equation. It will not be significantly easier to handle, since, due to the nonlinear dependence of the Hamiltonian on the field $\alpha_a$, it is not a closed equation in general. Instead, it is just a part of an infinite hierarchy of coupled equations, known as the Balitsky equations \cite{Balitsky:1995ub,Balitsky:2001gj}. Such equations are interpreted as projectile evolution, specifically they describe the eikonal scattering of a right-moving projectile off the target. As a result, the operator $O$ is naturally built using Wilson lines, with each line corresponding to a parton in the projectile. A Wilson line in the fundamental representation represents a quark (and its hermitian conjugate an antiquark), whereas a Wilson line in the adjoint stands for a gluon. For instance, the $S$-matrices for a color dipole made by a quark-antiquark pair in an overall color singlet state and a color quadrupole, i.e.~a system of two quarks and two antiquarks again in a color singlet, respectively read 
\begin{align}
\label{D_and_Q}
	D_{\bx \by} \equiv 1- T_{\bx\by} \equiv
	\frac{1}{\nc}\,
	\tr\big({V}^{\phantom{\dagger}}_{\bm{x}} 
	{V}_{\bm{y}}^{\dagger}\big)
	\qquad \mathrm{and} \qquad
	Q_{\bx \by \bz \bw} \equiv  
	\frac{1}{\nc}\,
	\tr\big(
	{V}^{\phantom{\dagger}}_{\bm{x}} 
	{V}_{\bm{y}}^{\dagger}
	{V}^{\phantom{\dagger}}_{\bm{z}} 
	{V}_{\bm{w}}^{\dagger}
	\big),
\end{align} 
where $\bx,\by,\dots$ are the parton transverse coordinates which, by definition, do not get modified by the scattering process. For later use, in the above we have also introduced the $T$-matrix for the scattering of the dipole off the nucleus. In Eq.~\eqref{D_and_Q} we have not yet averaged with the target weight-function. To denote such an average, sometimes we shall use the respective calligraphic letter instead of the usual expectation value brackets, for example $\mcal{D} = \langle D \rangle$, $\mcal{T} = \langle T \rangle$, $\mcal{Q} = \langle Q \rangle$ and so on. Physical observables can involve not only single-trace operators like those in Eq.~\eqref{D_and_Q}, but also multi trace ones like
\begin{align}	
	\label{multitrace} 
	O_{\bx_1 \by_1 \cdots \bx_2 \by_2 \cdots} = 
	\frac{1}{\nc}\,
	\tr\big({V}^{\phantom{\dagger}}_{\bm{x}_1} 
	{V}_{\bm{y}_1}^{\dagger}\cdots \big)
	\frac{1}{\nc}\,
	\tr\big({V}^{\phantom{\dagger}}_{\bm{x}_2} 
	{V}_{\bm{y}_2}^{\dagger}\cdots \big)
	\cdots.
 \end{align}
Using Eqs.~\eqref{H_JIM}, \eqref{LandR} and \eqref{dOdY_JIM} we can derive the evolution equations obeyed by the expectation values of the operators of interest. For instance, the scattering of the $q\mkern1mu\bar{q}$ pair in Eq.~\eqref{D_and_Q} satisfies
\begin{align}
	\label{Bal1}
	\frac{\dif \langle D_{\bx\by} \rangle}{\dif Y} = 
	\frac{\abar}{2\pi}
	\int_{\bz}
	\mcal{M}_{\bx\by\bz}
	\left[
	\langle D_{\bx\bz} D_{\bz\by} \rangle - \langle D_{\bx\by} \rangle
	\right],
\end{align}
where we introduced the dipole kernel \cite{Mueller:1993rr}
\begin{align}
	\label{dip_ker}
		\mcal{M}_{\bu\bv\bz} = \frac{(\bu-\bv)^2}
		{(\bu-\bz)^2 (\bz - \bv)^2}.
\end{align}  
We immediately see that the equation does not close since on the right hand side (r.h.s.)~a new quantity appears, the double trace $\langle D_{\bx\bz} D_{\bz\bx}\rangle$, for which one must write its evolution equation. It is not more difficult to consider all transverse coordinates to be different from each other and, with a notation convenient for our purposes later on, we find
\begin{align}
	\label{Bal2}
	\hspace*{-0.5cm}
	\frac{\dif \langle D_{\bx\by} D_{\bby\bbx} \rangle}{\dif Y} =\,&
	\frac{\abar}{2\pi}
	\int_{\bz}
	\mcal{M}_{\bx\by\bz}
	\left[
	\langle D_{\bx\bz} D_{\bz\by} D_{\bby\bbx} \rangle - 
	\langle D_{\bx\by} D_{\bby\bbx} \rangle
	\right]
	+ \bx\by \leftrightarrow \bby\bbx
	\nn*[0.2cm]
 	& \hspace*{-2.7
 	cm} + 
 	\frac{1}{2 
\nc^2}\,
 	\frac{\abar}{2\pi}
 	\int_{\bz}
	(\mcal{M}_{\bx\bbx\bz} \!+\! \mcal{M}_{\by\bby\bz} 
	\!-\!
	\mcal{M}_{\bx\bby\bz}
	\!-\!
	\mcal{M}_{\by\bbx\bz})
	\langle S_{\bx\bz \bby\bbx \bz \by} + S_{\bx\by\bz \bbx \bby \bz} - Q_{\bx\by\bby\bbx} - Q_{\bx\bbx \bby \by} \rangle.
\end{align}
The first line arises when the two derivatives of the Hamiltonian act on the same dipole, while the other dipole is just a spectator. Thus its structure is consistent with the Leibniz rule of differentiation and the first Balitsky equation \eqref{Bal1}. Not surprisingly, the new quantity is a triple dipole. Instead, the second line emerges when each derivative acts on a different dipole. This leads to a redistribution of color between the two dipoles and new multipole structures appear. More precisely these are quadrupoles and sextupoles, where the definition of the latter, denoted with $S$, follows trivially those in Eq.~\eqref{D_and_Q}. In both Eqs.~\eqref{Bal1} and \eqref{Bal2}, $\bz$ is viewed as the transverse momentum of the soft gluon radiated by the parent partons of the projectile system.

Thus, as anticipated, one arrives at an infinite hierarchy of equations, the Balitsky hierarchy, which is hard to deal with. Certain simplifications occur at large $\nc$, where one can neglect terms which are explicitly suppressed by $1/
\nc^2$, like those in the second line in Eq.~\eqref{Bal2}. Then the hierarchy admits the factorized solution
\begin{align}	
	\label{Oave_fact} 
	\langle O_{\bx_1 \by_1 \cdots \bx_2 \by_2 \cdots  } \rangle = 
	\Big \langle \frac{1}{\nc}\,
	\tr\big({V}^{\phantom{\dagger}}_{\bm{x}_1} 
	{V}_{\bm{y}_1}^{\dagger}\cdots \big) \Big \rangle 
	\Big \langle\frac{1}{\nc}\,
	\tr\big({V}^{\phantom{\dagger}}_{\bm{x}_2} 
	{V}_{\bm{y}_2}^{\dagger}\cdots \big) \Big \rangle
	\cdots
	\quad \mathrm{at \,\,large\,}\nc,
 \end{align}
if  such a property is satisfied in the initial condition at $Y_0$. Then Eq.~\eqref{Bal1} closes to give the BK equation  \cite{Balitsky:1995ub,Kovchegov:1999yj} and one can determine the dipole $\mcal{D}$. Regarding the quadrupole $\mcal{Q}$, one can show that it satisfies a linear integro-differential inhomogeneous equation whose coefficients 
depend on the already known dipole $\mcal{D}$ \cite{JalilianMarian:2004da}, and similarly for higher multipoles. But already for the quadrupole, it seems to be demanding to deal (numerically) with the large number of coordinates and the non-locality in the transverse space at the same time. For such observables, if they satisfy certain conditions, one can rely on the Gaussian approximation, which has the merit to hold at finite $\nc$ and will be briefly reviewed in the next section. 

If no approximation can be made, one can resort to an alternative formulation of the JIMWLK evolution as a Langevin equation \cite{Blaizot:2002np}, again valid at finite $\nc$, which can be solved on a lattice \cite{Rummukainen:2003ns,Lappi:2011ju,Dumitru:2011vk}. In such an approach the target color field is viewed as a random walk in the Wilson lines space, with the rapidity $Y$ being the analog of time. The expectation value of an observable, cf.~Eq.~\eqref{Oave}, can be obtained as an average over the noise term appearing in the corresponding Langevin equation. For example, for the fundamental dipole in Eq.~\eqref{D_and_Q} we have
\begin{align}
	\label{Dave_lang}
	\left\langle D_{\bx \by} \right\rangle_Y 
	=
	\frac{1}{\nc}
	\left\langle
	\tr\big({V}^{\phantom{\dagger}}_{N,\bm{x}} 
	{V}_{N,\bm{y}}^{\dagger}\big)
	\right\rangle_{\nu}
\end{align}
where the rapidity interval $\Delta Y \equiv Y-Y_0$ available for evolution is discretized according to $\Delta Y=\epsilon N_y$, with $N_y \to \infty$ and $\epsilon \to 0$. An evolution step, under which soft quantum fluctuations are integrated out, adds two new layers in $x^+$ in the support of the Wilson lines and leads to small, left and right, random rotations in color space. This results in the Langevin equation \cite{Iancu:2011nj,Lappi:2012vw}
\begin{align}
	\label{V_lang}
	V_{n,\bx} = 
	\exp(i \epsilon g \alpha^{L}_{n,\bx})
	V_{n-1,\bx}
	\exp(-i \epsilon g \alpha^{R}_{n,\bx}),
 \end{align}
with the left and right color matrix fields
\begin{align}
	\label{alpha_nu}
	\alpha^{L}_{n,\bx} = 
	\frac{1}{\sqrt{4\pi^3}}\int_{\bz} \
	\mcal{K}^i_{\bx\bz} \,\nu^{ia}_{n,\bz}\, t^a
	\qquad \mathrm{and} \qquad
 \alpha^{R}_{n,\bx} = 
 \frac{1}{\sqrt{4\pi^3}} \int_{\bz} \mcal{K}^i_{\bx\bz} \,\nu^{ia}_{n,\bz} U^{ab}_{n-1,\bz}\, t^b.
 \end{align}
Here $\nu_n$ is the noise which is Gaussian, white and local in rapidity and in transverse coordinate space, that is
\begin{align}
	\label{noise}
	\big\langle \nu^{ia}_{m,\bx}\, 
	\nu^{jb}_{n,\by} 
	\big\rangle
	=\frac{1}{\epsilon}\,
	\delta^{ij} \delta^{ab} 
	\delta_{mn} \delta_{\bx\by},
 \end{align} 
which determines the mechanism of averaging in Eq.~\eqref{Dave_lang}. To compute such averages or to derive the associated evolution equations, one must keep terms up to order $\epsilon$. Since the noise term is proportional to $1/\sqrt{\epsilon}$, one needs to expand the left and right rotations in Eq.~\eqref{V_lang} up to quadratic terms to achieve the required accuracy. To the order of interest, it helps noticing that these quadratic terms can be readily replaced by their average, because they cannot be multiplied by noise terms from other sources. When doing so, those originating from right rotations become independent of the Wilson line in the previous step (for left rotations this is trivially true, cf.~Eq.~\eqref{alpha_nu}) and represent virtual terms, more precisely
 \begin{align}
 	\label{rightquad}
 	-\frac{\epsilon^2 g^2}{2}\, 
 	\big\langle (\alpha^{R}_{n,\bx})^2 \big \rangle 
 	&= 
 	-\frac{\epsilon^2 g^2}{8\pi^3}\int_{\bz\bw}
 	\mcal{K}^i_{\bx\bz} \mcal{K}^j_{\bx\bw}
 	U^{ab}_{n-1,\bz}  U^{cd}_{n-1,\bw} 
 	T^b T^d
 	\big\langle \nu^{ia}_{n,\bz}\, 
 	\nu^{jc}_{n,\bw} \big\rangle
 	\nn & =
 	-\frac{\epsilon\abar }{2\pi} \int_{\bz} 
 	\mcal{K}^i_{\bx\bz}\mcal{K}^i_{\bx\bz}\,
 	=\, 
 	-\frac{\epsilon^2 g^2}{2}\, 
 	\big\langle (\alpha^{L}_{n,\bx})^2 \big \rangle.
 \end{align}
Last, but not least, in this Langevin formulation a Gaussian initial condition at the rapidity $Y_0$ means that the color charges are drawn from a Gaussian probability distribution. In turn, this determines the probability distribution for the initial value of $V_{0,\bx}$ which is necessary to build the Wilson lines via Eq.~\eqref{V_lang}.

\section{The Gaussian Approximation and a Valid Example}
\label{sec:GA}

The Gaussian Approximation (GA) to the JIMWLK evolution relies on the assumption that the target field configurations are described by a Gaussian distribution. Following \cite{Iancu:2011nj}, let us review the main features which are necessary for the current study. For our purposes, it is more suitable to start from an alternative definition of the GA, namely by formulating it in terms of a Hamiltonian which reads 
\begin{align}
\label{H_GA}
	H_\mathrm{GA} = - \frac{1}{2}
	\int_{\bm{u} \bm{v}}
	\gamma_{\bm{u}\bm{v}}(y)
	\left(
	L_{\bm{u}}^a 
	L_{\bm{v}}^a 
	+
	R_{\bm{u}}^a 
	R_{\bm{v}}^a 
	\right).
\end{align}
This Hamiltonian generates evolution in what we can call the ``parametrization rapidity'' $y$~\cite{Lappi:2016gqe}, so that in an analogy with Eq.~\eqref{dOdY_JIM}, the value of a 
generic observable is determined by
\begin{align}
	\label{dOdY}
	\frac{\dif \langle O \rangle}{\dif y} = \langle H_\mathrm{GA} O \rangle.
\end{align}
The terminology comes from the fact that $H_\mathrm{GA}$ generates a probability distribution for color fields $W_Y[\alpha]$ that  is a Gaussian in the field $\alpha_a$. 
The Hamiltonian in Eq.~\eqref{H_GA} is much simpler than the  JIMWLK Hamiltonian in Eq.~\eqref{H_JIM}, since it is local in the transverse space. The left and right derivatives are those already defined in Eq.~\eqref{LandR}. On the other hand, the kernel $\gamma_{\bm{u} \bm{v}}$ must be obtained from elsewhere to solve the equation. 

We start by observing that for the fundamental representation dipole one readily obtains
\begin{align}
	\label{gammaexp}
	\mcal{D}_{\bm{u}\bm{v}}
	= \exp \left[  - 2 g^2 \cf \int \dif y\mkern2mu \gamma_{\bm{u} \bm{v} }(y) \right] 
	\equiv  
	\exp \left( - \cf \Gamma_{\bm{u} \bm{v} } \right) 
\end{align}
with $\cf= (\nc^2-1)/2\nc$. Generally, for an arbitrary correlator the GA equation \eqref{dOdY} does not lead to a closed equation. Still, the form of the Hamiltonian in Eq.~\eqref{H_GA} is such that it mixes only correlators with the same total number of Wilson lines. For example, the sextupole $\langle S\rangle$ will mix with (certain permutations of) the dipole quadrupole product $\langle D Q \rangle$ and the three dipole product $\langle D D D \rangle$, so that one arrives at a linear system of unknown correlators. In many interesting cases in which the observable does not contain too many Wilson lines and/or if the configuration of the projectile charges is sufficiently simple, one can arrive at an analytic solution, where higher point functions of Wilson lines in a given representation are expressed as functions of $\Gamma_{\bm{u} \bm{v} }$ or, after we make use of Eq.~\eqref{gammaexp},  in terms of the dipole $ \mcal{D}_{\bm{u}\bm{v}}$. Thus, the GA equation~\eqref{dOdY} serves as a convenient way of generating Wilson lines from a Gaussian probability distribution, where all higher point functions are uniquely determined by just the two-point function. Notice that by the time this has been achieved, the parametrization rapidity $y$ has disappeared from the calculation: only various Wilson line correlators expressed as functions of the dipole remain. 

One should not forget that in such a construction non-linearities are still taken into account, since in general the dipole $ \mcal{D}_{\bm{u}\bm{v}}$ involves the color field to all orders. To properly determine the latter we can use the GA to truncate the Balitsky hierarchy already at the level of the first equation, that is Eq.~\eqref{Bal1} for the dipole. Expressing the higher-point correlator $\langle D D \rangle$ on the r.h.s.~in terms of the dipole $\mcal{D} = \langle D\rangle$, we obtain a closed equation for $\mcal{D}$ which is equivalent to the BK equation for the quantity $\mcal{D}^\mathrm{BK}$ defined via
\begin{align}
	\label{casimir_scaling}
	\mcal{D}_{\bm{u}\bm{v}}  
	= 
	\left(\mcal{D}^\mathrm{BK}_{\bm{u}\bm{v}} \right)^{2\cf/\nc}.
\end{align}
The above should not come as a surprise since, recalling that the BK equation is valid in the large-$\nc$ limit, it is a simple consequence of Casimir scaling.  This procedure allows us to obtain an improved version of BK evolution which contains finite-$\nc$ corrections by first solving the BK equation and then using Eq.~\eqref{casimir_scaling} to obtain the physical dipole in the fundamental representation. This removes most of the discrepancy between BK and JIMWLK evolution for the dipole operator \cite{Kovchegov:2008mk}. In the numerical studies in this paper we will, however, not use the BK equation as such, but rather directly solve the JIMWLK equation and compare higher Wilson line correlators obtained from the solution to the corresponding results in the GA.

As a side note, one often identifies the parametrization rapidity $y$, used to construct Wilson line correlators from a Gaussian distribution, with the evolution rapidity $Y$. While this identification does provide physical insight, it is not actually used in a practical calculation. In particular the equation~\eqref{dOdY} by itself does not determine  the dynamics, since  the rapidity (be it $y$ or $Y$) dependence of $\gamma_{\bm{u} \bm{v} }$ in Eq.~\eqref{dOdY} has to be input from the outside. The $y$ dependence of $\gamma_{\bm{u} \bm{v} }$ does not matter for physical observables since they only depend on the integral over $y$, while the $Y$ dependence has to be determined separately from an equation that actually contains the QCD dynamics.

To summarize so far, the dipole correlator whose evolution is given by the first equation in the Balitsky hierarchy, \emph{defines} the GA kernel $ \Gamma_{\bm{u} \bm{v} }$. Thus, the question of whether the GA and the full JIMWLK equation are consistent with each other, can only meaningfully be addressed by moving to the second level in the hierarchy, i.e~to the equations for the double dipole \eqref{Bal2} and the quadrupole. Even there, earlier investigations~\cite{Iancu:2011nj} already gave us some guidance as to where they can be expected to agree.

At saturation it is the virtual term of the JIMWLK Hamiltonian which drives the evolution and therefore it should not come as a surprise that a local (in transverse space) approximation, like the one in Eq.~\eqref{H_GA}, contains the relevant physics if the kernel is properly chosen. Moreover, a gauge invariant correlator, like a sum of products of traces, which involves the exchange of exactly two gluons at lowest order, can be written as a sum of dipole amplitudes in the regime of weak scattering. By definition the GA for these dipoles leads to just a mathematical identity and putting everything together we conclude that it is valid in both limiting regimes \cite{Iancu:2011nj}.

Thus there is a good chance that the GA could be accurate also in the transition region. This has turned out to be true for various observables, so let us review an example by considering 
\begin{align}
\label{O1234}
	O_{\bx\bbx\bby\by} =
 \frac{ \nc^2}{\nc^2 - 1}\,
 \bigg(Q_{\bx\bbx\bby\by}\,D_{\by\bby}
 - \frac{1}{
\nc^2}
 \,D_{\bx\bbx}
 \bigg),
 \end{align}
which appears in forward inclusive di-jet production in $pA$ collisions. The average value $\mcal{O}_{\bx\bbx\bby\by}$ of the scattering operator in Eq.~\eqref{O1234} depends on the four transverse positions  and the rapidity $Y$. To be definite, we take the simple but non-trivial configuration in which the two quarks of the quadrupole are at the same point, i.e.~$\bx=\bby$, and similarly $\bbx=\by$. Then the unique non-zero distance in this ``line'' configuration is $r \equiv r_{\bx\bbx}=r_{\bbx\bby}=r_{\bby\by}=r_{\by\bx}$, where we have introduced the compact notation $r_{\bx\by} \equiv |\bx-\by|$ for the transverse distance between two color charges. In the GA one finds that the evolution equation of $\langle Q_{\bx\by\bx\by}D_{\by\bx}\rangle$ couples only with that of $\langle D_{\bx\by} D_{\bx\by} D_{\by\bx} \rangle$ and solving a $2 \times 2$ inhomogeneous system of linear differential equations we obtain \cite{Iancu:2011nj}
\begin{align}
	\label{O1212_ave}
	\hspace{-0.8cm}
	\mcal{O}^\mathrm{GA}_{\bx\by\bx\by}(r,Y) = 
	\frac{\nc (\nc+2)}{2 (\nc+1)}\,
 	\mcal{D}(r,Y)^{\textstyle{\frac{3\nc-1}{\nc-1}}}
 	-
 	\frac{\nc (\nc-2)}{2 (\nc-1)}\,
 	\mcal{D}(r,Y)^{\textstyle{\frac{3\nc+1}{\nc+1}}}
 	-\frac{1}{
\nc^2-1}\,
 	\mcal{D}(r,Y).
 \end{align}
In order to arrive at the above, we have assumed that the GA is valid for the initial condition at the rapidity $Y_0$. It is rather important to notice that the expansion of $\mcal{O}_{\bx\by\bx\by}(r,Y)$ in Eq.~\eqref{O1212_ave} in the weak scattering regime reads
\begin{align}
	\label{op_hadron_ave}
	\mcal{O}_{\bx\by\bx\by}(r,Y) \simeq
	1 -  
	\frac{5 
\nc^2-1}{
\nc^2-1}\,\mcal{T}(r,Y)
	\quad \mathrm{for} \quad r^2\qs^2 \ll 1,
 \end{align}
that is, the lowest-order interaction involves the exchange of two gluons  or equivalently two powers of the color field $\alpha_a$. It is straightforward to verify that Eq.~\eqref{op_hadron_ave} is consistent with the weak scattering limit of Eq.~\eqref{O1234} for the configuration under consideration.

\begin{figure}
\begin{center}
\includegraphics[width=0.49\textwidth]{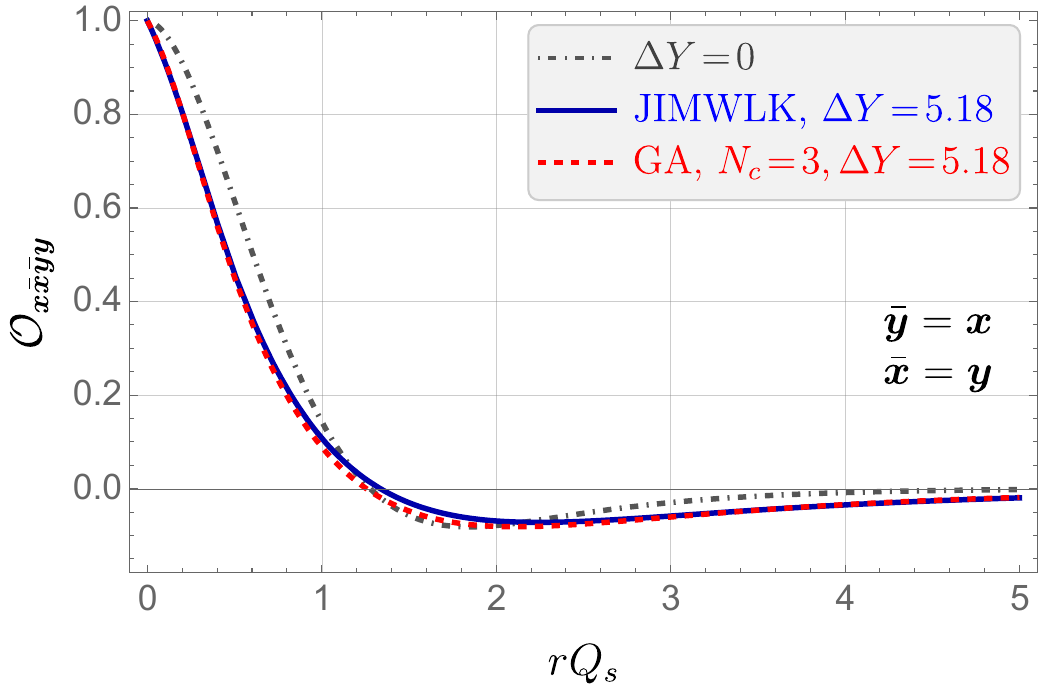}
\hspace*{0\textwidth}
\includegraphics[width=0.49\textwidth]{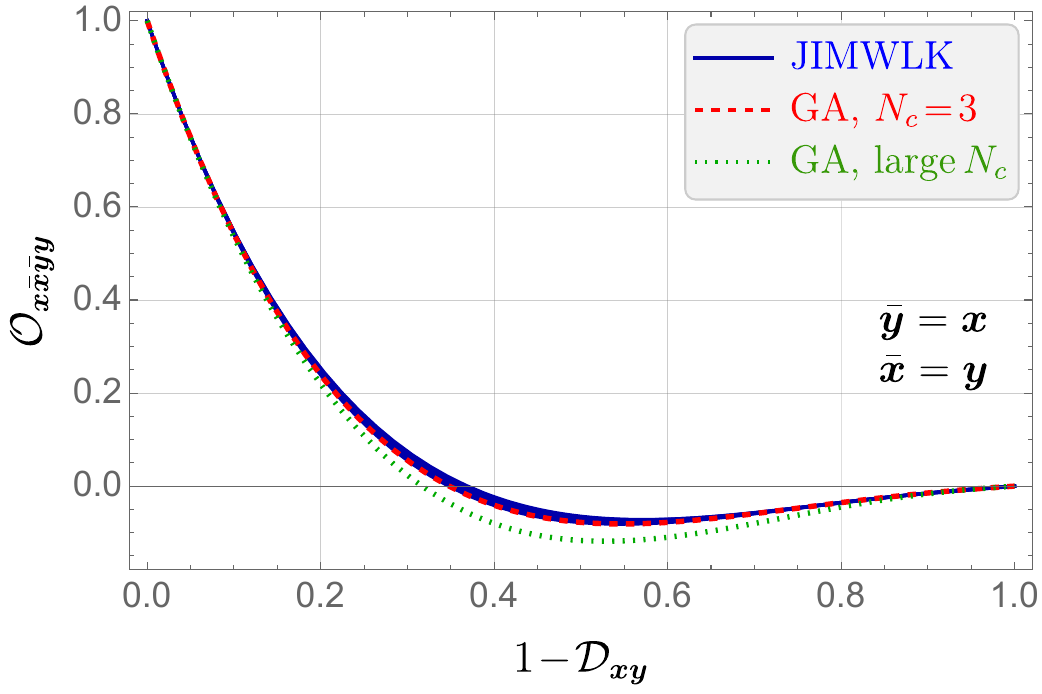}
\end{center}
\caption{\sl \small The expectation value of $O_{\bx\bbx\bby\by}$ for the ``line'' configuration using the JIMWLK equation and the GA. JIMWLK curves are built using the numerical solution given in \cite{Dumitru:2011vk}. GA curves are analytical expressions in terms of $\mcal{D}$ \cite{Iancu:2011nj}, cf.~Eq.~\eqref{O1212_ave}. Left panel: as a function of the scaling variable $r \qs$ at the initial rapidity and after evolution for $y=5.18$. Right panel: as a function of $1-\mcal{D}$ for six different values of rapidity from $y=0$ to $y=5.18$ (blue continuous lines). Figures have been adapted from \cite{Iancu:2011nj} to fit the current study.}
\label{fig:O_line}
\end{figure}

The two averages $\mcal{O}_{\bx\by\bx\by}(r,Y)$ and $\mcal{D}(r,Y)$ had already been evaluated for various values of the size $r$ by solving numerically the JIMWLK equation \cite{Dumitru:2011vk}. Using these numerical data the validity of Eq.~\eqref{O1212_ave} and thus of the GA for the particular observable in the given configuration were tested. It was found that the GA is perfect when $r \qs \ll 1$ (i.e.~when $\mcal{T}(r)\ll 1$) and $r \qs \gg 1$ (i.e.~when $\mcal{D}(r)\ll 1$) and moreover that it remains remarkably accurate even in the intermediate regime $r \qs \sim 1$, as exhibited in Fig.~\ref{fig:O_line}. Given such an outcome, not only in this example but also in other ones \cite{Alvioli:2012ba}, it is generally considered that the GA is a reliable approximation {\it globally}. We also stress that in our discussion we have not made any assumption on the number of colors $\nc$ which can be considered to be finite.

To conclude this section, the most important property of the GA to the JIMWLK equation in the present context, is that a single function, the kernel $\Gamma_{\bu\bv}(y)$ or equivalently the average dipole $S$-matrix $\mcal{D}_{\bu\bv}(Y)$, determines all the higher point correlators which start with a two-gluon exchange in the regime of weak scattering. We will always assume that the GA is valid in the initial condition at the rapidity $Y_0$, as for example this is indeed the case in the widely used MV model which is defined by a Gaussian weight-function with a specific kernel.

\section{Correlators with Four-gluon Exchange at Lowest order: Incoherent Diffraction}
\label{sec:W}

As we explained in Sect.~\ref{sec:GA}, the remarkable accuracy of the GA for arbitrary kinematics relies on its validity in the two limiting regimes of weak and strong scattering. While in the latter the GA seems to be generally robust, in the former a rather strong assumption has been made for the correlator under consideration: when expanded in a Taylor series for small coupling, its lowest order term must correspond to a two-gluon exchange or, equivalently, it must be quadratic in the gauge field $\alpha_a$. This is indeed true if we focus on inclusive multi-jet production in $\gamma A$ or $pA$ collisions, i.e.~for any process of the type discussed in Sect.~\ref{sec:GA}.

It is natural to ask what happens when we consider a correlator whose lowest order term corresponds to the exchange of four gluons. The expectation is that the GA will not be valid because it is very unlikely that an arbitrary 4-point function can be written in terms of 2-point ones due to the non-trivial color structure of QCD. This is further corroborated by the fact that after all the GA {\it is not exact} even for correlators which start with a two-gluon exchange, since small deviations are observed in the transition region. Let us consider $\mcal{O}_{\bx\by\bx\by}(r)$ defined in Sect.~\ref{sec:GA} and imagine that we start to move away from the weak scattering regime to approach the saturation momentum. It is not hard to be convinced that the difference between the JIMWLK and GA values should be attributed to $n$-point correlations, with $n > 2$, which are not captured by the full dipole $\mcal{D}(r)$.

\begin{figure}
\begin{center}
\includegraphics[width=0.6\textwidth]{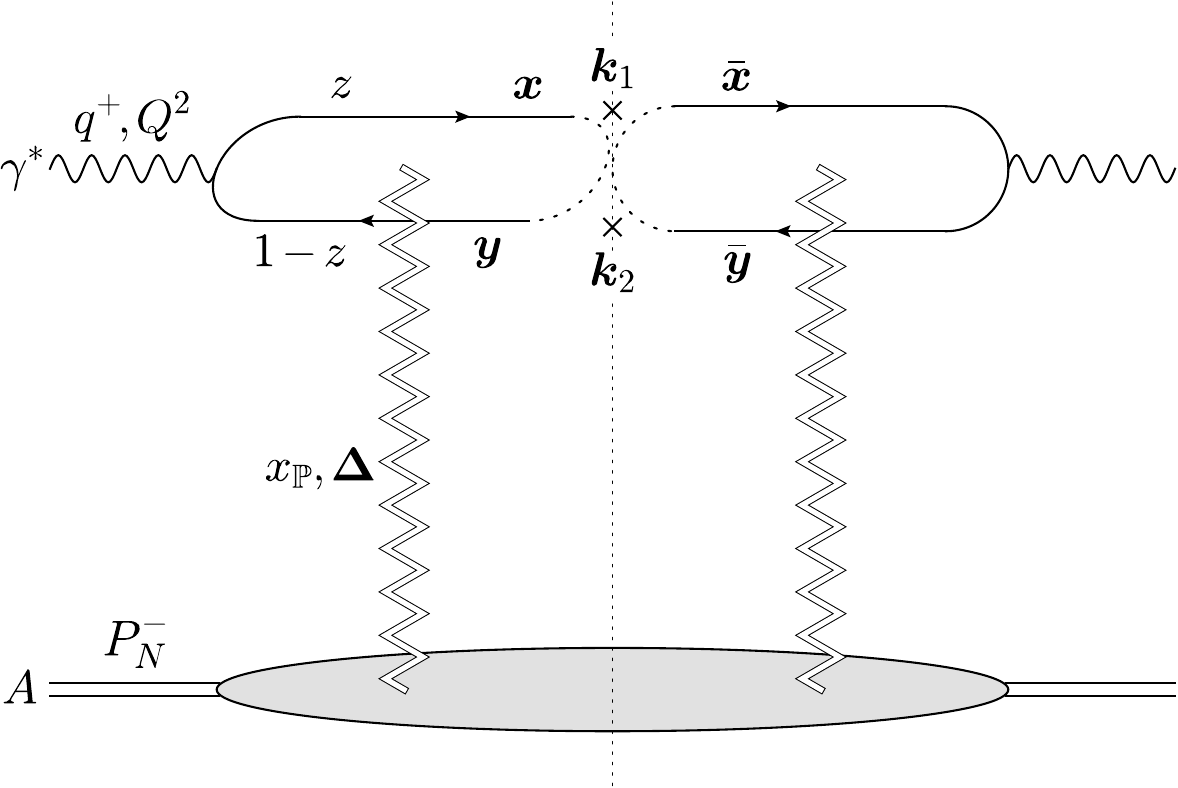}
\end{center}
\caption{\sl \small Incoherent diffractive production of two forward jets with transverse momenta $\bm{k}_1$ and $\bm{k}_2$ in $\gamma^* A$ collisions. A minus longitudinal momentum $\xP P_N^-$ and a transverse momentum $\bm{\Delta}=\bk_1+\bk_2$ are transferred from the nucleus to the projectile system.}
\label{fig:diff}
\end{figure}

Our interest to correlators whose perturbative expansion starts with an exchange of more than two gluons is not just academic, since interactions of this kind appear in certain types of diffractive processes, i.e.~when there is a large region void of particles in the final state or in other words, a rapidity gap.  In particular, let us consider diffractive dijet production in photon-nucleus interactions, where the photon could be either (quasi-)real like in ultra-peripheral heavy-ion collisions or virtual like in electron-proton deep inelastic scattering (DIS). To be specific, we focus on DIS as shown in Fig.~\ref{fig:diff}. The right moving virtual photon has four-momentum $(q^+, - Q^2/2q^+, \bm{0})$ and the left moving nucleus $(0,P_N^-,\bm{0})$ per nucleon, where the mass of the latter has been neglected. The virtual photon splits into a pair of a quark and an antiquark which carry fractions $z$ and $1-z$ of its longitudinal momentum $q^+$ and have transverse coordinates $\bx$ and $\by$ respectively and all these projectile variables remain unaltered during the interaction with the nucleus. The multiple, eikonal, scattering is described by the Wilson lines $V_{\bx}$ and $V_{\by}^{\dagger}$ and in a diffractive process we require that the multi-gluon exchange, represented by a zig-zag line in Fig.~\ref{fig:diff}, is colorless. This implies that the color lines must ``close'' separately in the direct amplitude (DA) and in the complex conjugate amplitude (CCA) and thus diffraction involves the correlator $\langle T_{\bx\by} T_{\bby\bbx} \rangle$, with $\bbx$ and $\bby$ the transverse coordinates in the CCA. When the target nucleus remains intact one speaks about coherent diffraction and normally it corresponds to an independent average in the DA and in the CCA, that is one takes $\langle T_{\bx\by} \rangle \langle T_{\bby\bbx}\rangle$. The difference 
\begin{align}
	\label{W}
	\mcal{W}_{\bx\by\bby\bbx} \equiv 
	\langle T_{\bx\by} T_{\bby\bbx} \rangle
	- \langle T_{\bx\by}  \rangle  \langle T_{\bby\bbx} \rangle
	 = 
	\langle D_{\bx\by} D_{\bby\bbx} \rangle 
	- \langle D_{\bx\by}  \rangle  \langle D_{\bby\bbx} \rangle
\end{align}
requires a color and momentum exchange between different constituents of the target which separately belong to the DA and the CCA in Fig.~\ref{fig:diff} and as a consequence leads to target breakup. We shall refer to it as incoherent diffraction \cite{Marquet:2010cf,Mantysaari:2019hkq,Rodriguez-Aguilar:2023ihz}. To produce two partons with the desired  transverse momenta one must take Fourier transforms in which $\bk_1$ and $\bk_2$ are respectively dual to $\bx-\bbx$ and $\by - \bby$. Assuming that $z \sim 1-z$ (so that neither $z$ nor $1-z$ is small), this $q\bar{q}$ contribution dominates the incoherent diffractive process when the transverse momentum imbalance $|\bk_1 + \bk_2|$ of the dijet is of the order of $|\bk_1| \equiv k_{1\perp}$ and $|\bk_2|\equiv k_{2\perp}$, while one must take into account higher order (in $\as$) partonic fluctuations of the virtual photon  when  $|\bk_1 + \bk_2| \ll k_{1\perp} \simeq k_{2\perp}$ \cite{Rodriguez-Aguilar:2023ihz,Rodriguez-Aguilar:2024efj}. The diffractive correlator in Eq.~\eqref{W} must be evaluated at the rapidity $Y_{\mathbb P} \equiv \ln 1/x_{\mathbb P}$, where $x_{\mathbb P}$ is the fraction of the longitudinal momentum $P_N^-$ transferred from the target to the projectile in order to put the latter ``on-shell''. One readily finds
\begin{align}
	\label{xP}
	x_{\mathbb{P}} = \frac{1}{s}
	\left(\frac{k_{1\perp}^2}{z} + \frac{k_{2\perp}^2}{1-z} + Q^2
	\right), 
\end{align}
with $s = 2 q^+ P_N^-$ the center of mass energy squared for the photon-nucleon system, therefore the rapidity gap $\YP$ generally depends on the kinematics of the outgoing partons. Still, with $z \sim 1-z$ and with $k_{i\perp}^2$ at most of the order of $Q^2$, $x_{\mathbb P}$ becomes quasi-independent of the final state and is approximately equal to $Q^2/s$. For simplicity we shall adopt such an approximation, since dealing with the details of the exact kinematics is not relevant to our main goals. Last, but not least, let us observe that $\mcal{W}$ is equal to the variance of the scattering amplitude and as such it is by definition determined by fluctuations in the target weight-function. These fluctuations can be of various types, for example they can arise due to the geometry of the system. It is well-known that ``hot-spots'' give a large cross section at low $|t|$, where $|t|=\Delta_{\perp}^2$ \cite{Lappi:2010dd,Mantysaari:2016ykx,Demirci:2022wuy} is the square of the momentum transferred from the target to the projectile. Particle fluctuations in the target would also contribute, but their analysis goes beyond the scope of our work since they are due to loops of Pomerons which are not contained in JIMWLK \cite{Iancu:2004iy,Iancu:2005nj,Kovner:2005nq,Hatta:2006hs,Bzdak:2015eii,McLerran:2015qxa,Le:2021afn,Le:2021qwx}. Here we shall only be concerned with color fluctuations which do exist both in the MV model and in JIMWLK evolution.

Let us now return to the correlator in Eq.~\eqref{W} and notice that terms linear in $T$ are absent, hence the lowest order term is expected to be quartic in the gauge field $\alpha_a$. The calculation in the GA for arbitrary coordinates and at finite $\nc$ is a straightforward exercise \cite{Dominguez:2008aa}. For convenience we give the exact expression in Appendix \ref{app:W}, since we shall repeatedly use it to discuss limiting cases of interest. To gain some insight, let us first focus on the limit of weak scattering, where we also expect the largest deviations of the GA from the JIMWLK result according to our considerations above. Using Eq.~\eqref{appeq:DD} we find that 
\begin{align}
	\label{W_GA_weak}
	\mcal{W}_{\bx\by\bby\bbx}^\mathrm{GA} \simeq
	\frac{1}{2 (\nc^2-1)}\,
	\left( 
	\mcal{T}_{\bx\bby} + \mcal{T}_{\by\bbx} - \mcal{T}_{\bx\bbx} - \mcal{T}_{\by\bby} 
	\right)^2
	\quad \mathrm{for} \quad \mcal{T} \ll 1,
\end{align}
which is indeed of order $\mcal{T}^2 \sim \alpha_a^4$. It goes without saying that the amplitudes appearing on the r.h.s.~are already averaged over all possible target configurations. Even though $\mcal{W}_{\bx\by\bby\bbx}$ is suppressed by $1/\nc^2$, when compared for example to the disconnected piece $\langle T_{\bx\by}  \rangle  \langle T_{\bby\bbx} \rangle$ (which, as already pointed out, determines coherent diffraction), we stress that this is the leading order term in incoherent diffraction. More generally, there are also other correlators that are of leading order in $\nc$, but do not contain interactions with two-gluon exchanges. In fact, we shall discuss such an example in Sect.~\ref{sec:break}.
 
One of the main tasks in this work is to explore the non-validity of the GA for the incoherent diffractive correlator in Eq.~\eqref{W}. First we shall show analytically that the weak field expression in Eq.~\eqref{W_GA_weak} is not consistent with the linearized Balitsky equations. Then we shall proceed to a numerical calculation of Eq.~\eqref{W} using the Langevin form of the JIMWLK equation and we will demonstrate that that the result is sizeably different from the GA one in a wide kinematic regime, both in rapidity and in transverse coordinate space.  

\section{Breakdown of the GA for weak scattering}
\label{sec:break}

In this section we will show that the GA for the incoherent diffractive correlator $\mcal{W}_{\bx\by\bby\bbx}$ is not compatible with the Balitsky equations for weak scattering. In order to simplify the calculation we shall reduce the number of independent coordinates, and hence of possible sizes, by considering the less general case of $\mcal{W}_{\bx\by\by\bbx}$. This still depends on three independent variables, assuming as usual that the target is invariant under translations and rotations in the two-dimensional transverse plane. Trivially, Eq.~\eqref{W_GA_weak} becomes
\begin{align}
	\label{W_GA_1332_weak}
	\mcal{W}_{\bx\by\by\bbx}^\mathrm{GA} \simeq
	\frac{1}{2 (\nc^2-1)}\,
	\left( 
	\mcal{T}_{\bx\bbx} - \mcal{T}_{\bx\by} - \mcal{T}_{\bbx\by} 
	\right)^2
	\quad \mathrm{for} \quad \mcal{T} \ll 1.
\end{align} 

If the GA was valid for $\mcal{W}_{\bx\by\by\bbx}$, then Eq.~\eqref{W_GA_1332_weak} should hold at an arbitrary rapidity and not only at $Y_0$ where the initial condition is imposed, so let us examine the evolution of such an expression. When we differentiate its r.h.s.~with respect to $Y$, it is sufficient and self-consistent to use the BFKL equation for $\mcal{T}$, since Eq.~\eqref{W_GA_1332_weak} is correct only to order quartic order in the gauge field\footnote{It is useful to recall that the JIMWLK evolution of a correlator of $n$ fields, involves $m$ fields with $m \geq n$. To our purposes here it will be enough to keep only the $m=n$ terms and neglect all unitarity corrections, i.e.~terms with $m>n$.}. We find
\begin{align}
	\label{dW_GA_1332_weak_dY}
	\hspace{-0.4cm}
	\frac{\dif\mcal{W}_{\bx\by\by\bbx}^\mathrm{GA}}{\dif Y}\simeq
	\frac{1}{\nc^2 - 1}\,
	\frac{\abar}{2\pi}
	&\left( 
	\mcal{T}_{\bx\bbx} - \mcal{T}_{\bx\by} - \mcal{T}_{\bbx\by} 
	\right)
	\bigg[
	\int_{\bm{z}}
	\mcal{M}_{\bx\bbx\bm{z}}
	\left( 
 	\mcal{T}_{\bx\bm{z}} + \mcal{T}_{\bbx\bm{z}}  - \mcal{T}_{\bx\bbx}
	\right)
	\nn
	&-\int_{\bm{z}}
	\mcal{M}_{\bx\by\bm{z}}
	\left( 
 	\mcal{T}_{\bx\bm{z}} + \mcal{T}_{\by\bm{z}}  - \mcal{T}_{\bx\by}
	\right)
	-\int_{\bm{z}}
	\mcal{M}_{\bbx\by\bm{z}}
	\left( 
 	\mcal{T}_{\bbx\bm{z}} + \mcal{T}_{\by\bm{z}}  - \mcal{T}_{\bbx\by}
	\right)
	\bigg].
\end{align}

On the other hand we do know how to write the proper evolution equation for $\mcal{W}_{\bx\by\by\bbx}$ by making use of the two Balitsky equations \eqref{Bal1} and \eqref{Bal2}. We can split the final result in two terms according to
\begin{align}
	\label{dW_1332_dY_split}
	\frac{\dif\mcal{W}_{\bx\by\by\bbx}}{\dif Y} = 
	\frac{\dif\mcal{W}_{\bx\by\by\bbx}^\mathrm{D}}{\dif Y}	+
	\frac{\dif\mcal{W}_{\bx\by\by\bbx}^\mathrm{ND}}{\dif Y} 
\end{align}
where
\begin{align}
	\label{dW_1332_dY_D}
	\hspace{-0.7cm}
	\frac{\dif\mcal{W}_{\bx\by\by\bbx}^\mathrm{D}}{\dif Y}
	=\,\, &
	\frac{\abar}{2\pi} \int_{\bz}
	\mcal{M}_{\bx\by\bz} 
	\left[
	\langle D_{\bx\bz} D_{\bz \by} D_{\by\bbx} \rangle
	- \langle D_{\bx\bz} D_{\bz \by} \rangle \langle D_{\by\bbx} \rangle
	-\langle D_{\bx\by} D_{\by\bbx} \rangle
	+\langle D_{\bx\by}  \rangle  \langle D_{\by\bbx} \rangle
	\right]
	\nn
	+ &\, 
	\frac{\abar}{2\pi} \int_{\bz}
	\mcal{M}_{\bbx\by\bz} 
	\left[
	\langle D_{\bx\by} D_{\by\bz} D_{\bz \bbx} \rangle
	- \langle D_{\bx\by} \rangle \langle D_{\by\bz} D_{\bz \bbx} \rangle
	-\langle D_{\bx\by} D_{\by\bbx} \rangle
	+\langle D_{\bx\by}  \rangle  \langle D_{\by\bbx} \rangle
	\right]
\end{align}
and 
\begin{align}
	\label{dW_1332_dY_ND}
	\frac{\dif\mcal{W}_{\bx\by\by\bbx}^\mathrm{ND}}{\dif Y}
	= \frac{1}{2 \nc^2}\,\frac{\abar}{2\pi}
	\int_{\bz}
	(\mcal{M}_{\bx\bbx\bz}-
	\mcal{M}_{\bx\by\bz}-\mcal{M}_{\bbx\by\bz} )
	\langle
	S_{\bx\bz\by\bbx\bz\by} +S_{\bx\by\bz\bbx\by\bz}
	- 2 D_{\bx\bbx}
	\rangle.
\end{align}
The ``dipole'' term, indexed with D, received contributions from both Balitsky equations and evidently involves only dipoles. As discussed in Sect.~\ref{sec:JIM}, in JIMWLK evolution at large $\nc$ the average of a product reduces to the product of the averages. As a consequence, the leading in $\nc$ term of the scattering correlator in the r.h.s.~of Eq.~\eqref{dW_1332_dY_D} should be proportional to $1/\nc^2$, in agreement with the $\nc$-dependence of $\mcal{W}_{\bx\by\by\bbx}$ on the left hand side (l.h.s.). The ``non-dipole'' term, indexed with ND, receives contributions only from the second Balitsky equation, i.e.~the one for $\langle D_{\bx\by} D_{\by\bbx} \rangle$, and it involves sextupoles. Now the necessary factor proportional to $1/\nc^2$ is explicit in the r.h.s.~of Eq.~\eqref{dW_1332_dY_ND}, whereas the correlators appearing there in the integrand are of $\mcal{O}(1)$ in the $\nc$ counting.

As expected, the evolution equation for $\mcal{W}_{\bx\by\by\bbx}$ doesn't close: only the last two terms in each of the two lines in Eq.~\eqref{dW_1332_dY_D}, i.e.~those originating from virtual diagrams, can be rewritten in terms of $\mcal{W}_{\bx\by\by\bbx}$. Thus one would have to deal with the full hierarchy as we explained in Sect.~\eqref{H_JIM}, but this is not really necessary for what we want to achieve. To make a comparison of the above with the GA in Eq.~\eqref{dW_GA_1332_weak_dY}, we should take the limit of weak scattering in Eqs.~\eqref{dW_1332_dY_D} and \eqref{dW_1332_dY_ND}. For the dipole term this is rather straightforward, since it suffices to write $D=1-T$ and neglect terms of order $T^3$. Then one readily discovers that all terms can be combined  in such a way to form $\mcal{W}$, more precisely
\begin{align}
	\label{dW_1332_dY_D_lin}
	\frac{\dif\mcal{W}_{\bx\by\by\bbx}^\mathrm{D}}{\dif Y}
	\simeq
	\frac{\abar}{2\pi} \int_{\bz}
	\big[&\mcal{M}_{\bx\by\bz} 
	\left(
	\mcal{W}_{\bx\bz\by\bbx}+ \mcal{W}_{\bz\by\by\bbx} -\mcal{W}_{\bx\by\by\bbx}
	\right)
	\nn
	&+ 
	\mcal{M}_{\bbx\by\bz} 
	\left(
	\mcal{W}_{\bx\by\by\bz}+ \mcal{W}_{\bx\by\bz\bbx} -\mcal{W}_{\bx\by\by\bbx}
	\right)
	\big].
\end{align}
Still, it is doubtful that the sextupole correlator appearing in the non-dipole term in Eq.~\eqref{dW_1332_dY_ND} can be written in terms of $\mcal{W}$, even if we expand to quartic order in the field $\alpha_a$. 

\subsection{One evolution step: general considerations}
\label{sub:evol}

In the remaining part of the current section we shall focus on performing only one evolution step for the diffractive correlator $\mcal{W}$. Assuming, as usual, that the GA describes accurately the initial condition, both the sextupole on the r.h.s.~of Eq.~\eqref{dW_1332_dY_ND} and the correlator $\mcal{W}$ on the r.h.s.~of Eq.~\eqref{dW_1332_dY_D_lin} are completely determined by the dipole amplitude $\mcal{T}$. The dipole contribution is straightforward and inserting Eq.~\eqref{W_GA_weak} in Eq.~\eqref{dW_1332_dY_D_lin} we get
\begin{align}
	\label{dW_1332_dY_D_T}
	\frac{\dif\mcal{W}_{\bx\by\by\bbx}^\mathrm{D}}{\dif Y}
	\simeq 
	\frac{1}{\nc^2 - 1}\,
	\frac{\abar}{2\pi} 
	\int_{\bz}
	\mcal{M}_{\bx\by\bz} 
	\big[ &
	(\mcal{T}_{\by\bz} - \mcal{T}_{\bbx\bz})^2 
	+ (\mcal{T}_{\by\bz} - \mcal{T}_{\bbx\bz})
	(\mcal{T}_{\bx\bbx} +\mcal{T}_{\bbx\by} -\mcal{T}_{\bx\by})
	\nn
	&+\mcal{T}_{\bbx\by} ( \mcal{T}_{\bx\bbx} - \mcal{T}_{\bx\by})
	\big] + \bx \leftrightarrow \bbx.
\end{align}
The dilute limit of the desired sextupole in the GA is calculated in Appendix \ref{app:W} and given in Eq.~\eqref{appeq:sext_dip_4g}. It is symmetric under the exchange $\bm{z} \leftrightarrow \by$, hence the two sextupoles in the integrand in Eq.~\eqref{dW_1332_dY_ND} contribute equally and we finally arrive at
\begin{align}
	\label{dW1332_dY_ND_T}
	\hspace{-0.4cm}
	\frac{\dif \mcal{W}_{\bx\by\by\bbx}^\mathrm{ND}}{\dif Y}= 
	\frac{1}{\nc^2 - 1}\, \frac{\abar}{2\pi} 
	\int_{\bm{z}} &
	(\mcal{M}_{\bx\bbx\bz}-\mcal{M}_{\bx\by\bz}-\mcal{M}_{\bbx\by\bz})
	\big[
	(\mcal{T}_{\bx\bz} - \mcal{T}_{\by\bm{z}}) 
	(\mcal{T}_{\bbx\bz} - \mcal{T}_{\by\bm{z}})
	\nn
	&
	+2 \mcal{T}_{\bx\bbx} \mcal{T}_{\by\bm{z}}
	-\mcal{T}_{\bx\by}
	(\mcal{T}_{\bbx\bz}  + \mcal{T}_{\by\bz})
	-\mcal{T}_{\bbx\by}
	(\mcal{T}_{\bx\bz}  + \mcal{T}_{\by\bz}) 
	+\mcal{T}_{\bx\by} \mcal{T}_{\bbx\by}
	\big]. 
\end{align}
The dipole and non-dipole contributions given in Eqs.~\eqref{dW_1332_dY_D_T} and \eqref{dW1332_dY_ND_T} are proportional to the exact same color factor. One could put the two terms together, but the final expression wouldn't simplify considerably. In particular, the dipole kernel $\mcal{M}_{\bx\bbx\bz}$ appears only in the non-dipole term and thus it would retain its form in the sum. But despite the fact that the color factor in Eqs.~\eqref{dW_1332_dY_D_T} and \eqref{dW1332_dY_ND_T} is also the same with the one appearing in the equation derived in the GA, cf.~\eqref{dW_GA_1332_weak_dY}, a direct inspection shows that the two results are different. This is already an indication, if not a proof, that the GA breaks down for the incoherent diffractive correlator under study. 

Before closing this subsection, we would like to make a couple of pertinent comments. (i) At a first glance one may think that the violation of the GA in the example under consideration should not come as a surprise, since in order to evolve the GA expression in Eq.~\eqref{W_GA_1332_weak} we used only the first Balitsky equation \eqref{Bal1}, whereas in the ``exact'' treatment we also used the second Balitsky equation \eqref{Bal2} which encodes non-trivial color exchanges and as a consequence involves a more complicated color configuration, the sextupole. However the explanation cannot be that simple, because the general expression for $\langle D_{\bx\by} D_{\by\bbx} \rangle - \langle D_{\bx\by} \rangle \langle D_{\by\bbx} \rangle$ and its weak-scattering limit shown in Eq.~\eqref{W_GA_1332_weak} capture aspects of the full Balitsky hierarchy. In fact, if we consider $\mcal{W}_{\bx\by\by\bx}$, i.e.~the more special configuration in which we identify $\bbx$ with $\bx$, we find that Eqs.~\eqref{dW_1332_dY_D_T} and \eqref{dW1332_dY_ND_T} combine in an apparently non-trivial way to give the GA result in Eq.~\eqref{dW_GA_1332_weak_dY}. (ii) This last fact does not mean at all that the GA is accurate for $\mcal{W}_{\bx\by\by\bx}$; it is so only for just one evolution step. More generally one should evaluate the r.h.s.~in Eqs.~\eqref{dW_1332_dY_D} and \eqref{dW_1332_dY_ND} exactly (and not in the GA) and we have already seen that the correlator $\mcal{W}_{\bx\by\by\bz}$ appearing in the r.h.s.~of Eq.~\eqref{dW_1332_dY_D_lin} (which is the weak scattering limit of the dipole contribution in Eq.~\eqref{dW_1332_dY_D}) does not satisfy the GA. The same conclusion is expected to hold for $\mcal{S}_{\bx\bm{z}\by\bx\bm{z}\by}-1$ which appears in the corresponding non-dipole contribution. This particular correlator and its less special version $\mcal{S}_{\bx\bm{z}\by\bbx\bm{z}\by}-\mcal{D}_{\bx\bbx}$ start at order $\alpha_a^4$ and are of leading order in $\nc$, but still, in the physics problem under study, there is an explicit prefactor $1/\nc^2$ as manifest in Eq.~\eqref{dW_1332_dY_ND}. In Sect.~\ref{sec:num} we shall confirm numerically that $\mcal{W}_{\bx\by\by\bx}$ (and $\mcal{S}_{\bx\bm{z}\by\bx\bm{z}\by}-1$) cannot be computed accurately using the GA.
 
\subsection{One evolution step: an analytical calculation}
\label{sub:analytical}

As we have seen the results of one evolution step using the GA, cf.~Eq.~\eqref{dW_GA_1332_weak_dY}, and using the Balitsky equations, cf.~Eqs.~\eqref{dW_1332_dY_D_T} and Eq.~\eqref{dW1332_dY_ND_T}, do not have the same functional form. The purpose of the present subsection is to consider a certain expression for the averaged dipole amplitude $\mcal{T}$, perform the integration over the transverse coordinates of the daughter gluon and explicitly verify that the results are indeed different.  Moreover, this will provide us with a first indication about the difference between the two results at the quantitative level.

Perhaps the most reasonable choice for $\mcal{T}$ would have been that corresponding to the MV model but in order to simplify the analytical calculation we shall instead consider the GBW model
\begin{align}
	\label{T_GBW}
	\mcal{T} = 1 -  \exp\bigg(\!-\frac{r^2 \qs^2}{4}\bigg)
	\quad\Longrightarrow \quad
	\mcal{T} \simeq \frac{r^2 \qs^2}{4}
	\quad \mathrm{for} \quad
	r^2  \qs^2 \ll 1.
\end{align}
We need only the weak scattering limit given by the approximate equality in the above. When we insert it into Eq.~\eqref{dW_GA_1332_weak_dY} we readily see that the integration diverges logarithmically for large $z$ since the dipole kernel falls like $1/z^4$. This is hardly a problem, since we know that unitarity corrections will eventually lead to a finite result and in order to take this into account it suffices to introduce an appropriate cutoff. A convenient way to do so is to perform the integration in a large disk with radius $R \sim 1/\qs$ and with center located at the vicinity of the three color sources at $\bx,\bbx$ and $\by$. Such a calculation is done in Appendix \ref{app:int} and gives the change in the GA correlator in one evolution step as
\begin{align}
	\label{dW_GA_1332_GBW}
	\frac{\dif\mcal{W}_{\bx\by\by\bbx}^\mathrm{GA}}{\dif Y}\simeq
	\frac{\abar}{\nc^2 - 1}\,
	\frac{\qs^4}{16}
	\left( 
	r_{\bx\bbx}^2 - r_{\bx\by}^2 - r_{\bbx\by}^2
	\right)\!
	\left(
	r_{\bx\bbx}^2 \ln \frac{R^2}{r_{\bx\bbx}^2}
	-r_{\bx\by}^2 \ln \frac{R^2}{r_{\bx\by}^2}
	-r_{\bbx\by}^2 \ln \frac{R^2}{r_{\bbx\by}^2}
	\right),
\end{align}
with the terms neglected being of the order of $r^6 \qs^4/ R^2 \sim r^6 \qs^6$, i.e.~being power suppressed, where $r$ stands for any of the three intercharge distances. This means that, for a given $R$, the above contains all terms of the order of $r^4 \qs^4$ and not only those which are logarithmically enhanced.

Similarly, we wish to compute one step of the JIMWLK evolution for the same operator, i.e. the r.h.s.~of Eqs.~\eqref{dW_1332_dY_D_T} and~\eqref{dW1332_dY_ND_T} for the GBW amplitude in Eq.~\eqref{T_GBW}. It is important to notice that the amplitudes involving the radiated gluon at $\bz$ appear combined in such a way that the integration is diverging at most logarithmically in $R$ and not as a power. The details of the calculation are again left for Appendix \ref{app:int} and here we present only the final results. The dipole and non-dipole contributions respectively give
\begin{align}
	\label{dW_1332_GBW_D}
	\hspace{-0.9cm}
	\frac{\dif\mcal{W}_{\bx\by\by\bbx}^\mathrm{D}}{\dif Y}
	\simeq 
	\frac{\abar}{\nc^2 - 1}\,
	\frac{\qs^4}{16}
	\left[r_{\bx\by}^2 r_{\bbx\by}^2
	\left(\ln \frac{R^2}{r_{\bx\by}^2}
	+\ln \frac{R^2}{r_{\bbx\by}^2}
	+2
	\right)
	-\left( r_{\bx\bbx}^2 - r_{\bx\by}^2 - r_{\bbx\by}^2 \right)^2
		\right],
\end{align}
\begin{align}
	\label{dW_1332_GBW_ND}
	\hspace{-1cm}
	\frac{\dif\mcal{W}_{\bx\by\by\bbx}^\mathrm{ND}}{\dif Y}
	\simeq 
	\frac{\abar}{\nc^2 \!-\! 1}
	\frac{\qs^4}{32}
	\bigg[
	\big(
	r_{\bx\bbx}^2 \!-\! r_{\bx\by}^2 \!-\! r_{\bbx\by}^2
	\big)\!
	\bigg(\!
	r_{\bx\bbx}^2 \ln \frac{R^2}{r_{\bx\bbx}^2}
	\!-\!r_{\bx\by}^2 \ln \frac{R^2}{r_{\bx\by}^2}
	\!-\!r_{\bbx\by}^2 \ln \frac{R^2}{r_{\bbx\by}^2}
	\bigg)\!
	+4 r_{\bx\by}^2 r_{\bbx\by}^2
	\bigg],
\end{align}
with corrections which are power suppressed as in Eq.~\eqref{dW_GA_1332_GBW}. A simple inspection reveals that the sum of the last two is not equal to the GA result in Eq.~\eqref{dW_GA_1332_GBW}. In fact the disagreement appears already at the level of the logarithmic term on which we now focus. To this end, let us further assume that all three intercharge distances are of the same order, say $r^2$. Then, to logarithmic accuracy, we can replace them with $r^2$ in the argument of all the logarithms (and only there). It is also more instructive and convenient to use the sizes $r_{\bx\by}$, $r_{\bbx\by}$ and the angle $\theta$ between the two corresponding vectors as independent variables, i.e.~we replace the dependence on $r_{\bx\bbx}^2$ by making use of 
\begin{align}
	\label{theta}
	r_{\bx\bbx}^2 = r_{\bx\by}^2+r_{\bbx\by}^2 - 2 r_{\bx\by} r_{\bbx\by} \cos \theta.
\end{align}
Then in the GA we find
\begin{align}
	\label{dW_GA_1332_log}
	\frac{\dif\mcal{W}_{\bx\by\by\bbx}^\mathrm{GA}}
	{\dif Y}\simeq
	\frac{\abar}{\nc^2 - 1}\,
	\frac{\qs^4}{4}\,
	r_{\bx\by}^2 r_{\bbx\by}^2 
	\cos^2\mkern-2mu\theta\,
	\ln \frac{R^2}{r^2}
	\qquad \mathrm{(leading \,\, log)},
\end{align}
while the JIMWLK result reads
\begin{align}
	\label{dW_1332_log}
	\frac{\dif\mcal{W}_{\bx\by\by\bbx}}{\dif Y}\simeq
	\frac{\abar}{\nc^2 - 1}\,
	\frac{\qs^4}{4}\,
	r_{\bx\by}^2 r_{\bbx\by}^2 
	\frac{1+ \cos^2\mkern-2mu\theta}{2}\,
	\ln \frac{R^2}{r^2}
	\qquad \mathrm{(leading \,\, log)},
\end{align} 
where the dipole and non-dipole contributions led correspondingly to the 1 and $\cos^2\mkern-2mu\theta$ terms appearing in the fraction in Eq.~\eqref{dW_1332_log}. While the two results exhibit the same functional dependence on the sizes $r_{\bx\by}^2$ and $r_{\bbx\by}^2$, the angular one is different. They are equal only for very special configurations that have $\theta=0$ or $\theta=\pi$, i.e.~when all the charges lie on the same line\footnote{This is valid only for the leading logarithmic term. The complete results in Eqs.~\eqref{dW_GA_1332_GBW}, \eqref{dW_1332_GBW_D} and \eqref{dW_1332_GBW_ND} are in agreement if and only if $\bbx=\bx$.}. On the contrary, the difference maximizes when $\theta=\pi/2$, i.e.~when the charges are located on the vertices of a right triangle. In this case both expressions take their minimum value, but while the GA result vanishes, the JIMWLK one remains non-zero.

Perhaps the most important property to observe in Eqs.~\eqref{dW_GA_1332_log} and \eqref{dW_1332_log} is that the JIMWLK result is systematically larger than the GA one. It is not obvious how this will be transmitted to the cross section for incoherent diffractive dijet production, since we need to compute the more general correlator $\mcal{W}_{\bx\by\bby\bbx}$ and then perform the required Fourier transforms. However, we cannot help but point out that if we average in the angle $\theta$ the JIMWLK answer is 50\% larger.

\begin{figure}
\begin{center}
\includegraphics[width=0.505\textwidth]{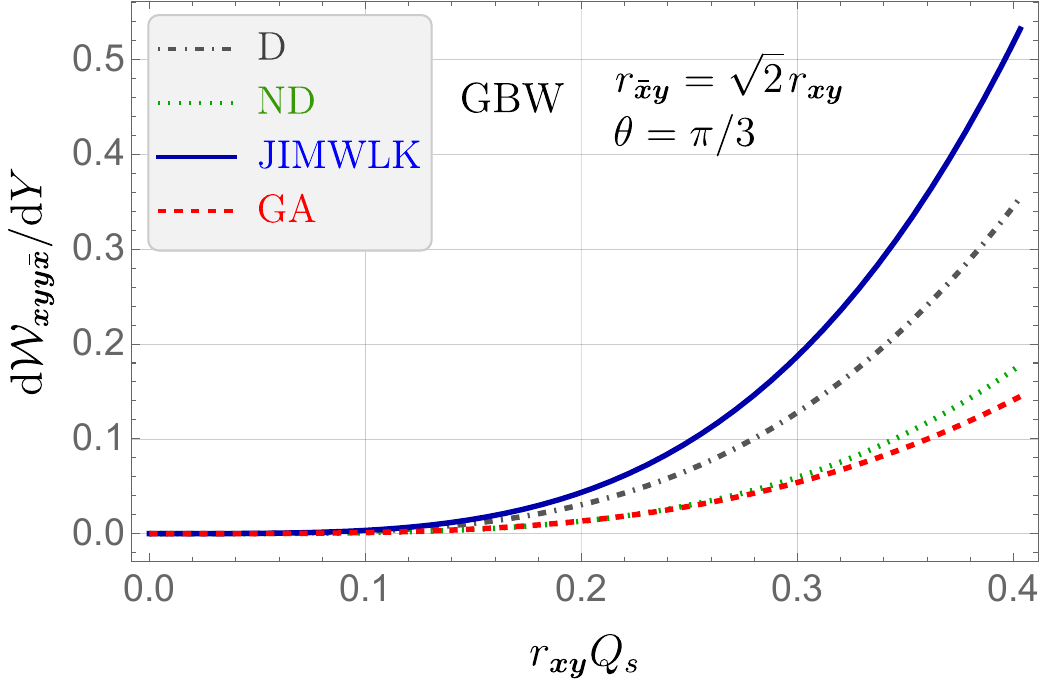}
\hspace*{0\textwidth}
\includegraphics[width=0.475\textwidth]{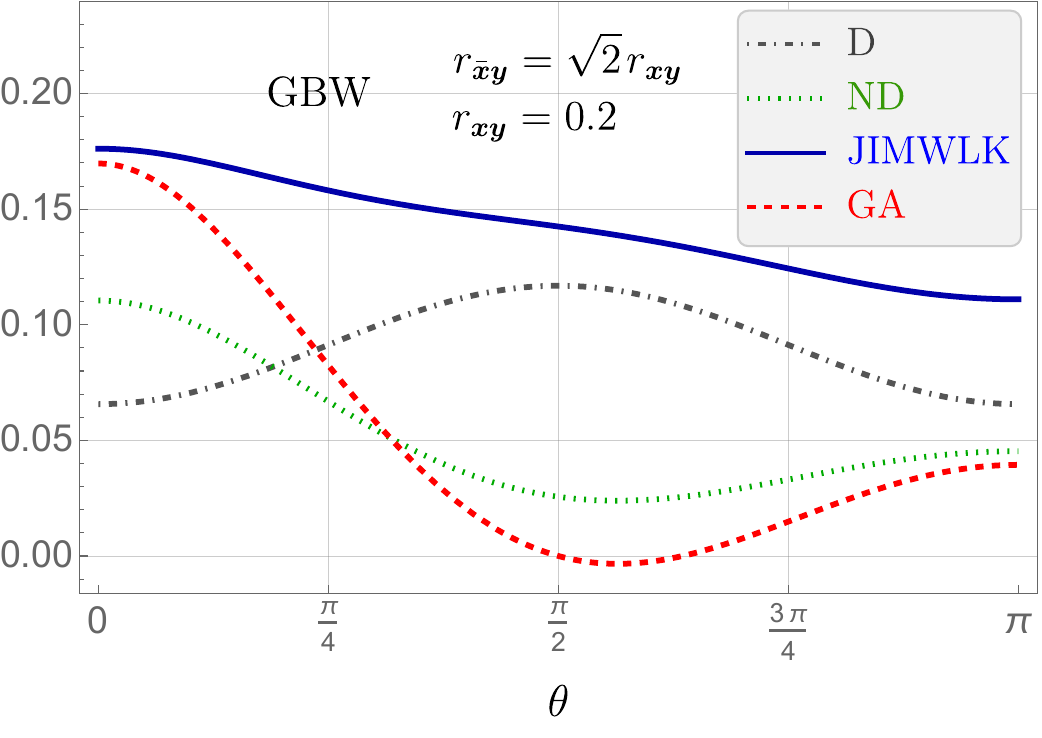}
\end{center}
\caption{\sl \small Analytic results, cf.~Eqs.~\eqref{dW_GA_1332_GBW}, Eq.~\eqref{dW_1332_GBW_D} and \eqref{dW_1332_GBW_ND}, for the one step rapidity evolution of the incoherent diffractive correlator $\mcal{W}_{\bx\by\by\bbx}$ for $r_{\bbx\by} =\sqrt{2}\mkern2mu r_{\bx\by}$ as a function of the scaled size $r_{\bx\by} \qs$ for fixed angle $\theta=\pi/3$ (left panel) and as a function of $\theta$ for fixed $r_{\bx\by}=0.2$ (right panel) in the weak scattering regime, where $\theta$ is the angle between $\br_{\bx\by}$ and $r_{\bbx\by}$. The initial condition is given by the (linearized) GBW model with $\qs^2=2\mkern1.5mu\mathrm{GeV}^2$ and the cutoff is $R=2/\qs$. A factor $\abar/[16(\nc^2-1)]$ has been neglected.}
\label{fig:W_GBW}
\end{figure}

As an illustration, in Fig.~\ref{fig:W_GBW} we plot the analytic expressions in Eqs.~\eqref{dW_GA_1332_GBW}, Eq.~\eqref{dW_1332_GBW_D} and \eqref{dW_1332_GBW_ND} and we verify the large discrepancy between the GA and JIMWLK results\footnote{Unlike in Eqs.~\eqref{dW_GA_1332_log} and \eqref{dW_1332_log}, the plots in the right panel except the one corresponding to the dipole contribution are not symmetric around $\theta=\pi/2$, due to the asymmetry of the non-logarithmic pieces.}. Such an unambiguous conclusion is independent of the initial condition at $Y_0$. Indeed, using the (weak scattering limit of the) MV model 
\begin{align}
	\label{T_MV}
	\mcal{T} \simeq 
	1 - \exp \left(\!
	-\frac{r^2 Q_\mathrm{A}^2}{4}\,
	\ln \frac{4}{r^2 \Lambda^2}
	\right),
	\end{align}
where $Q_\mathrm{A}^2 = \qs^2/\ln(\qs^2/\Lambda^2)$ with $\Lambda$ being of the order of the QCD scale, we can easily integrate numerically the r.h.s.~in Eqs.~\eqref{dW_GA_1332_weak_dY}, \eqref{dW_1332_dY_D_T} and \eqref{dW1332_dY_ND_T} to arrive at results which are qualitatively the same as those exhibited in Fig.~\ref{fig:W_GBW} for the GBW model.

\begin{figure}
\begin{center}
\includegraphics[width=0.49\textwidth]{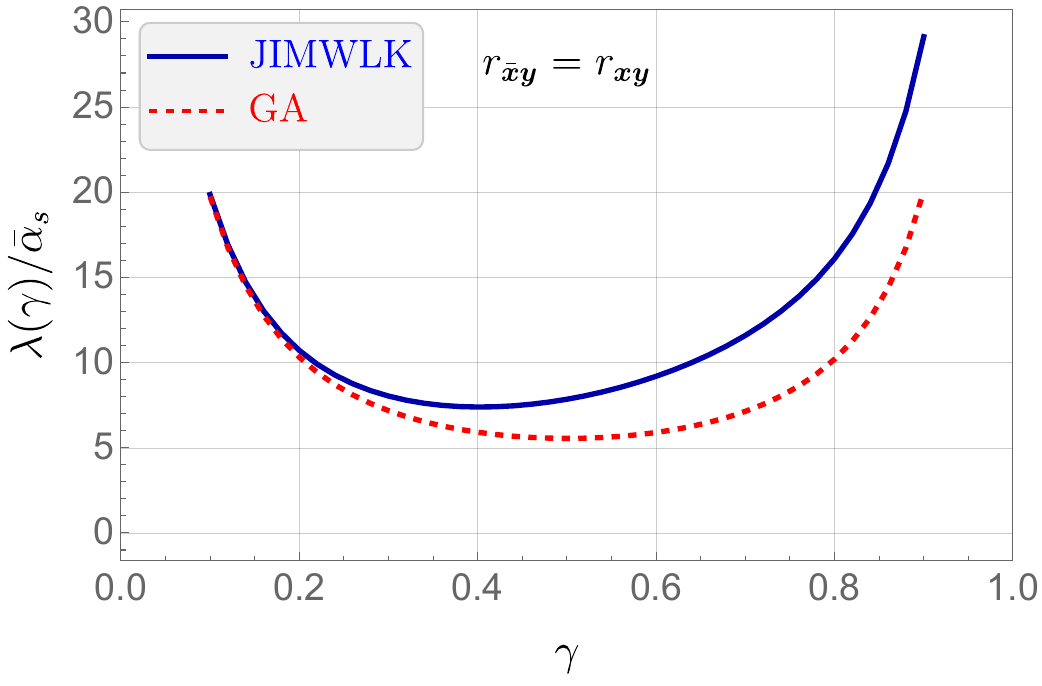}
\hspace*{0\textwidth}
\includegraphics[width=0.49\textwidth]{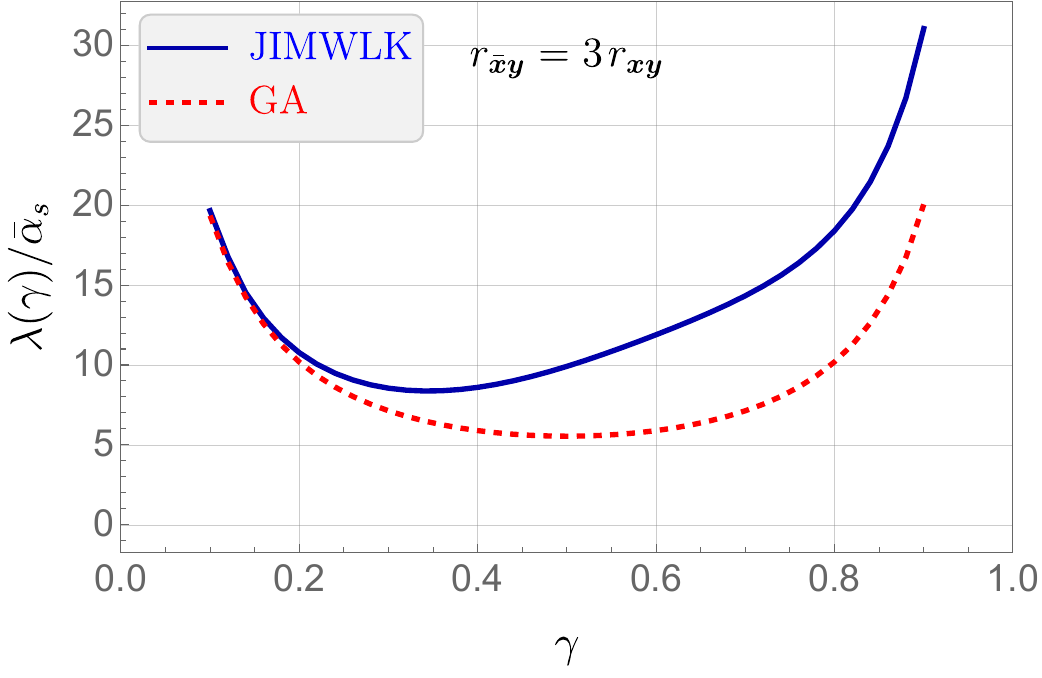}
\end{center}
\caption{\sl \small The effective JIMWLK eigenvalue $\lambda_\mathrm{eff}(\gamma)/\abar$ for the parton configurations with $r_{\bar{\bx}\by} = r_{\bx\by}$ (left panel) and $r_{\bar{\bx}\by}=3 r_{\bx\by}$ (right panel) compared to the corresponding GA eigenvalue $\lambda_\mathrm{GA}(\gamma)/\abar = 2\chi(\gamma)$.}
\label{fig:lambda}
\end{figure}

Ideally, in order to have a complete picture in the regime of weak scattering, one should solve the related eigenvalue problem. In the GA this is straightforward since $\mcal{W}$ is explicitly given in terms of $\mcal{T}^2$. When $\mcal{T}_{\bx\by} = r_{\bx\by}^{2\gamma}$, with $0 < \Re(\gamma) <1 $, we trivially arrive at the eigenvalue $\lambda_\mathrm{GA}(\gamma) = 2 \abar \chi(\gamma)$, where $\chi(\gamma) = 2 \psi(1) - \psi(\gamma) - \psi(1-\gamma)$ is the leading order BFKL characteristic function \cite{Kuraev:1977fs,Balitsky:1978ic}. Solving the ``exact'' problem is significantly more involved, because $\mcal{W}_{\bx\by\bar{\by}\bar{\bx}}$ depends on many coordinates and because it couples to other operators. This goes beyond the scope of our present work, however we can provide a hint about what the result may look like. Following the setup of the current section let us first choose parton configurations such that $\by = \bar{\by}$ and $r_{\bar{\bx}\by} = r_{\bx\by} \equiv r$ and let us assume the GA on the r.h.s.~of the (linearized) JIMWLK evolution equations for $\mcal{W}_{\bx\by\by\bar{\bx}}$. Now we notice that one can integrate over $\bz$ without imposing any restrictions since $0 < \Re(\gamma) <1 $. We shall not delve into details here, but only say that after the integration we do not recover a result proportional to the input function. Still, we do so if we average both sides over the angle $\theta$ between the vectors $\br_{\bx\by}$ and $\br_{\bbx\by}$. Thus we can define an ``effective'' eigenvalue via the relation $K \otimes \overline{\mcal{W}}_\gamma(r)= \lambda_\mathrm{eff}(\gamma) \overline{\mcal{W}}_\gamma(r)$ where the bar stands for the angular averaging and $K$ simply denotes the operation of one evolution step. Such a $\lambda_\mathrm{eff}(\gamma)$ can be easily constructed by numerical integration and is shown in the left panel in Fig.~\ref{fig:lambda} together with $\lambda_\mathrm{GA}(\gamma)$, whereas in the right panel we repeat the same exercise with $r_{\bar{\bx}\by} = 3 r_{\bx\by}$.  We point out that in both cases $\lambda_\mathrm{eff}(\gamma)$ is 40-60\% larger than $\lambda_\mathrm{GA}(\gamma)$ for $1/2<\gamma <1$ (which are the potentially relevant values for the saturation problem) in agreement with our earlier findings. Even though our analysis has been admittedly very crude, we believe that a precise treatment of the problem will most likely reveal that the evolution of the incoherent diffractive correlator is indeed driven by an eigenvalue larger than $2\abar \chi(\gamma)$. We recall that the latter determines the evolution of the corresponding correlator in coherent diffraction.  

\section{Numerical calculation for arbitrary kinematics}
\label{sec:num}

\subsection{Numerical method}

Now we proceed to the calculation of the incoherent diffractive correlator   in a kinematic regime which includes not only large rapidities, but also large distances which correspond to scattering sensitive to saturation. We shall make use of a Langevin equation which is a slightly modified version of the one presented earlier in Sect.~\ref{sec:JIM} in such a way to include running coupling effects.
Overall the numerical setup follows Ref.~\cite{Lappi:2012vw}, which we refer to for a more detailed discussion. We work on a coordinate space lattice of $N_\perp^2$ points in the transverse plane with periodic boundary conditions, with a lattice spacing $a$ and a physical size $L_\perp= N_\perp a$.

The starting point of the evolution are initial configurations of Wilson lines, which are obtained from the MV model~\cite{McLerran:1993ni,McLerran:1993ka,McLerran:1994vd}, with a numerical  procedure described in more detail in Ref.~\cite{Lappi:2007ku}. We obtain the Wilson lines from color charges that depend on the longitudinal coordinate $x^+$ in the path ordered exponential as \eqref{Wilson_U} by discretizing the longitudinal direction into $N_y$ rapidity slices as
\begin{align}
	V_{n,\bx} = \prod_{n=1}^{N_y}\exp\left[ i g\alpha_n^a(\bx) t^a\right]
\end{align}
where the index $n$ corresponds to a discretization of $x^+$.
The covariant gauge color field $\alpha_n^a$ is obtained from the color charge by solving the Poisson equation
\begin{align}\label{eq:Poisson}
\nabt^2 \alpha_n^a = \rho^a_n(\bx),
\end{align}
The distribution of Wilson lines follows from the probability distribution  of the color charges. In the MV model the charge correlator is assumed to be local in longitudinal and transverse coordinates and Gaussian. The locality in the longitudinal coordinate ($x^+$, or $n$ in the discretized form) makes the infinitesimal steps building up the path ordered exponential independent, and allows analytical calculations in spite of the nonlinearity of the exponentiation in going from the color fields to the Wilson lines. This is the discretized analogue of the locality in the parametrization rapidity $y$ of the GA equation \eqref{dOdY}. The Gaussianity implies that the full probability distribution of color charges is determined purely by their 2-point function. These two features are similar in the MV model and in the GA, and allow for analytical calculations of Wilson line operators. The MV model thus provides an initial condition for JIMWLK evolution where the GA is valid by construction.  
The third feature of the MV model is that the color charges $\rho^a_n(\bx)$ are also taken to be independent at each transverse coordinate, which, through the Laplace operator, translates into the small distance behavior of the dipole amplitude $\mcal{T} (r) \sim r^2\ln(1/r)$ for $r\to 0$. For the Fourier transform of the amplitude this corresponds to $\mcal{T}(k) \sim 1/k^4$, i.e. an anomalous dimension $\gamma=1$, which will then be modified during the high energy evolution. 

In terms of equations these features are summarized as 
\begin{align}
\left< \rho^a_m(\bx)\rho^b_n(\by)\right> = \frac{1}{N_y}g^2\mu^2 \delta_{mn} \delta^{ab} \delta(\bx-\by),
\end{align}
where the (parametrization) rapidity discretization is normalized so that
\begin{align}
\left< \rho^a(\bx)\rho^b(\by)\right> = \sum_{m,n}\left< \rho^a_m(\bx)\rho^b_n(\by)\right> = g^2\mu^2 \delta_{mn} \delta^{ab} \delta(\bx-\by)
\end{align}
is independent of $N_y$. On a transverse lattice with lattice spacing $a$ the delta function becomes
$\delta(\bx-\by) \to \delta_{\bx,\by}/a^2$ with a discrete dimensionless Kronecker delta for lattice sites $\delta_{\bx,\by}$.
The Poisson equation \eqref{eq:Poisson} is solved by going to momentum space. Here we regularize the singularity at $k=0$ by leaving out the zero momentum mode; physically this corresponds to imposing as an additional constraint the overall color neutrality of the system
\begin{align}
\int \dif^2\bx \rho^a_m(\bx) =0.
\end{align}
In terms of momentum scales this means  that the finite size of the target provides an infrared cutoff. However, since saturation screens the color fields at distances $\gtrsim 1/\qs$,  physically meaningful quantities are not sensitive to this lattice IR cutoff.

The Wilson line distribution in the MV model is determined by the value of the color charge density  parameter $g^2\mu$, which provides a typical characteristic scale that acts as the saturation scale at the initial condition of the JIMWLK evolution. Once the Wilson lines have been constructed, we can then use them to extract the saturation scale $\qs$. For this one has to choose an explicit, model-independent definition for $\qs$ that can also be used for the Wilson line configurations after the evolution has changed the value of $\qs$. We adopt here the convention that $\qs$ is defined as the solution of the following equation
\begin{align}
\mcal{D}\big(r= \sqrt{2}/\qs \big) = \exp\left(-1/2\right),
\end{align}
which is consistent with the convention for the GBW model,  Eq.~\eqref{T_GBW}.

The JIMWLK equation is solved numerically using the Langevin formulation \eqref{V_lang}. The fixed coupling equation has an unrealistically fast speed of evolution, and a very poor convergence to the continuum limit \cite{Rummukainen:2003ns,Triantafyllopoulos:2005cn}. Instead, here we will introduce a running coupling whose precise implementation follows the ``noise $\as$'' prescription introduced in Ref.~\cite{Lappi:2012vw}.  This amounts to replacing the correlation function of the noise, which is a delta function in coordinates, i.e. an integral over momenta, with a function where this momentum integral is weighted by the momentum-dependent running coupling. This can be justified by the observation that the coordinates of the noise actually correspond to the coordinates of the gluon  emitted in one evolution step, and this prescription corresponds to taking the scale in the running coupling to be the momentum of this gluon. This corresponds to modifying Eq.~\eqref{noise} in the following way:
\begin{align}
	\as \big\langle \nu^{ia}_{m,\bx}\, 
	\nu^{jb}_{n,\by} 
	\big\rangle
	=\frac{\as}{\epsilon}\,
	\delta^{ij} \delta^{ab} 
	\delta_{mn} \delta_{\bx\by}
\quad 
\longrightarrow
\quad
	\frac{1}{\epsilon}\,
	\delta^{ij} \delta^{ab} 
	\delta_{mn} 
\int\frac{\dif^2\bk}{(2 \pi)^2}e^{i \bk \cdot(\bx-\by)} \as(\bk).
\end{align}
Because of the integral over momenta $\bk$, we need to regularize the running coupling. This is done by taking it as
\begin{align}\label{eq:kspaceas}
  \as(\bk) = \frac{4\pi}{
\beta \ln\left\{ \left[
       \left(\frac{\mu_0^2}{\lqcd^2}\right)^{\frac{1}{c}}
      +\left( \frac{\bk^2}{\lqcd^2} \right)^{\frac{1}{c}} \right]^{c}
\right\} 
},
\end{align}
so that for small momenta $k\lesssim \mu_0$ the coupling freezes to a maximal value $\alpha_0$. 

For the studies presented here we use a $N_\perp^2 = 1024^2$-lattice, with the MV model parameter taken as $g^2\mu L= 80$ and $N_y=100$. This results in a saturation scale $\qs L_\perp=71.8$ or $\qs a=0.0697$, meaning that our initial saturation scale is  well separated both from the lattice IR and UV cutoffs, $\qs \gg 2\pi/L_\perp$ and $\qs \ll 1/a$.
In solving the JIMWLK equation,  we take $\lqcd L_\perp =6$, $\mu_0 L= 15$ and $c=0.2$, which correspond to $\alpha_0 = \as(k=0)= 0.762$,  but since already our initial saturation scale is $\qs > \mu_0$, the simulation happens fully in the running coupling regime and the effective coupling remains much smaller than this. The results presented here are obtained by making 2000 rapidity steps of length $0.0002$ in the rescaled rapidity  $\alpha_0 y /\pi^2$ used in the code, thus corresponding to an evolution by $5.18$ units in rapidity. Since the purpose here is not to study asymptotic features of JIMWLK evolution, but violations of the GA starting from a Gaussian initial condition, this amount of evolution is plenty. During this evolution, the saturation scale grows from $\qs a = 0.0697$ at the initial condition to $\qs a = 0.143$ after 5.18 units of rapidity, staying still safely away from the lattice IR and UV cutoffs. This growth of the saturation scale would correspond to an average  $\lambda\equiv \dif \ln \qs^2 /\dif y = 0.28$, which is not unreasonable phenomenologically in the kinematical regime of current experiments~\cite{GolecBiernat:1998js,Triantafyllopoulos:2002nz}.

\subsection{Numerical results}

\begin{figure}[tb!]
\begin{center}
\includegraphics[width=0.32\textwidth]{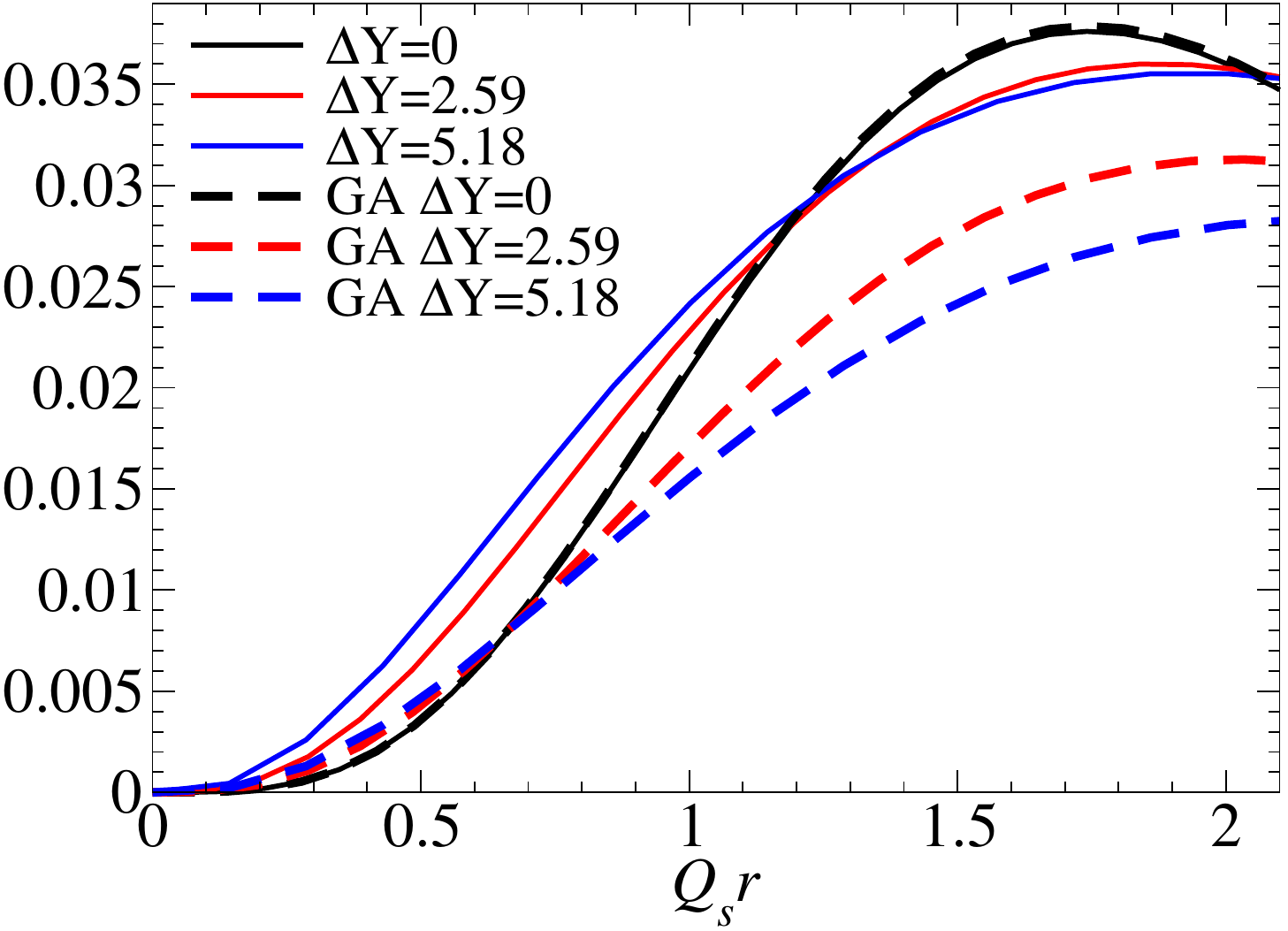}
\hspace*{0\textwidth}
\includegraphics[width=0.32\textwidth]{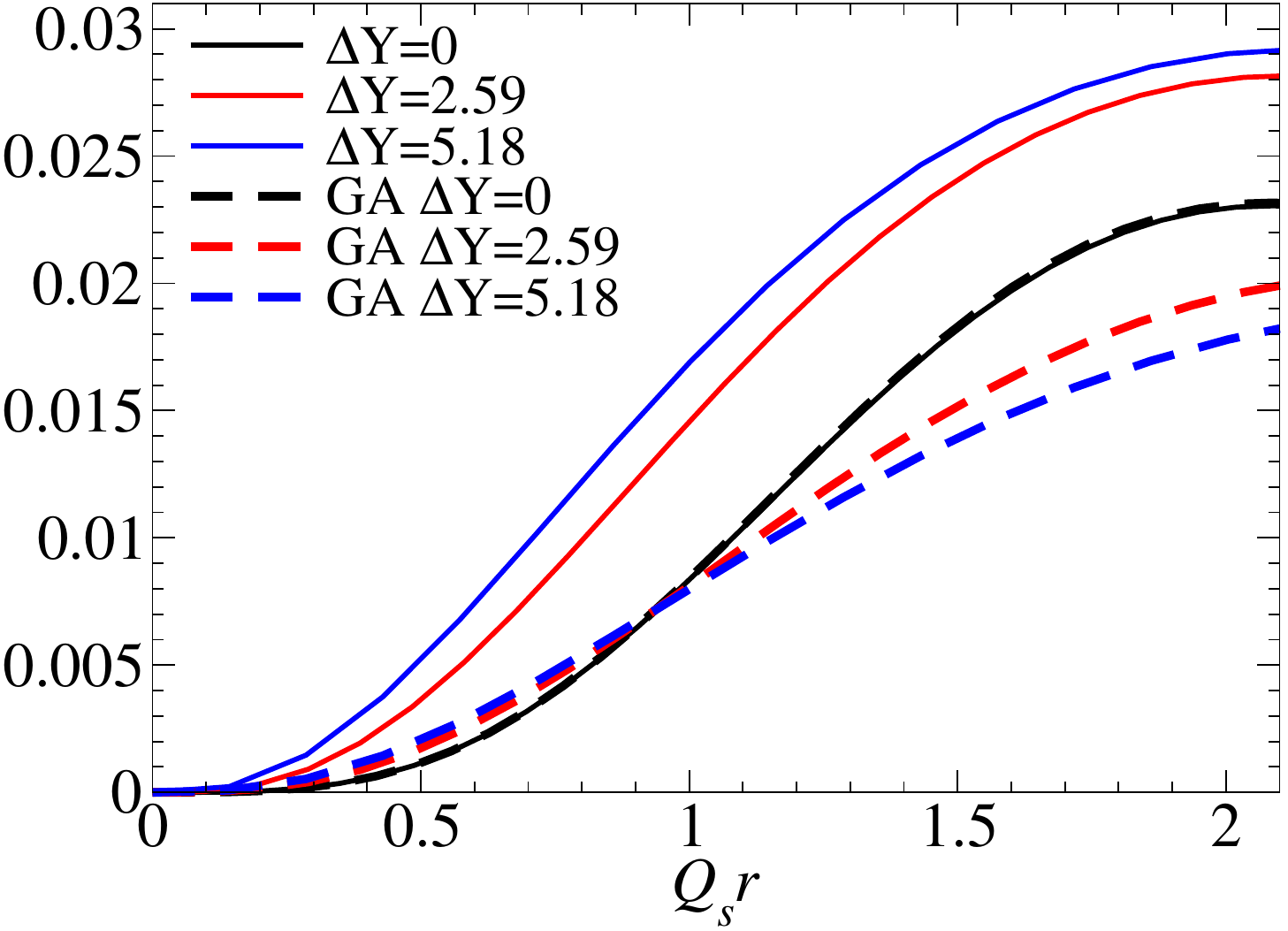}
\hspace*{0\textwidth}
\includegraphics[width=0.32\textwidth]{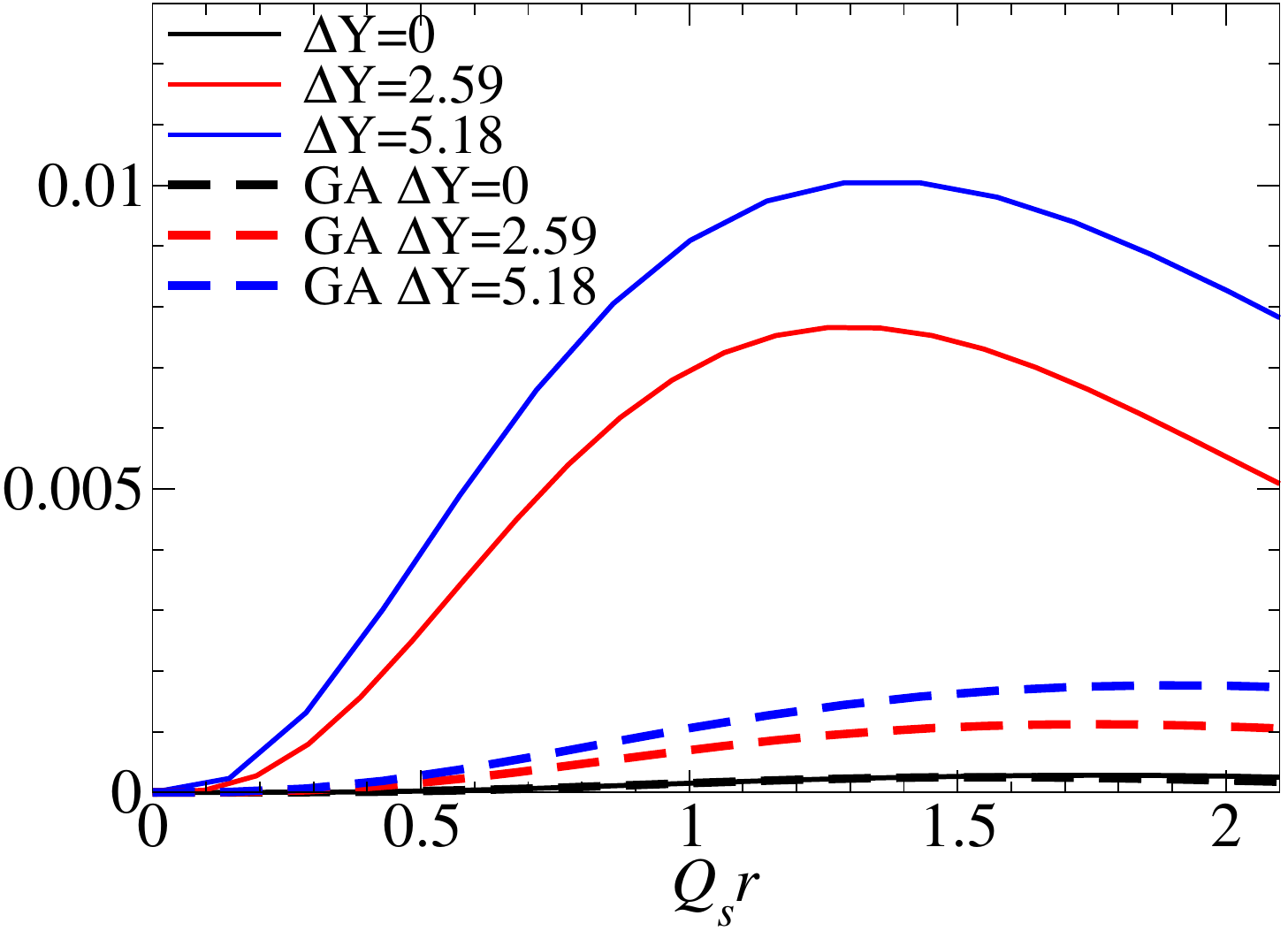}
\\
\hfill
 \raisebox{-.5\height}{\includegraphics[scale=0.32]{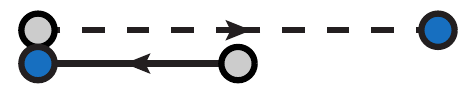}
\begin{tikzpicture}[overlay]
\node[anchor=east] at(-2.5,0.2) {$\by$};
\node[anchor=north] at(-1.3,0.1) {$\bx$};
\node[anchor=south] at(-0.3,0.35) {$\bbx$};
\node[anchor=south] at(-1.3,0.35) {$2r$};
\node[anchor=north] at(-1.9,0.1) {$r$};
\end{tikzpicture}
}
\hfill \rule{0pt}{1pt}\hfill
\raisebox{-.5\height}{\includegraphics[scale=0.32]{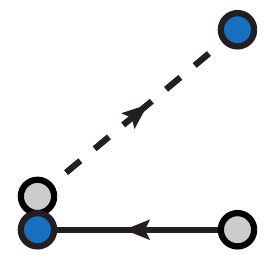}
\begin{tikzpicture}[overlay]
\node[anchor=east] at(-1.4,0.2) {$\by$};
\node[anchor=north] at(-0.3,0.1) {$\bx$};
\node[anchor=south] at(-0.3,1.25) {$\bbx$};
\node[anchor=south east] at(-0.8,0.7) {$\sqrt{2}r$};
\node[anchor=north] at(-0.8,0.1) {$r$};
\end{tikzpicture}
}
\hfill \rule{0pt}{1pt}\hfill
\raisebox{-.5\height}{\includegraphics[scale=0.32]{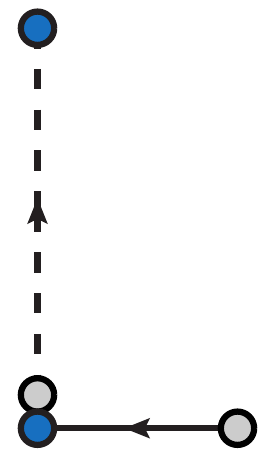}
\begin{tikzpicture}[overlay]
\node[anchor=east] at(-1.4,0.2) {$\by$};
\node[anchor=north] at(-0.3,0.1) {$\bx$};
\node[anchor=north east] at(-1.4,2.2) {$\bbx$};
\node[anchor=west] at(-1.3,1.2) {$2r$};
\node[anchor=north] at(-0.8,0.1) {$r$};
\end{tikzpicture}
}
\hfill \rule{0pt}{1pt}
\end{center}
\caption{\sl \small Numerical results for the rapidity evolution of the incoherent diffractive correlator $\mcal{W}_{\bx\by\by\bbx}$ as a function of the scaled size $r\qs \equiv r_{\bx\by} \qs$ for  $r_{\bbx\by} =2\mkern2mu r_{\bx\by}$ at a  fixed angle $\theta = 0$ (left panel), $r_{\bbx\by} =\sqrt{2} \mkern2mu r_{\bx\by}$  at a  fixed angle $\theta = \pi/4$ (middle panel) and for  $r_{\bbx\by} =2\mkern2mu r_{\bx\by}$ at a fixed angle $\theta = \pi/2$ (right panel). 
In terms of unit vectors $\ihat$ and $\jhat$ the configurations can be expressed as  $\bx = \by + r \ihat $ with $\bbx = \by + 2r\ihat$ (left panel), $\bbx = \by + r(\ihat+\jhat)$ (middle panel) and 
$\bbx = \by + 2r\jhat$ (right panel).
The coordinate configurations are demonstrated below the plots, with a light circle corresponding to a Wilson line and a dark dark one to a Hermitian conjugate, the arrows showing which Wilson lines are color contracted to be in the same dipole, and the solid line always having a length $r$.
All the distances are expressed in units of the saturation scale corresponding to each evolution rapidity.}
\label{fig:W1332_num}
\end{figure}

We start by presenting results for the correlator that one needs for incoherent diffraction, the connected part of the dipole-dipole correlator $\mcal{W}_{\bx\by\bby\bbx}$ defined by \eqref{W}. Due to translational invariance, the general correlator depends on three independent coordinate differences. We will here quantify it in different coordinate combinations in such a way that we simultaneously vary the length of all these coordinate separations, keeping the directions and relative lengths same. When interpreting the values of the correlators, $\mcal{W}_{\bx\by\bby\bbx}$ is parametrically of order $1/\nc^2$, so we expect both the full JIMWLK and GA results to be of the order $0.01$ to $0.05$.

Let us consider the case where two of the coordinates are taken to be the same, $\mcal{W}_{\bx\by\by\bbx}$ studied before in Sect.~\ref{sec:break}. In Fig.~\ref{fig:W1332_num} we present the JIMWLK results for $\mcal{W}_{\bx\by\by\bbx}$ in comparison with the GA. The analytic expression of $\mcal{W}_{\bx\by\by\bbx}$ in the GA in terms of the dipole (with the latter determined numerically from the same numerical solution of the JIMWLK equation) is rather simple and given in Eq.~\eqref{appeq:DD1332} in the Appendix.  We use the same independent variables as in Sect.~\ref{sec:break}, that is, the sizes $r_{\bx\by}$ and $r_{\bbx\by}$ and the angle $\theta$ between the two corresponding vectors. In both cases we keep the ratio $r_{\bbx\by}/r_{\bx\by}$ and the angle $\theta$ fixed, while we vary the size $r_{\bx\by}$. These coordinate configurations are also demonstrated in the figure. As we evolve in rapidity we observe growing deviations between the JIMWLK and the GA results, not only for weak scattering (as already confirmed in Sect.~\ref{sec:break}), but in a much wider region which includes distances of the order of $1/\qs$. Perhaps not surprisingly, when $\theta=\pi/2$ we see strong {\it relative} deviations, since for this particular setup we expect the GA result to be extremely small (recall that it exactly vanishes for the GBW model, cf.~Eq.~\eqref{W_GA_1332_weak}), but in both cases the {\it absolute} deviations are roughly the same. Thus, we can safely draw the conclusion that the difference between the two results is an effect of order one, with the JIMWLK result being systematically larger than the GA one. 

\begin{figure}[tb!]
\begin{center}
\includegraphics[width=0.32\textwidth]{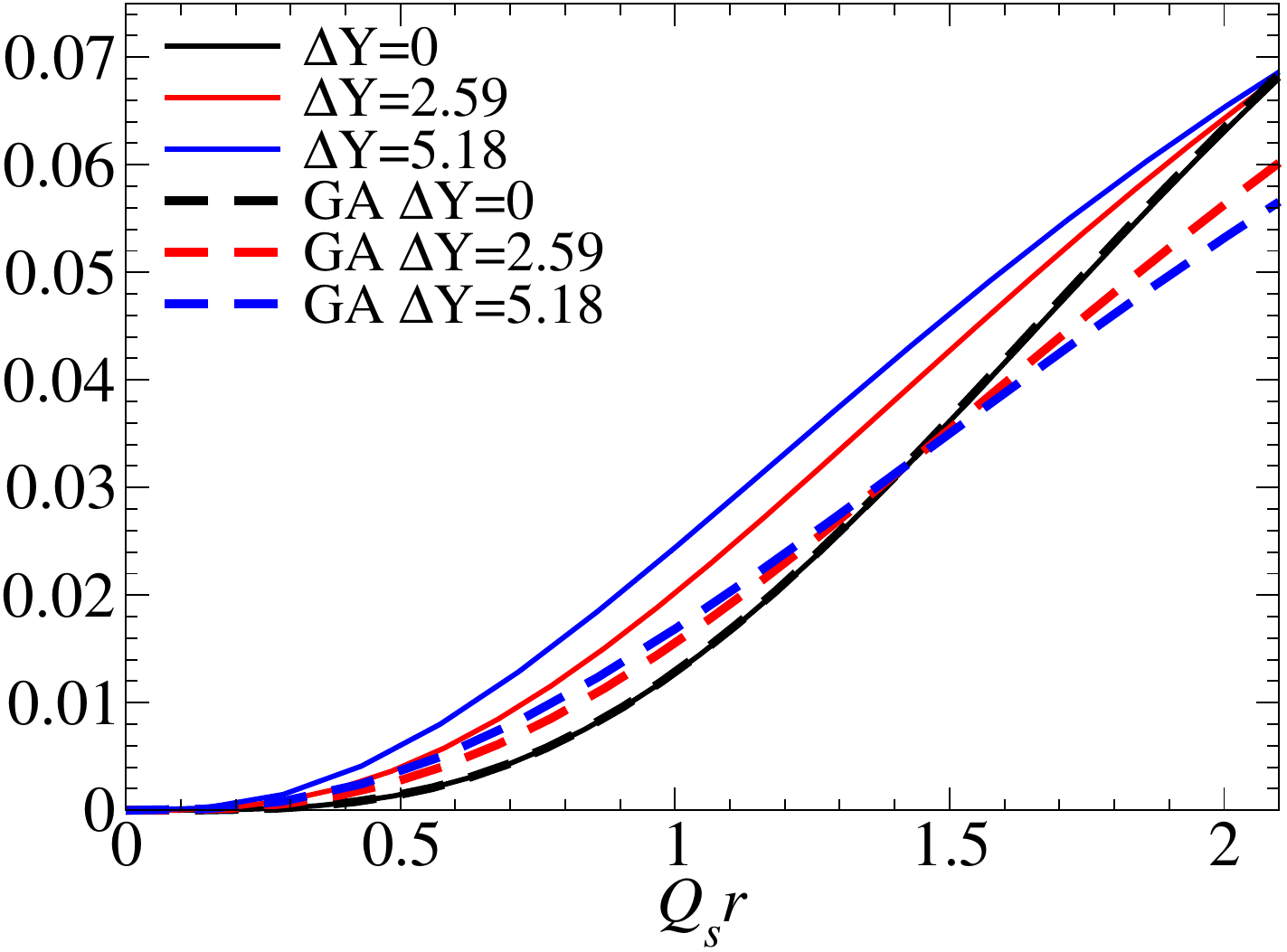}
\hspace*{0\textwidth}
\includegraphics[width=0.32\textwidth]{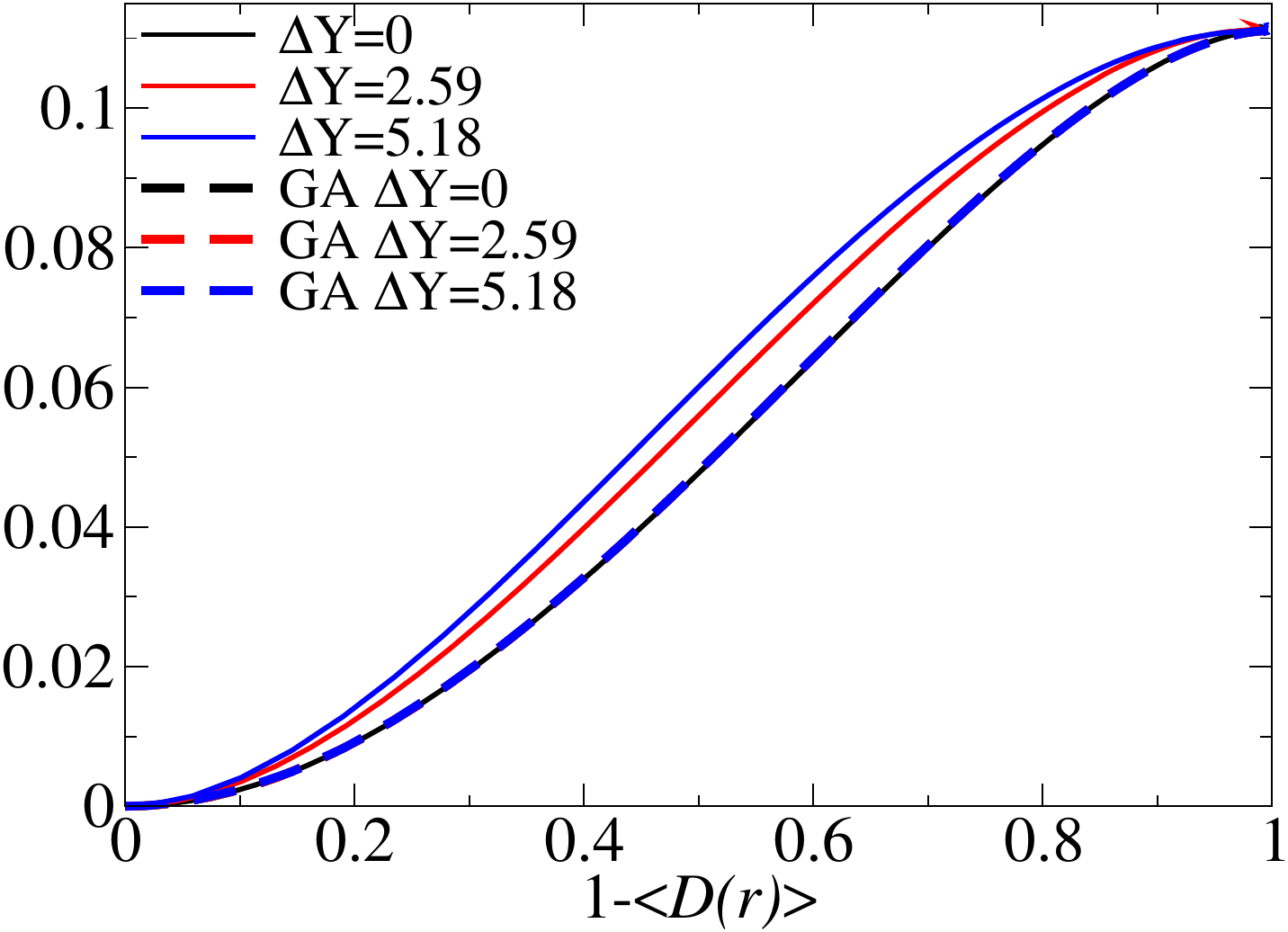}
\hspace*{0\textwidth}
\includegraphics[width=0.32\textwidth]{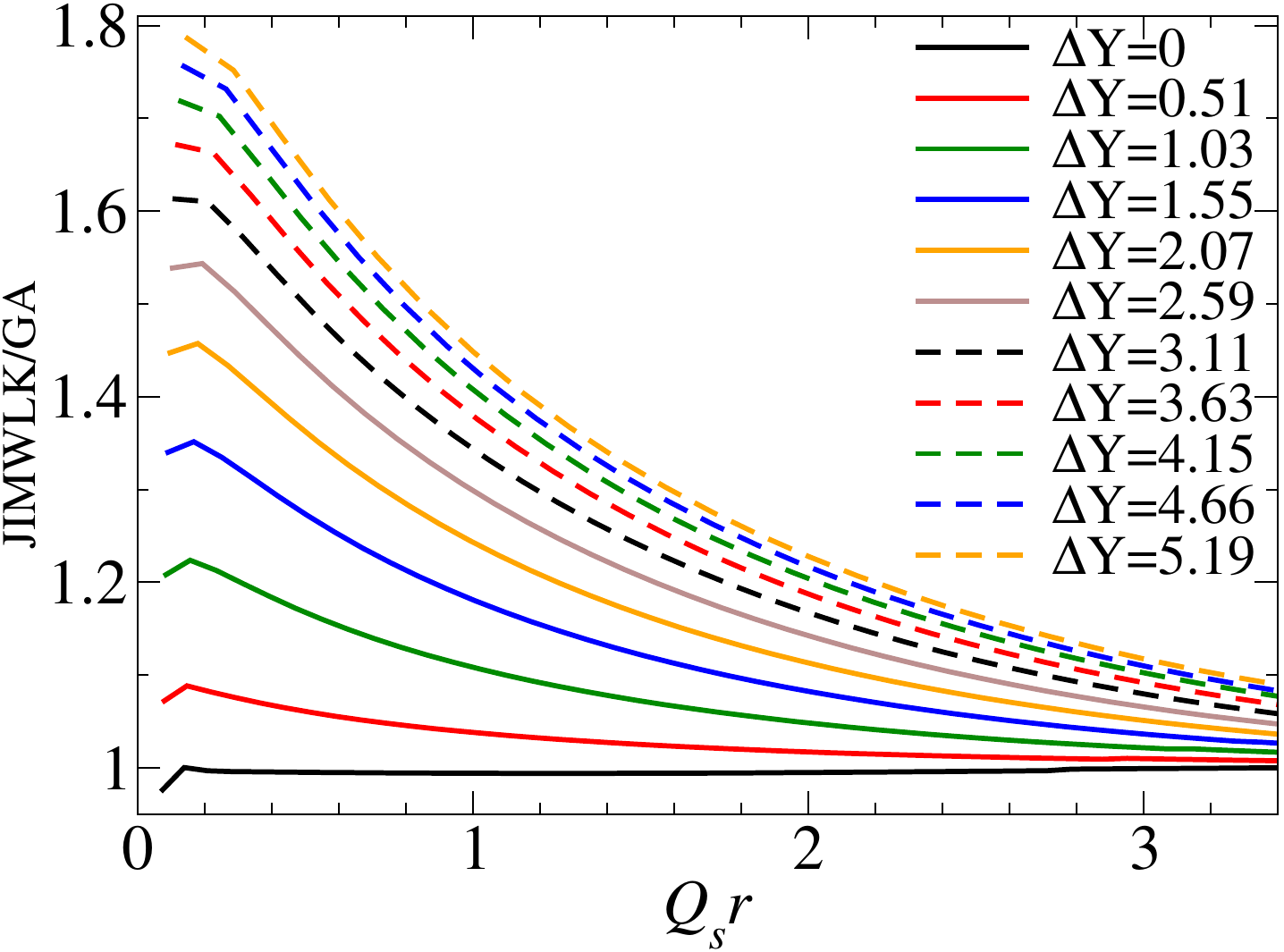}
\end{center}
\caption{\sl \small Numerical results for the rapidity evolution of the incoherent diffractive correlator $\mcal{W}_{\bx\by\by\bx}$.   Left and middle panel: JIMWLK and GA, as a function of of the scaled size $r_{\bx\by} \qs$ (left) and of $1-\langle D(r) \rangle$ (middle), similarly to Fig.~\ref{fig:O_line}. Right panel:  ratio of JIMWLK to GA vs. $r_{\bx\by} \qs$.}
\label{fig:W1221_num}
\end{figure}

In Fig.~\ref{fig:W1221_num} we show the results for the special configurations where the quark and antiquark coordinates in both dipoles are the same, corresponding to $\mcal{W}_{\bx\by\by\bx}$ which depend only one variable, the size $r_{\bx\by}$. Here, similarly to Fig.~\ref{fig:O_line}, we also show the correlator as a function of the dipole  scattering amplitude $1-\langle D \rangle$: in this way of plotting the GA curves overlap by definition, but the JIMWLK evolved one deviates from it. As anticipated in the discussion at the end of Sect.~\ref{sub:evol}, one should not be misled by the fact that in this case the GA happens to be valid for just one evolution step. Indeed, we see sizeable violations which are better exhibited on the right panel where we show the ratio of the JIMWLK to GA results. After a few units in rapidity evolution the difference is very large in the weak scattering regime, and it remains significant, roughly 50\%, for distances of the order of $1/\qs$. Only when approaching the unitarity limit, the GA seems to become valid.

\begin{figure}
\begin{center}
\includegraphics[width=0.32\textwidth]{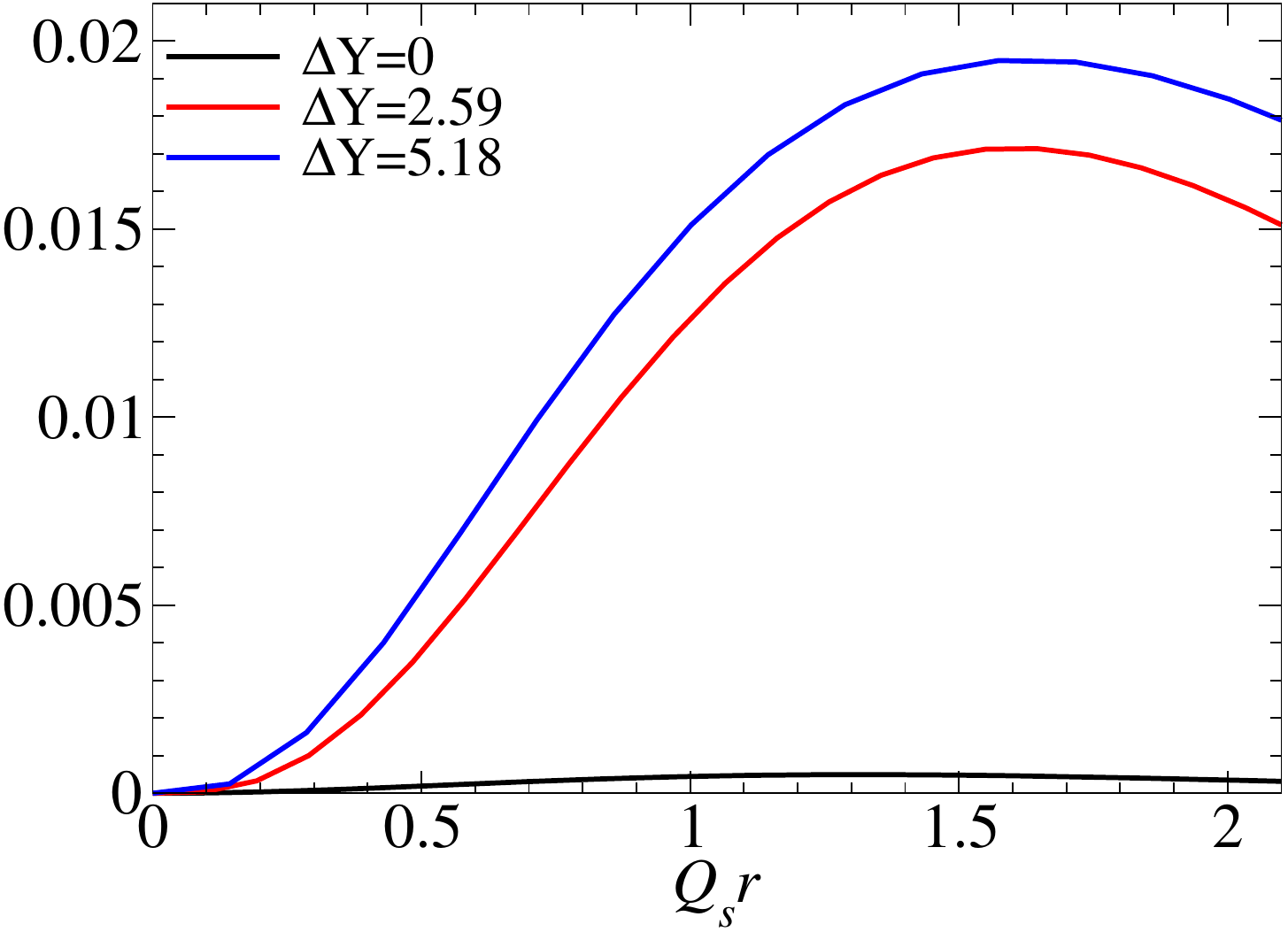}
\hspace*{0\textwidth}
\includegraphics[width=0.32\textwidth]{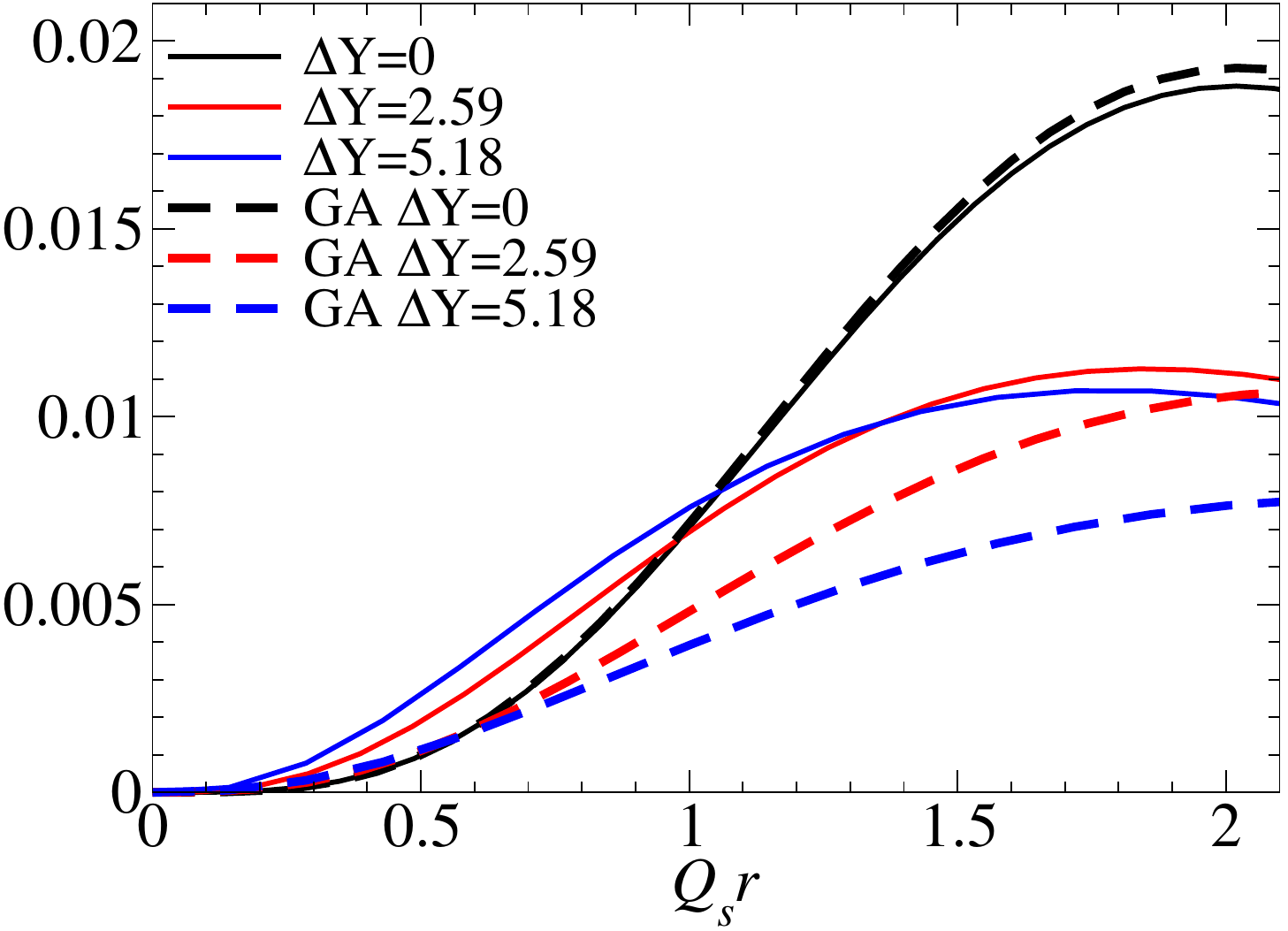}
\hspace*{0\textwidth}
\includegraphics[width=0.32\textwidth]{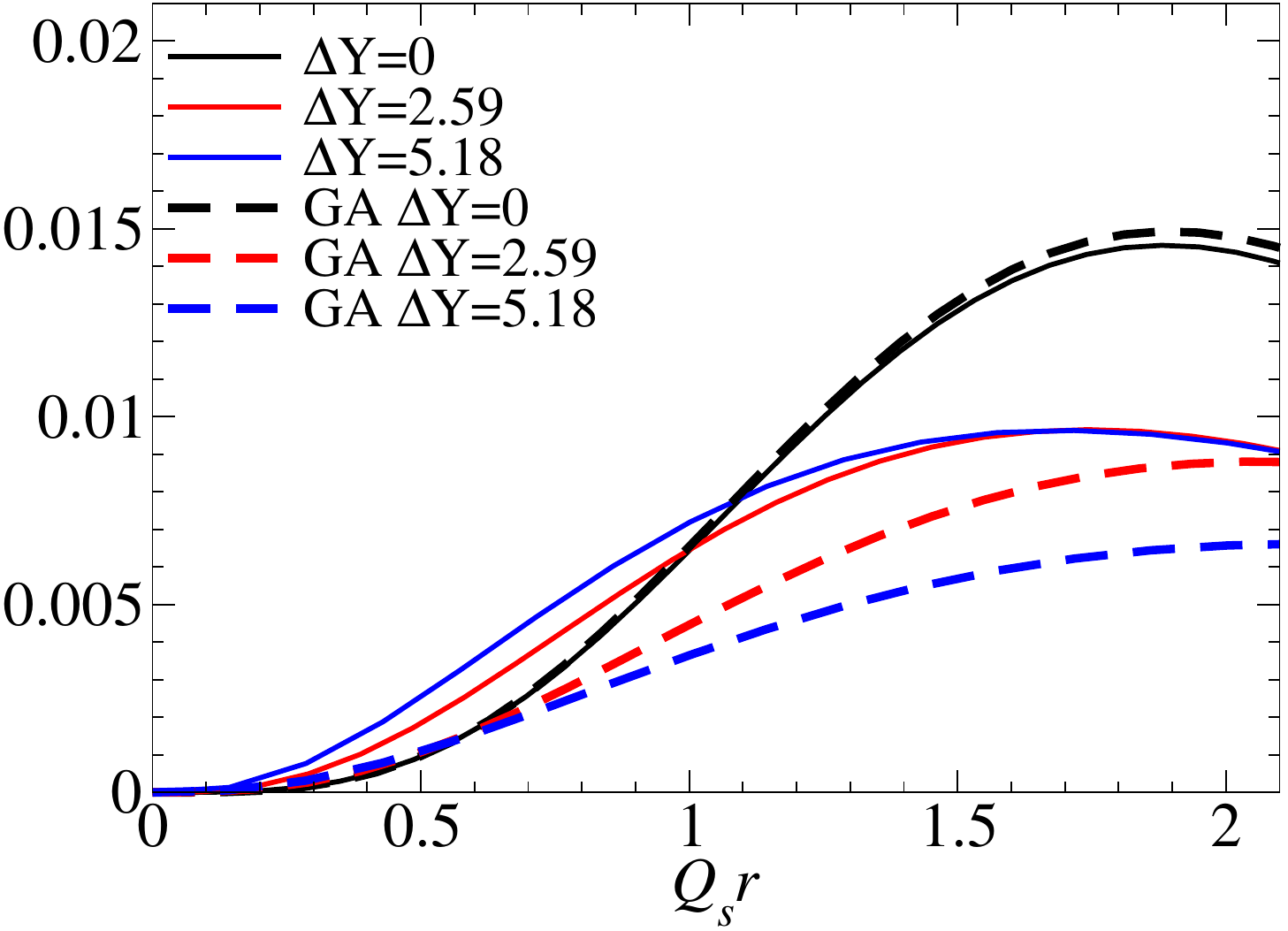}
\\
\hfill
 \raisebox{-.5\height}{\includegraphics[scale=0.4]{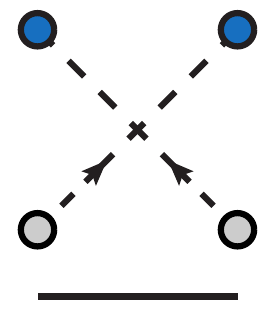}
\begin{tikzpicture}[overlay]
\node[anchor=east] at(-1.8,0.5) {$\bx$};
\node[anchor=west] at(-0.3,0.5) {$\bby$};
\node[anchor=east] at(-1.9,1.9) {$\bbx$};
\node[anchor=west] at(-0.3,1.9) {$\by$};
\node[anchor=north] at(-1.0,0.1) {$r$};
\end{tikzpicture}
}
\hfill \rule{0pt}{1pt}\hfill
\raisebox{-.5\height}{\includegraphics[scale=0.4]{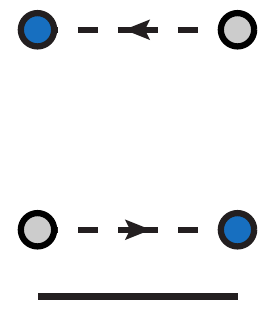}
\begin{tikzpicture}[overlay]
\node[anchor=east] at(-1.8,0.5) {$\bx$};
\node[anchor=west] at(-0.3,0.5) {$\by$};
\node[anchor=east] at(-1.9,1.9) {$\bbx$};
\node[anchor=west] at(-0.3,1.9) {$\bby$};
\node[anchor=north] at(-1.0,0.1) {$r$};
\end{tikzpicture}
}
\hfill \rule{0pt}{1pt}\hfill
\raisebox{-.5\height}{\includegraphics[scale=0.4]{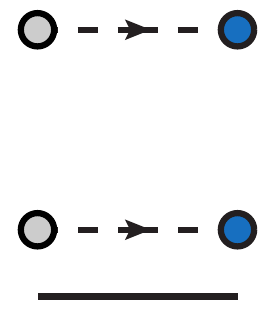}
\begin{tikzpicture}[overlay]
\node[anchor=east] at(-1.8,0.5) {$\bx$};
\node[anchor=west] at(-0.3,0.5) {$\by$};
\node[anchor=east] at(-1.9,1.9) {$\bby$};
\node[anchor=west] at(-0.3,1.9) {$\bbx$};
\node[anchor=north] at(-1.0,0.1) {$r$};
\end{tikzpicture}
}
\hfill \rule{0pt}{1pt}
\end{center}
\caption{\sl \small Numerical results for the rapidity evolution of the incoherent diffractive correlator $\mcal{W}_{\bx\by\bby\bbx}$, with the coordinates located at the corners of a square with radius $r$, as a function of the scaled distance $r\qs$. The different coordinate assignments are demonstrated in the plot. The configurations can be expressed as  
$\by = \bx + r(\ihat+\jhat)$, $\bby = \bx + r\ihat$, 
$\bbx = \bx + r\jhat$ (left panel),
$\by = \bx + r\ihat$, $\bby = \bx + r(\ihat+\jhat)$, 
$\bbx = \bx + r\jhat$ (middle panel)
and
$\by = \bx + r\ihat$, $\bby = \bx + r\jhat$, 
$\bbx = \bx + r(\ihat+\jhat)$ (right panel).
For the configuration in the left panel, the GA, and the initial condition for JIMWLK, are zero and we interpret the small nonzero value in the MV model (black line) as a discretization artefact.}
\label{fig:Wsq_num}
\end{figure}

In Fig.~\ref{fig:Wsq_num} we show the correlator $\mcal{W}_{\bx\by\bby\bbx}$ in configurations where all the coordinates are different, and located at the corners of a square with side length $r$. The first of these where the two dipoles ``cross'' is curious, because the GA (and thus the initial condition for JIMWLK evolution) are zero. After JIMLWK evolution this correlator reaches values of $\sim 0.02$, which is quite significant remembering that the maximal value for such a $\sim 1/\nc^2$ correlator is $\lesssim 0.1$. In both other cases the correlator differs from the GA value (determined with help of Eqs.~\eqref{appeq:Wsq_m} and \eqref{appeq:Wsq_r}) by a similar amount, which translates into a factor of $\sim 2$ relative difference.

\begin{figure}
\begin{center}
\includegraphics[width=0.32\textwidth]{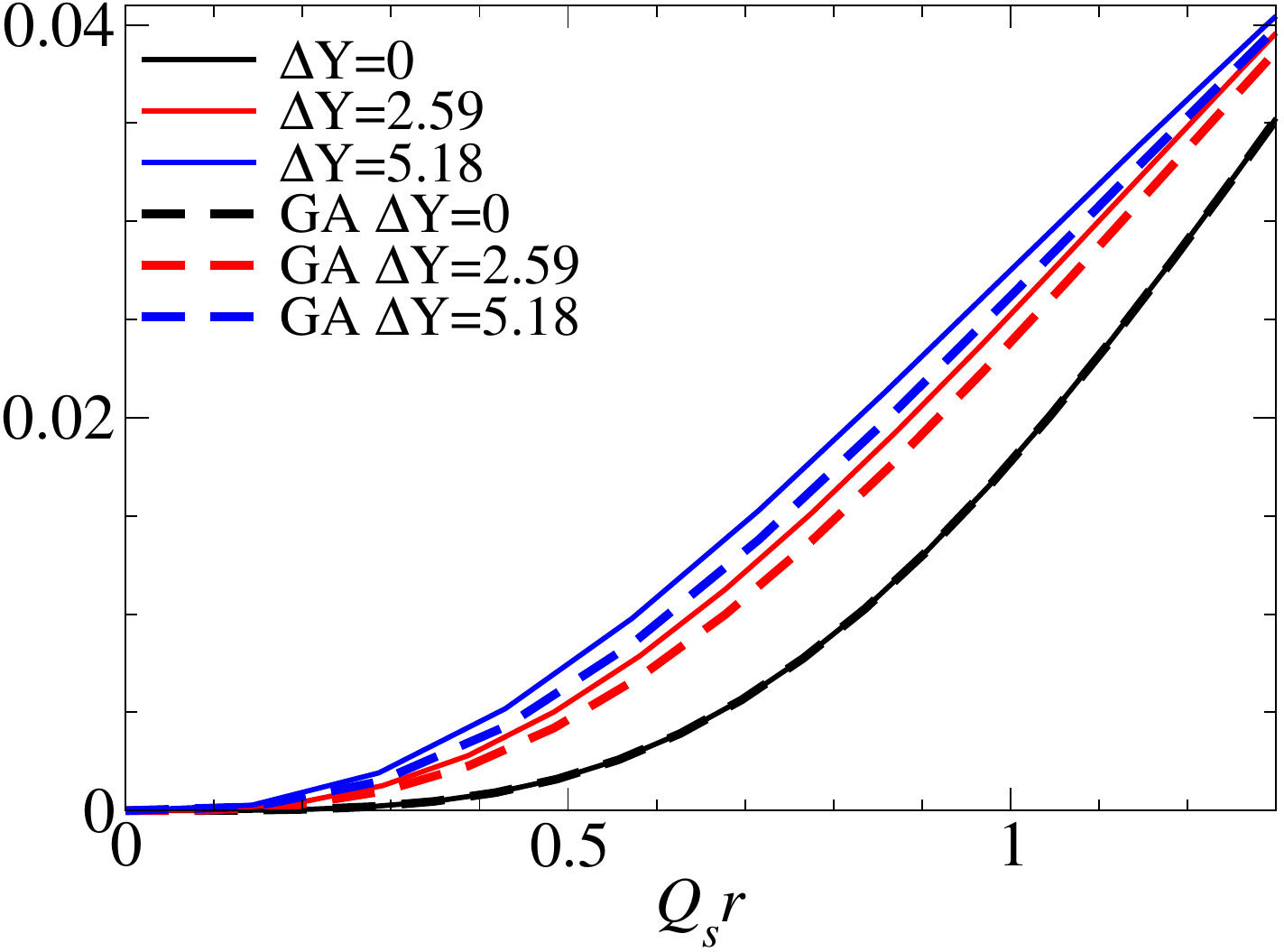}
\hspace*{0\textwidth}
\includegraphics[width=0.32\textwidth]{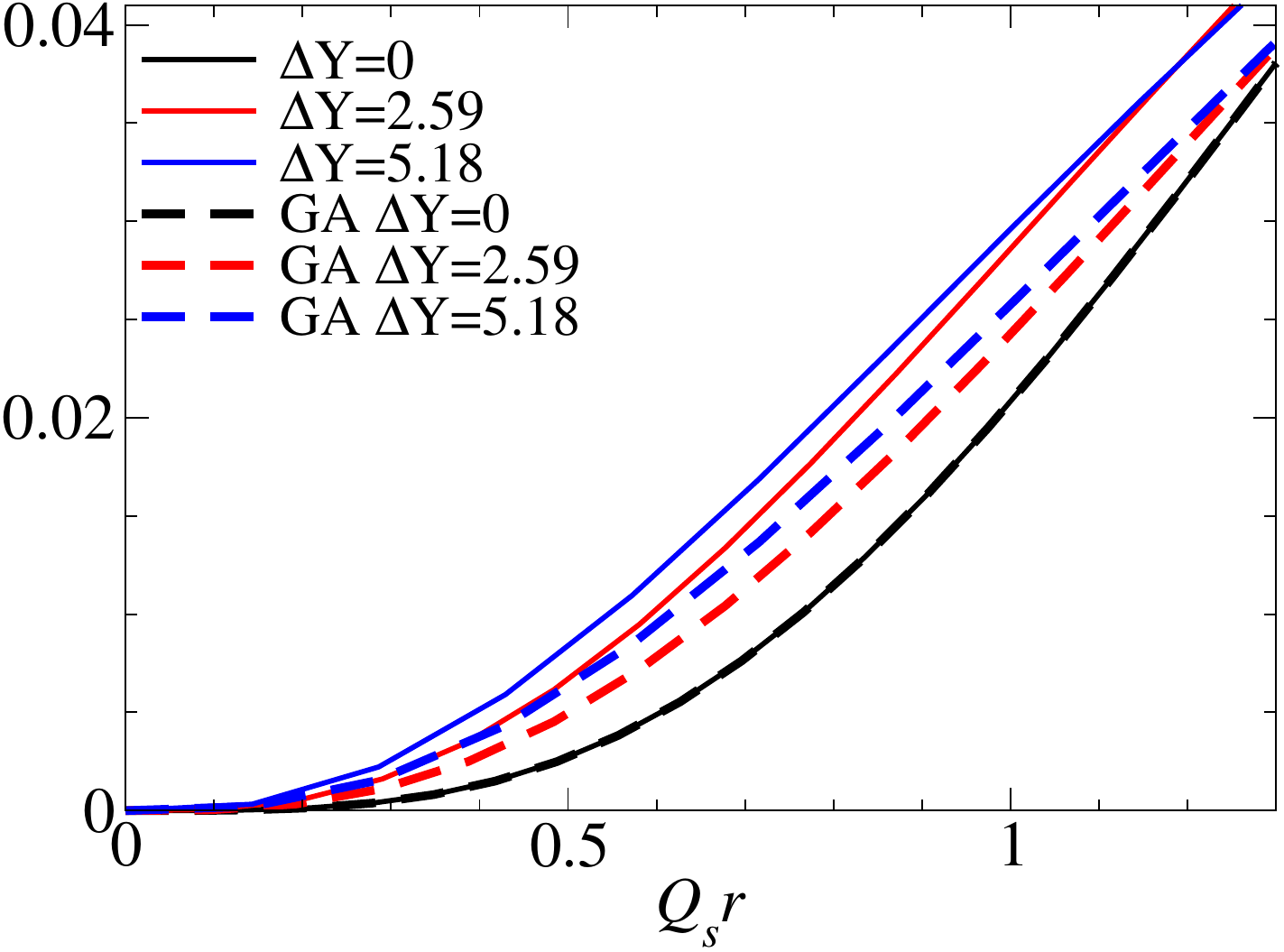}
\hspace*{0\textwidth}
\includegraphics[width=0.32\textwidth]{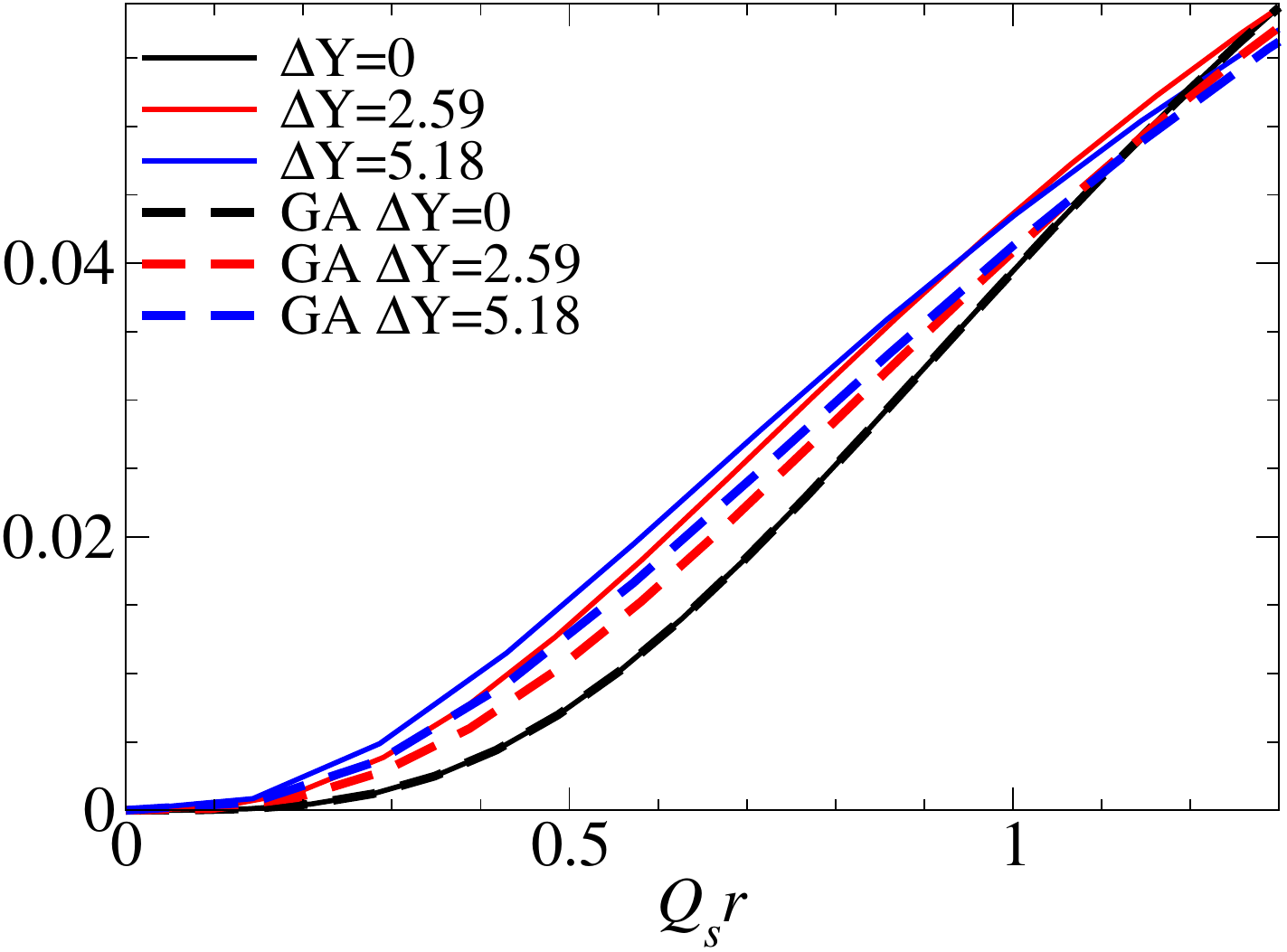}
\\
\hfill
 \raisebox{-.5\height}{\includegraphics[scale=0.3]{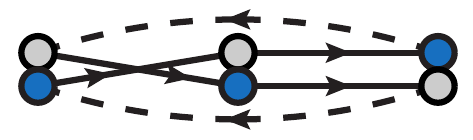}
\begin{tikzpicture}[overlay]
\node[anchor=east] at(-2.5,0.35) {$\by$};
\node[anchor=north] at(-1.3,0.1) {$\bx$};
\node[anchor=west] at(-0.3,0.35) {$\bz$};
\node[anchor=south] at(-1.3,0.6) {$2r$};
\node[anchor=north] at(-1.9,0.1) {$r$};
\end{tikzpicture}
}
\hfill \rule{0pt}{1pt}\hfill
\raisebox{-.5\height}{\includegraphics[scale=0.3]{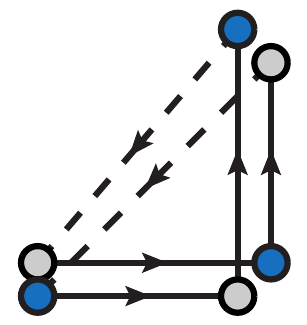}
\begin{tikzpicture}[overlay]
\node[anchor=east] at(-1.5,0.2) {$\by$};
\node[anchor=north west] at(-0.3,0.3) {$\bx$};
\node[anchor=west] at(-0.3,1.4) {$\bz$};
\node[anchor=south east] at(-0.9,0.8) {$\sqrt{2}r$};
\node[anchor=north] at(-0.9,0.1) {$r$};
\node[anchor=west] at(-0.2,0.8) {$r$};
\end{tikzpicture}
}
\hfill \rule{0pt}{1pt}\hfill
\raisebox{-.5\height}{\includegraphics[scale=0.3]{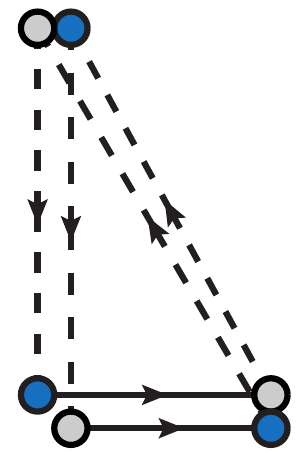}
\begin{tikzpicture}[overlay]
\node[anchor=east] at(-1.5,0.2) {$\by$};
\node[anchor=west] at(-0.2,0.2) {$\bx$};
\node[anchor=east] at(-1.5,2.2) {$\bz$};
\node[anchor=east] at(-1.5,1.2) {$2r$};
\node[anchor=west] at(-0.7,1.3) {$\sqrt{5}r$};
\node[anchor=north] at(-0.8,0.1) {$r$};
\end{tikzpicture}
}
\hfill \rule{0pt}{1pt}
\end{center}
\caption{\sl \small Numerical results for the sextupole $(1-\mcal{S}_{\bx\bz\by\bx\bz\by})/\nc^2$ as a function of the scaled size $r_{\bx\by} \qs$ for $r_{\bz\by} =2 r_{\bx\by}$  (left and right panels) at fixed angles $\theta=0$ (left panel) and $\theta = \pi/2$ (right panel),   and for $r_{\bz\by} =\sqrt{2} r_{\bx\by}$ at fixed angle $\theta = \pi/4$ (middle panel).  These are the same configurations as in Fig.~\ref{fig:W1332_num} if we let $\bz\to\bbx$. The coordinate configurations are demonstrated below the plots, with a light circle corresponding to a Wilson line and a dark one to a Hermitian conjugate and the arrows showing the order of multiplication within the sextupole. The solid lines have length $r$.}
\label{fig:sext_num}
\end{figure}
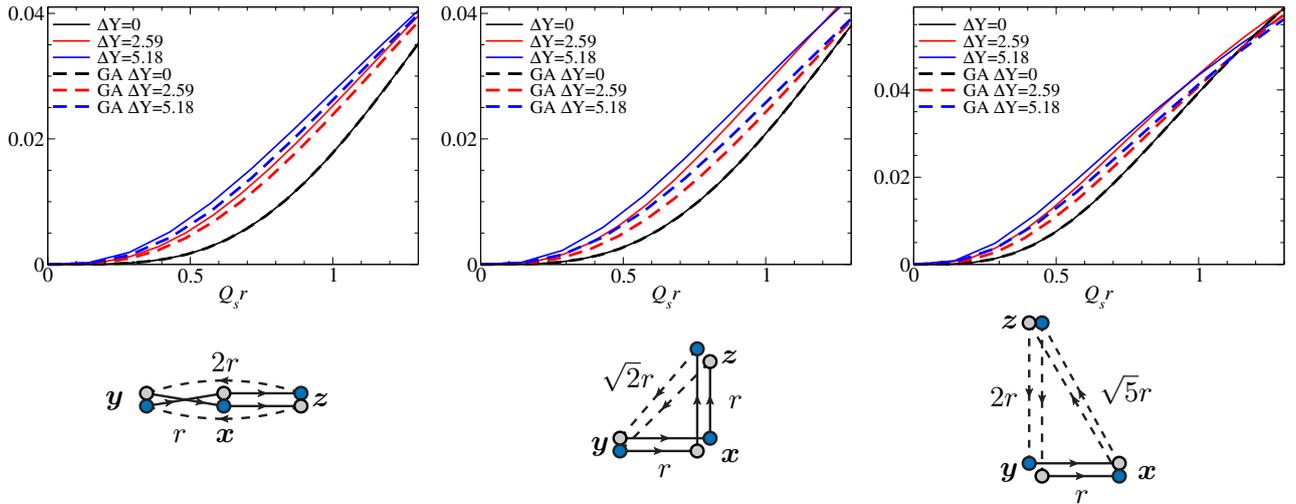

Finally in Fig.~\ref{fig:sext_num} we present results for the sextupole $(1-\mcal{S}_{\bx\bz\by\bx\bz\by})/\nc^2$, for which the analytic expression in the GA can be readily obtained from Eq.~\eqref{appeq:S123123}. We explicitly divide the result by $\nc^2$ as dictated from the $\nc$-\,counting in Eq.~\eqref{dW_1332_dY_ND}. We notice that the GA is not failing as dramatically as for $\mcal{W}$, with deviations of some tens of \% observed, but still it does not capture accurately the JIMWLK result. In fact, due to ``operator mixing'', these moderate deviations have an impact on the precise evaluation of $\mcal{W}$.  

The correlators that we have studied in this section demonstrate that differences between the GA and full JIMWLK can be large, up to a factor 2. They differ between different coordinate configurations. Thus the effect on specific cross sections must be studied on a case by case basis. In the following section we will quantify the effect on a relatively straightforward observable, a total incoherent diffractive cross section.  

\section{Cross Section for Incoherent Diffraction}
\label{sec:sigma}

Even though we do not aim to undertake a precise phenomenological study in the present work, let us now study a quantity which is closer to a physical observable. Cross sections for incoherent diffraction are most often measured differentially in $|t|$, i.e.~in the square of the momentum transferred from the target to the projectile. In a realistic phenomenological setting, at small values of $|t|$ this is dominated by physics of ultimately nonperturbative origin:~the fluctuations of nucleon positions in a nucleus, or of hot spot structures in the nucleon~\cite{Lappi:2010dd,Mantysaari:2016ykx,Demirci:2022wuy}. It would also be interesting to consider the effects of particle fluctuations in the high energy evolution due to loops of Pomerons \cite{Iancu:2004iy,Iancu:2005nj,Hatta:2006hs,Le:2021afn}, but these are higher order effects from the point of view of the standard CGC power counting and perhaps are not very significant for a large nuclear target. The color charge fluctuations, which are thoroughly analyzed in the present work and are incorporated both in the MV model and in the JIMWLK evolution for Wilson lines, are the dominant effect only at high $|t|$. In order to not be distracted by a modelling of the impact parameter dependence, here we will just consider the contribution of the color charge fluctuations integrated over $|t|$. We will quantify it in terms of  the ratio of the full JIMWLK calculation to the GA. But we emphasize that our results should be, for phenomenological purposes, considered as important corrections for the normalization of the large $|t|$-part of the incoherent cross section rather than for the whole range in $|t|$.

To this end, recall from Sect.~\ref{sec:W} that the jet momenta $\bk_1$ and $\bk_2$ are dual to $\bx-\bbx$ and $\by - \bby$ via Fourier transforms. Hence in the inclusive diffractive cross section an integration over the transverse momenta leads to identifying the CCA coordinates with the DA ones, that is $\bbx= \bx$ and $\bby=\by$. 
For the $t$-integrated diffractive cross section in a homogenous target, the integration over the impact parameter trivially gives its transverse area $S_{\perp}$ and finally one is left with just one integration in the transverse space, the one over the separation $\br \equiv \bx-\by$. As a result, the $q\bar{q}$ contribution to the total incoherent diffractive cross section for the $\gamma^*A$ process with transversely (T) or  longitudinally (L) polarized virtual photons reads
\begin{align}
	\label{sigma_inc}
	\sigma^\mathrm{inc}_\mathrm{T,L}(Q,\YP) = \frac{\alpha_\mathrm{em} \nc S_{\perp}
	\sum e_f^2}{\pi^2}
	\int \dif z\, \dif^2 \br \mkern1mu
	P_\mathrm{T,L}(z)\mkern1mu \bar{Q}^2 K_{1,0}^2(\bar{Q}r) \mcal{W}(r,\YP), 
\end{align}
while the coherent one is
\begin{align}
	\label{sigma_coh}
	\sigma^\mathrm{coh}_\mathrm{T,L}(Q,\YP) = \frac{\alpha_\mathrm{em} \nc S_{\perp}
	\sum e_f^2}{\pi^2}
	\int \dif z\, \dif^2 \br
	\mkern1mu P_\mathrm{T,L}(z)\mkern1mu 
	\bar{Q}^2 K_{1,0}^2(\bar{Q}r) [\mcal{T}(r,\YP)]^2, 
\end{align}
where 
\begin{align}
	\label{PLT}
	P_\mathrm{T}(z) = \frac{z^2 +(1-z)^2}{2}
	\quad \mathrm{and}  \quad
	P_\mathrm{L}(z) = 2 z (1-z).
\end{align}
Here quarks are taken to be massless, $\alpha_\mathrm{em}$ is the electromagnetic coupling constant, $e_f$ is the fractional electric charge of the flavor $f$ and $\bar{Q}^2 \equiv z(1-z) Q^2$. The  modified Bessel functions of the second kind  $K_{1,0}(\bar{Q}r)$  encode the $\gamma^*_{\mathrm{T},\mathrm{L}}\to q\bar{q}$ splitting, with the indices $n=1,0$ corresponding to the transverse and longitudinal polarizations respectively. 

\begin{figure}
\begin{center}
\includegraphics[width=0.49\textwidth]{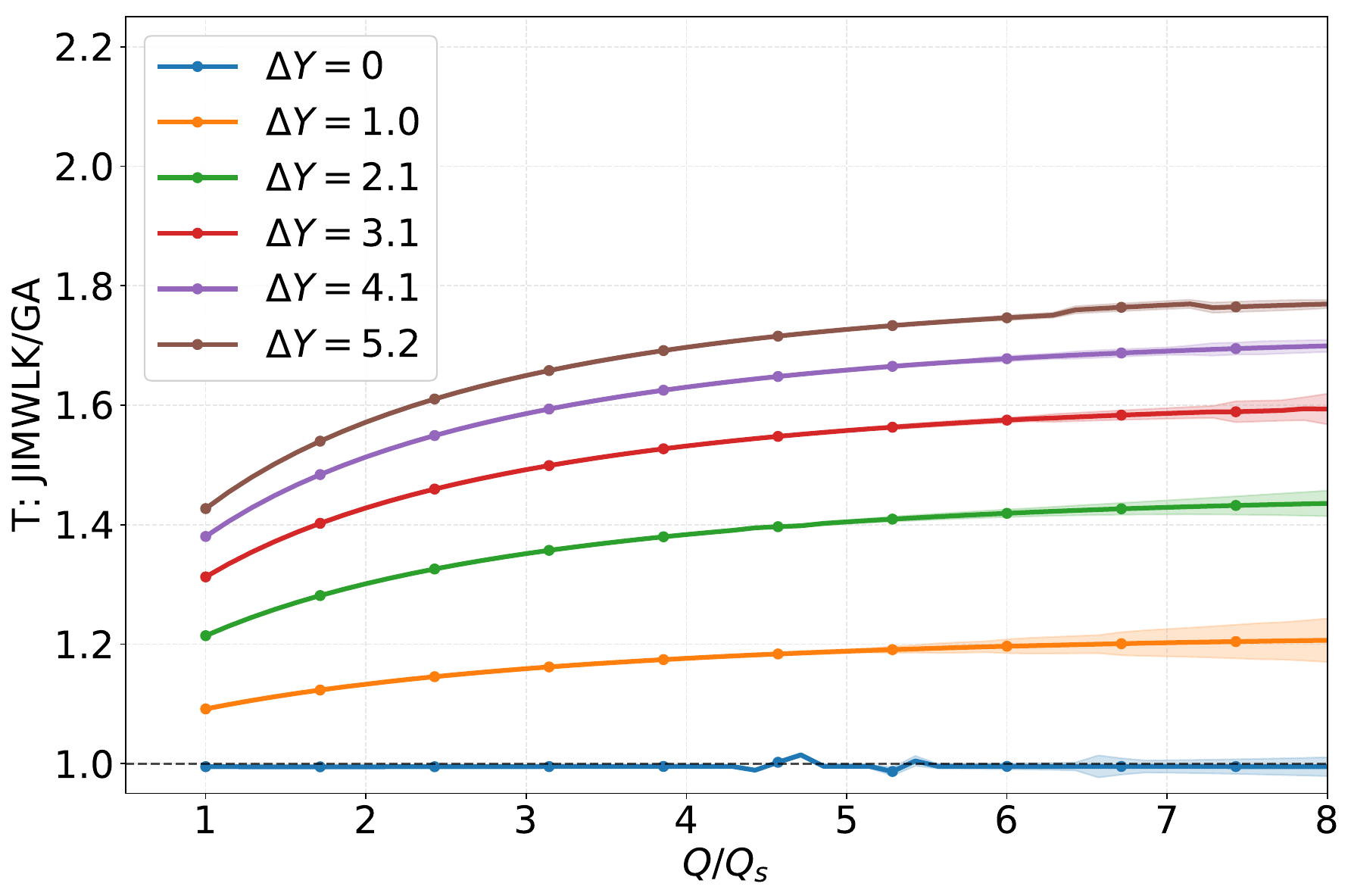}
\hspace*{0\textwidth}
\includegraphics[width=0.49\textwidth]{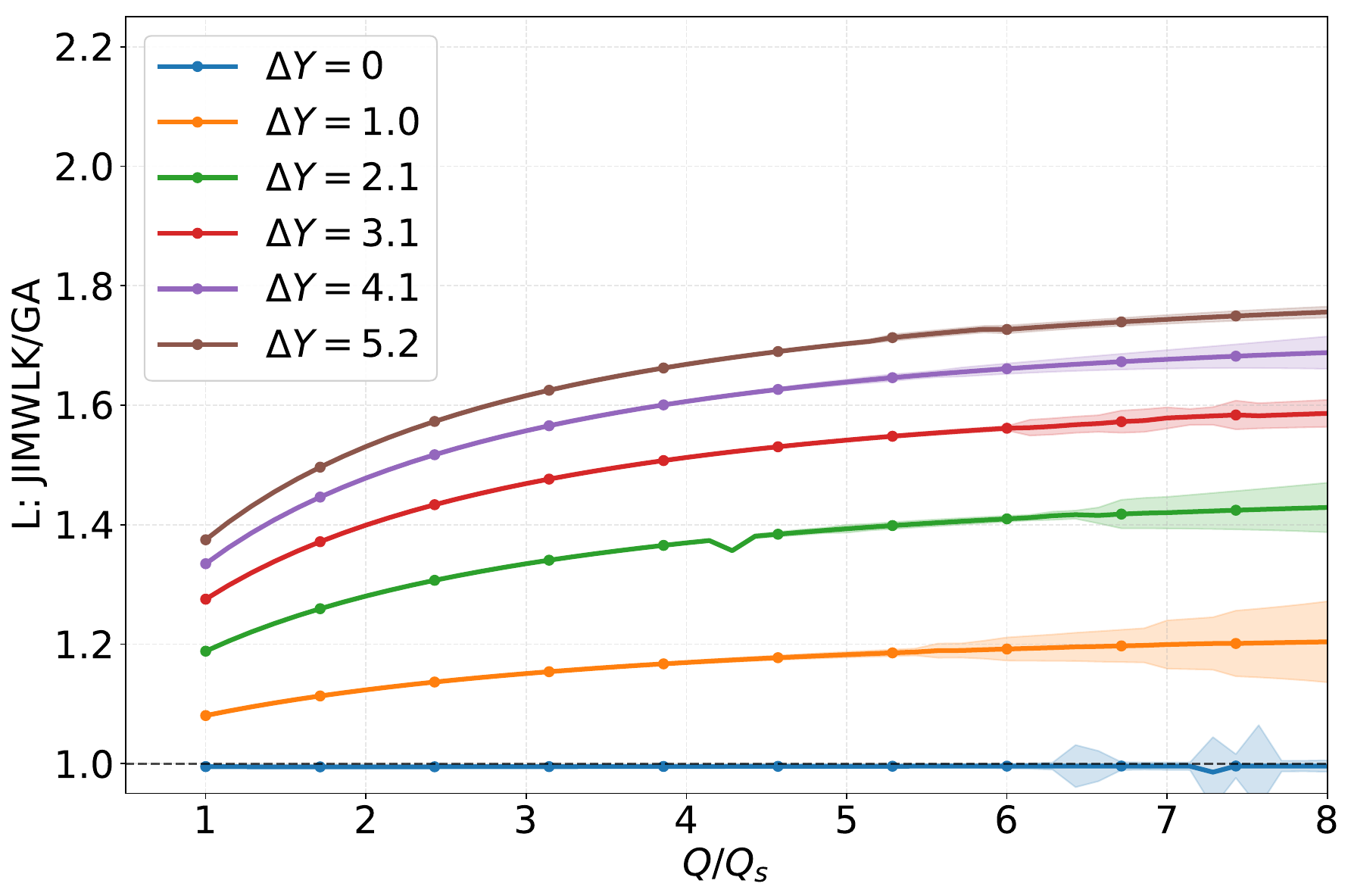}
\end{center}
\caption{\sl \small Numerical results for the transverse (left panel) and longitudinal (right panel) differential incoherent diffractive cross sections $\dif \sigma_\mathrm{T,L}^\mathrm{inc}/\dif z$: ratio of JIMWLK to GA results as a function of the scaled variable $Q/\qs(\YP)$ for various values of the rapidity gap $\YP$ at fixed fraction $z=1/2$.}
\label{fig:sigma_ratio}
\end{figure}

The correlator $\mcal{W}(r,\YP)$ appearing in Eq.~\eqref{sigma_inc} is precisely $\mcal{W}_{\bx\by\by\bx}(\YP)$ which has been numerically calculated in Sect.~\ref{sec:num}, cf.~Fig.~\ref{fig:W1221_num}. It is not difficult to further perform the integration over the separation $\br$, while we have opted to not integrate over the longitudinal fraction $z$ and in what follows we have assumed a fixed value $z=1/2$. In Fig.~\ref{fig:sigma_ratio} we show the evolution with $\YP$ of the ratio of the JIMWLK to GA results for the differential cross sections $\dif \sigma^\mathrm{inc}_\mathrm{T,L}/\dif z$. We confirm that the results for the incoherent diffractive correlator, c.f.~the right panel in Fig.~\ref{fig:W1221_num}, are transmitted to the corresponding cross sections. Indeed, with increasing $\YP$ the JIMWLK result is sizeably larger than the GA not only when $Q^2$ is high but also in the vicinity of the saturation scale $\qs^2(\YP)$.    

\begin{figure}
\begin{center}
\includegraphics[width=0.49\textwidth]{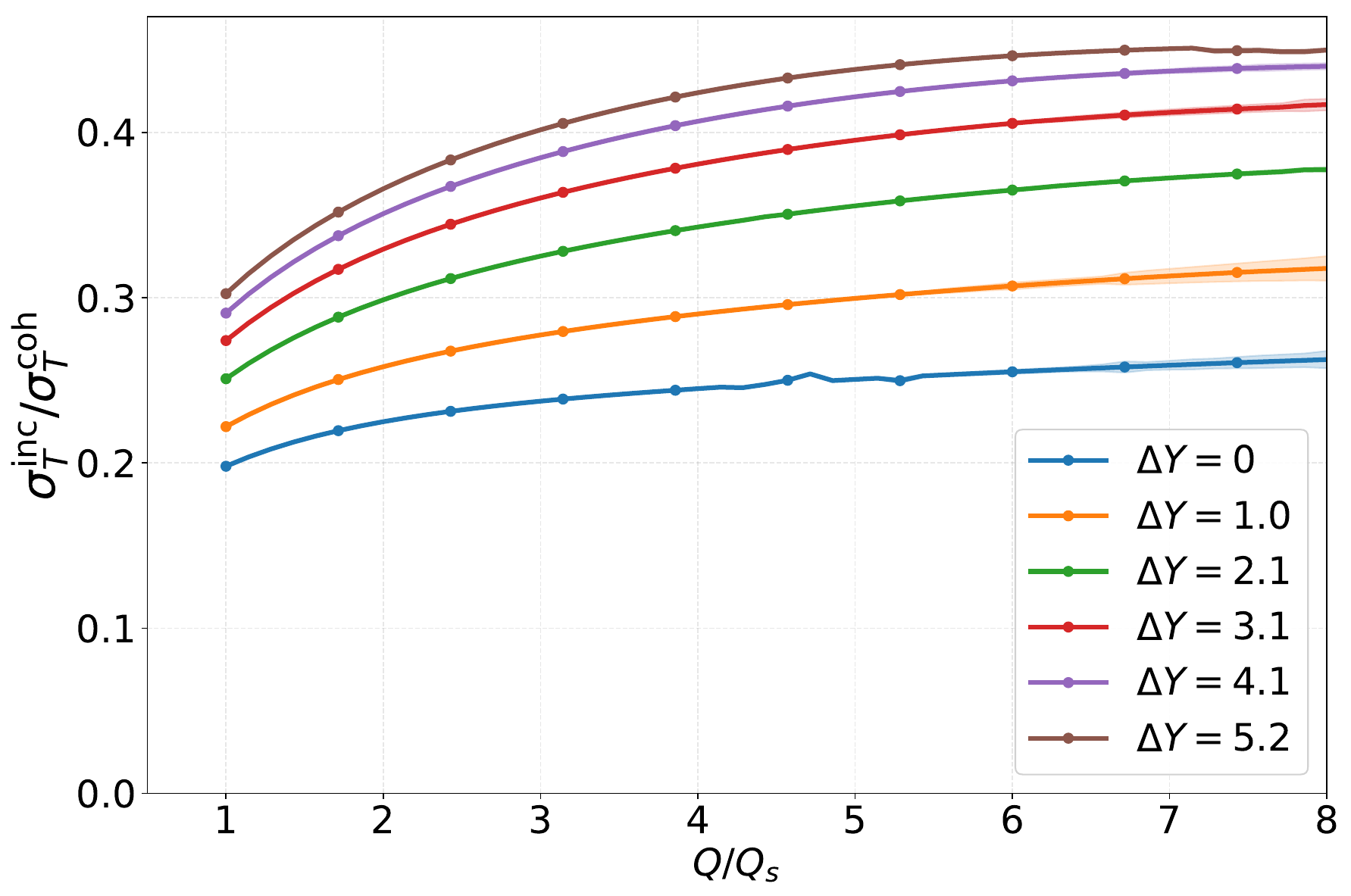}
\hspace*{0\textwidth}
\includegraphics[width=0.49\textwidth]{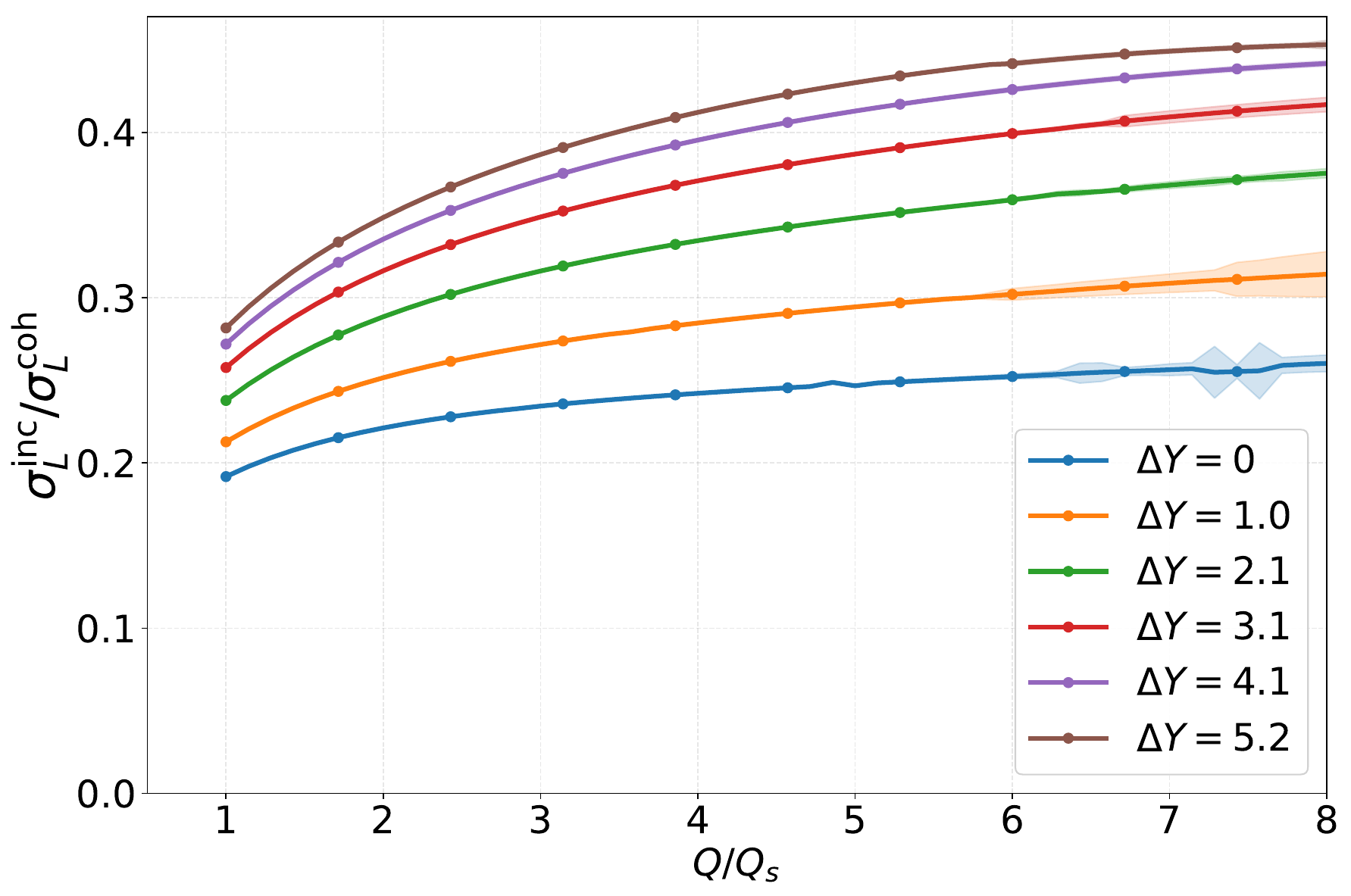}
\end{center}
\caption{\sl \small JIMWLK results for the ratio of incoherent (only color charge fluctuation contribution) to coherent diffractive cross sections $(\dif \sigma^\mathrm{inc}_\mathrm{T,L}/\dif z)/(\dif \sigma^\mathrm{coh}_\mathrm{T,L}/\dif z)$  as a function of the scaled variable $Q/\qs(\YP)$ for various values of the rapidity gap $\YP = Y_0+\Delta Y$ at fixed fraction $z=1/2$.  Left panel: transverse. Right panel: longitudinal.}
\label{fig:inc_coh}
\end{figure}

Finally, we can get an estimate for the overall size of the incoherent cross section compared to the coherent one. To this end we make a comparison with the coherent one in Fig.~\ref{fig:inc_coh}, by showing the ratio $(\dif \sigma^\mathrm{inc}_\mathrm{T,L}/\dif z)/(\dif \sigma^\mathrm{coh}_\mathrm{T,L}/\dif z)$ as obtained from JIMWLK evolution using  Eqs.~\eqref{sigma_inc} and \eqref{sigma_coh}. Such a ratio scales with $1/\nc^2$, however we find that its numerical value is larger than this parametrical estimate (in the dilute limit the incoherent cross section in fact has a prefactor $2/(\nc^2-1)$ compared to the coherent one) and therefore the incoherent cross section is more sizeable than what one may naively expect. We recall that, as discussed above, additional fluctuations visible at low $t$ values will further increase the incoherent cross section, which will  further increase the ratio~\cite{Mantysaari:2017dwh}. 

\section{Conclusion and perspectives}
\label{sec:conc}

In the framework of the Color Glass Condensate we have considered the ``connected'' piece of the double dipole (in the fundamental representation) scattering amplitude, a correlator appearing in incoherent diffraction in photon-nucleus collisions. This is a non-trivial example of a quantity which starts with a four gluon exchange in the limit of weak scattering. We have performed a detailed study of this double dipole correlator,  and a sextupole correlator that similarly starts at four gluon exchange, in the transverse plane and as a function of rapidity. We have explicitly shown both analytically and numerically that the Gaussian Approximation does not capture accurately enough the results obtained from JIMWLK evolution (despite the fact that such an approximation is reliable for correlators which start with a two gluon exchange) and in fact it systematically underestimates them. Hence, it is clear that for observables which involve such kind of correlators, using JIWMLK evolution (in its Langevin form) becomes inevitable.

From a theoretical and conceptual point of view, it would be interesting to find out if there are certain regimes in the transverse space which lead to the observed discrepancy between the JIMWLK and GA results. It would be also useful to understand the kind of special correlations which are generated by the noise in the Langevin equation but are missed when using the GA. Moreover, one could study in more detail the linear evolution of the incoherent diffractive correlator, e.g. similarly to Ref.~\cite{Lappi:2019kif} or  by trying to recast it into an eigenvalue problem and eventually exploring if there is a particular solution which determines the growth with energy.

Eyeing a bit towards more detailed phenomenological studies, one could study the implications of our results for the problem of dijet production in incoherent diffraction in $\gamma^*A$ collisions. Here, an important additional aspect for phenomenology that we have not considered here is that  when the dijet transverse imbalance is much smaller than the transverse momentum of each of the two jets, one must take into account the $q\bar{q}g$ contribution in which one of the partons is integrated over~\cite{Iancu:2022lcw,Iancu:2023lel,Hauksson:2024bvv,Rodriguez-Aguilar:2024efj}. When integrating over the gluon, one encounters the connected piece of a double dipole in the {\it adjoint} representation, which will have to be calculated using the JIMWLK equation. Clearly, one does not need to restrict oneself to $\gamma^*A$ collisions, but can study similar  problems in, for example, $pA$ collisions\footnote{Notice that in this case the diffractive projection on the proton side (the ``projectile'') will depend on non-perturbative physics.}. Before closing, let us also note that four gluon exchanges at lowest order also appear in the so-called ``glasma graphs'' \cite{Dusling:2009ni,Dusling:2013oia}  in the context of unequal rapidity correlations in particle production at high energy, although setting up the JIMWLK evolution for  such a case \cite{Iancu:2013uva,Lappi:2019kif} is significantly more complicated.

\begin{acknowledgments}
TL work was supported by the Research Council of Finland, the Centre of Excellence in Quark Matter (project 346324 and 364191) and by the European Research Council (ERC, grant agreement No. ERC-2018-ADG-835105 YoctoLHC). The content of this article does not reflect the official opinion of the European Union and responsibility for the information and views expressed therein lies entirely with the authors. TL also acknowledges support from grant NSF PHY-2309135 to the Kavli Institute for Theoretical Physics (KITP). 
\end{acknowledgments}
 
\appendix

\section{The double dipole correlator and simple sextupoles in the GA}
\label{app:W}

\subparagraph{Double dipole:\!\!} Here we first give the expression for the average of a product of two fundamental dipole $S$-matrices in the GA. For simplicity we further assume that the average dipole $S$-matrix depends only on the magnitude of the dipole separation, i.e.~$\mcal{D}_{\bx\by} = \mcal{D}(r_{\bx\by})$ in our compact notation. Then, in a form convenient to our purposes, we have \cite{Dominguez:2008aa}
\begin{align}
	\label{appeq:DD}
	\left \langle 
	D_{\bx\by} D_{\bby\bbx} 
	\right \rangle^\mathrm{GA}
	= \,&  
	\mcal{D}_{\bx\by}  
	\mcal{D}_{\bby\bbx}\, 
	e^{\textstyle -\frac{F}{2} + \frac{f}{\nc^2}}
	\nn
	& \times 
	\left[
	\left(
	\frac{\sqrt{\Delta} + F}{2 \sqrt{\Delta}}
	-\frac{f}{\nc^2 \sqrt{\Delta}}
	\right)
	e^{ \textstyle \frac{\sqrt{\Delta}}{2}}
	+\left(
	\frac{\sqrt{\Delta} - F}{2 \sqrt{\Delta}}
	+\frac{f}{\nc^2 \sqrt{\Delta}}
	\right)
	e^{ \textstyle -\frac{\sqrt{\Delta}}{2}}
	\right],
\end{align}
where we have defined
\begin{align}
	\label{appeq:DeltaFf}
	\Delta = F^2 - \frac{4}{\nc^2}\, 
	f (F-f),
	\quad
	F= \frac{\nc}{2 \cf} 
	\ln \frac{\mcal{D}_{\bx\by}
	\mcal{D}_{\bby\bbx}}
	{\mcal{D}_{\bx\bbx}
	\mcal{D}_{\by\bby}}
	\quad \mathrm{and}
	\quad
	f= 
	\frac{\nc}{2 \cf}
	\ln \frac{\mcal{D}_{\bx\bby} 
	\mcal{D}_{\by\bbx}}
	{\mcal{D}_{\bx\bbx}
	\mcal{D}_{\by\bby}}.
\end{align}
To construct the incoherent diffractive correlator $\mcal{W}_{\bx\by\bby\bbx}^{\mathrm{GA}}$ defined in Eq.~\eqref{W}, ones simply subtracts the ``disconnected'' piece $\mcal{D}_{\bx\by} \mcal{D}_{\bby\bbx}$ from Eq.~\eqref{appeq:DD}. When we identify the transverse positions of the antiquark of the first dipole and of the quark of the second dipole the above simplifies considerably and gives
\begin{align}
	\label{appeq:DD1332}
	\left \langle 
	D_{\bx\by} D_{\by\bbx} 
	\right \rangle^\mathrm{GA}
	 = \frac{\nc^2-1}{\nc^2}\,
	 \frac{(\mcal{D}_{\bx\by}
	 \mcal{D}_{\by\bbx})^
	 {\nc^2/(\nc^2-1)}}
	 {\mcal{D}_{\bx\bbx}^
	 {1/(\nc^2-1)}}
	 +\frac{\mcal{D}_{\bx\bbx}}{\nc^2}.
\end{align}
Simple expressions can also be given for the ``square'' configurations studied in Sect.~\ref{sec:num} and presented in Fig.~\ref{fig:Wsq_num}. For the one in the left panel one finds that $\left \langle D_{\bx\by} D_{\bby\bbx} \right \rangle^\mathrm{GA}$ factorizes exactly and hence $\mcal{W}_{\bx\by\bby\bbx}^{\mathrm{GA}}$ vanishes. For the other two we have
\begin{align}
	\hspace{-0.7cm}
	\label{appeq:Wsq_m}
	\left \langle 
	D_{\bx\by} D_{\bby\bbx} 
	\right \rangle^\mathrm{GA}_{\square,\mathrm{middle}}
	=
	\mcal{D}^2(r)
	\left\{
	\frac{\nc+1}{2\nc}
	\left[\frac{\mcal{D}^2(r)}{\mcal{D}^2(\sqrt{2} \hor r)}\right]^
	{1/(\nc+1)}
	+
	\frac{\nc-1}{2\nc}
	\left[\frac{\mcal{D}^2(\sqrt{2}\hor r)}{\mcal{D}^2(r)}\right]^
	{1/(\nc-1)}
	\right\}
\end{align}
and 
\begin{align}
	\hspace{-0.7cm}
	\label{appeq:Wsq_r}
	\left \langle 
	D_{\bx\by} D_{\bby\bbx} 
	\right \rangle^\mathrm{GA}_{\square,\mathrm{right}}
	=
	\mcal{D}^2(r)
	\left\{
	\frac{\nc^2-1}{\nc^2}
	\left[\frac{\mcal{D}^2(r)}{\mcal{D}^2(\sqrt{2} \hor r)}\right]^
	{1/(\nc^2-1)}
	+
	\frac{1}{\nc^2}\,
	\frac{\mcal{D}^2(\sqrt{2}\hor r)}{\mcal{D}^2(r)}
	\right\}.
\end{align}

It is also instructive to further take the large-$\nc$ limit of Eq.~\eqref{appeq:DD} and find
\begin{align}
	\label{appeq:W_large_Nc}
	\mcal{W}_{\bx\by\bby\bbx}^\mathrm{GA}
	\simeq
	\frac{\mcal{D}_{\bx\by}  
	\mcal{D}_{\bby\bbx}}
	{\nc^2}\, 
	\frac{f^2}{F^2}
	\left(
	e^{-F} -1 + F
	\right)
	\quad \mathrm{for} \quad\nc \gg 1,
\end{align}
where, to the order of accuracy, one should replace $2 \cf \to 
\nc$ for $F$ and $f$ in Eq.~\eqref{appeq:DeltaFf}. Now one sees explicitly the $1/\nc^2$ dependence of $\mcal{W}_{\bx\by\bby\bbx}$ independently of the configuration and the sizes of the projectile partons.  Still, we point out that throughout this work we make use of the finite\,-$\nc$ expression as determined by Eq.~\eqref{appeq:DD}.

\subparagraph{A sextupole at weak scattering:\!\!} In Sect.~\ref{sec:break} we were in need of calculating the weak scattering limit of the sextupole $\mcal{S}_{\bx\bm{z}\by\bbx\bm{z}\by}$ in the GA. To do so, we first write its ``evolution equation'' using the Gaussian Hamiltonian in Eq.~\eqref{H_GA} and we obtain
\begin{align}
	\label{appeq:dsext_dY}
	\frac{\dif \mcal{S}^\mathrm{GA}_{\bx\bm{z}\by\bbx\bm{z}\by}}{\dif y} 
	= 
	& - \frac{g^2 \nc}{2} 
	(\gamma_{\bx\by} + \gamma_{\bbx\by}
	+\gamma_{\bx\bm{z}} + \gamma_{\bbx\bm{z}} + 2\gamma_{\by\bm{z}})\,
	\mcal{S}_{\bx\bm{z}\by\bbx\bm{z}\by}
	+\frac{g^2}{\nc}\,\gamma_{\bx\bbx}\,
	\mcal{S}_{\bx\bm{z}\by\bbx\bm{z}\by}
	\nn
	& +\frac{g^2 \nc}{2}
	(\gamma_{\bx\by} + \gamma_{\bbx\by} 
	-\gamma_{\bx\bm{z}} - \gamma_{\bbx\bm{z}} + 2 \gamma_{\by\bm{z}}) 
	\langle D_{\bx\bm{z}} D_{\bm{z}\bbx} \rangle
	\nn
	& +\frac{g^2 \nc}{2}
	(-\gamma_{\bx\by} - \gamma_{\bbx\by} 
	+\gamma_{\bx\bm{z}}
	+ \gamma_{\bbx\bm{z}}
	+ 2\gamma_{\by\bm{z}} 
	) 
	\langle D_{\bx\by} D_{\by\bbx} \rangle
	\nn
	& +\frac{g^2 \nc}{2}
	(\gamma_{\bbx\by}- \gamma_{\bx\bbx}
	+ \gamma_{\bx\bm{z}} -\gamma_{\by\bm{z}} )
	\langle D_{\bm{z}\by} Q_{\bx\bm{z}\by\bbx} \rangle
	\nn
	& +\frac{g^2 \nc}{2}
	(\gamma_{\bx\by}- \gamma_{\bx\bbx}
	+ \gamma_{\bbx\bm{z}} -\gamma_{\by\bm{z}} )
	\langle D_{\by \bm{z}} Q_{\bx\bbx\bm{z}\by} \rangle. 
\end{align}
Since we are working at order $\alpha_a^4$ and since the $\gamma$'s are already of order $\alpha_a^2$, we can greatly simplify the r.h.s.~of the above equation by making use of the expansions (and of those with the necessary permutations of coordinates)
\begin{align}
	\label{appeq:expansions}
	&\hspace{-0.5cm}
	\langle D_{\bx\bz} D_{\bz\bbx} \rangle
	= \mcal{D}_{\bx\bz} \mcal{D}_{\bz\bbx}
	+ \mcal{O}\left(\alpha_a^4\right) = 
	1 - \mcal{T}_{\bx\bz} - \mcal{T}_{\bz\bbx}
	+ \mcal{O}\left(\alpha_a^4\right)
	\nn[0.1cm]
	&\hspace{-0.5cm}\langle D_{\bm{z}\by} Q_{\bx\bm{z}\by\bbx} \rangle =
	\mcal{D}_{\bm{z}\by} \mcal{Q}_{\bx\bm{z}\by\bbx} 
	+ \mcal{O}\left(\alpha_a^4\right)
	=  
	1 - \mcal{T}_{\bx\bz} + \mcal{T}_{\bbx\bz} -2 \mcal{T}_{\by\bz}
	- \mcal{T}_{\bx\bbx} + \mcal{T}_{\bx\by} - \mcal{T}_{\bbx\by} + 
	\mcal{O}\left(\alpha_a^4\right)
	\nn[0.1cm]
	&\hspace{-0.4cm}\mcal{S}_{\bx\bm{z}\by\bbx\bm{z}\by} = 1 
	+ \mcal{O}\left(\alpha_a^4\right).
\end{align}
It worths emphasizing that the first two of these equations are in fact special cases of a more general property: at weak scattering, one can always factorize an average of a product of operators to the product of the individual averages with the error being of order $\alpha_a^4$. Now, when the $\gamma$'s in Eq.~\eqref{appeq:dsext_dY} multiply the unity in equations \eqref{appeq:expansions} all terms cancel. Moreover, when the $\gamma$'s multiply the $\mcal{T}$'s, we can make use of
\begin{align}
	\label{appeq:gamma_T}
	2 g^2 \cf \mkern1mu \gamma_{\bx\by} = \frac{\dif \mcal{T}_{\bx\by}}{\dif Y} +\mcal{O}\left(\alpha_a^4\right)
\end{align}
and we eventually discover that Eq.~\eqref{appeq:dsext_dY} can be conveniently written as a total derivative. Assuming that in the initial condition the scattering vanishes and integrating up to $Y$ we obtain
\begin{align}
	\label{appeq:sext_dip_4g}
	\hspace{-0.4cm}
	\mcal{S}^\mathrm{GA}_{\bx\bm{z}\by\bbx\bm{z}\by} 
	- \mcal{D}_{\bx\bbx} =  
	\frac{\nc^2}{\nc^2 - 1}
	\big[&
	(\mcal{T}_{\bx\bz} - \mcal{T}_{\by\bm{z}}) 
	(\mcal{T}_{\bbx\bz} - \mcal{T}_{\by\bm{z}})
	+2 \mcal{T}_{\bx\bbx} \mcal{T}_{\by\bm{z}}
	\nn
	&-\mcal{T}_{\bx\by}
	(\mcal{T}_{\bbx\bz}  + \mcal{T}_{\by\bz})
	-\mcal{T}_{\bbx\by}
	(\mcal{T}_{\bx\bz}  + \mcal{T}_{\by\bz}) 
	+\mcal{T}_{\bx\by} \mcal{T}_{\bbx\by}
	\big]
	+\mcal{O}\left(\alpha_a^6\right).   
\end{align}

\subparagraph{Another sextupole in the full GA:\!\!} Finally, in Sections \ref{sec:break} and \ref{sec:num} we also came across $\mcal{S}_{\bx\bz\by\bx\bz\by}-1$ which starts at order $\alpha_a^4$ while it is leading in $\nc$. It can be analytically computed in the GA including unitarity corrections. Indeed, a quick inspection of Eq.~\eqref{appeq:dsext_dY}, after we identify $\bbx$ with $\bx$, reveals that the r.h.s.~involves only $\mcal{S}_{\bx\bz\by\bx\bz\by}$ and $\langle D_{\bx\bz} D_{\bz\bx} \rangle$ and its two permutations. Such a double dipole is readily evaluated in the GA as a special case of Eq.~\eqref{appeq:DD1332}. We further employ Eq.~\eqref{gammaexp} to replace the kernels  in terms of the dipoles and we find that Eq.~\eqref{appeq:dsext_dY} reduces to a closed first order inhomogeneous differential equation, which after careful integration leads to
\begin{align}
	\label{appeq:S123123}
	1-\mcal{S}_{\bx\bz\by\bx\bz\by}
	= 
	\frac{2 \cf}{\nc}
	\left[1- 
	\mcal{D}_{\bx\bz}^{\textstyle \frac{\nc}{\cf}}
	-
	\mcal{D}_{\bx\by}^{\textstyle \frac{\nc}{\cf}}
	-
	\mcal{D}_{\bz\by}^{\textstyle \frac{\nc}{\cf}}
	+2 (\mcal{D}_{\bx\bz} \mcal{D}_{\bx\by} \mcal{D}_{\bz\by})
	^{\textstyle \frac{\nc}{2\cf}}
	\right].
\end{align}

\section{Integrations in the GBW model}
\label{app:int}

Here we would like to derive the analytic results for the GBW model given in Sect.~\ref{sec:break}. The most convenient way to perform the integration over the gluon transverse coordinate $\bz$ is by using a Feynman parametrization. 
\subparagraph{Gaussian approximation:\!\!} Let us start from the expression in the GA and consider the first term in the square bracket in Eq.~\eqref{dW_GA_1332_weak_dY}. We first let $\bz \to \bz + \bx$ and then we integrate for $z \leq R$. At the moment we focus only on the dipole kernel for which we have
\begin{align}
	\label{appeq:feynman}
	\int
	\frac{\dif^2 \bz}{2\pi }\, \mcal{M}_{\bx\bbx\bz} 
	\left[\cdots \right]
	&= 
	\int_{z \leq R}
	\frac{\dif^2 \bz}{2\pi }\, \frac{r_{\bx\bbx}^2}{z^2 (\bz - \br_{\bbx\bx})^2}
	\left[\cdots \right]
	\nn[0.1cm]
	&= \int_{0}^1
	\dif \lambda 
	\int_{z \leq R}
	\frac{\dif^2 \bz}{2\pi }\,
	\frac{r_{\bx\bbx}^2}{\big[(1-\lambda)z^2 
	+\lambda(\bz-\br_{\bbx\bx})^2
	 \big]^2}  
	 \left[\cdots \right]
	\nn[0.1cm]
	 & =
	\int_{0}^1
	\dif \lambda 
	\int_{z \leq R}
	\frac{\dif^2 \bz}{2\pi }\,
	\frac{r_{\bx\bbx}^2}
	{\big[(\bz - \lambda \br_{\bbx\bx})^2 
	+\lambda(1-\lambda) r_{\bx\bbx}^2\big]^2} 
	\left[\cdots \right]
	\nn[0.1cm]
	& 
	\simeq
	\int_{0}^1
	\dif \lambda 
	\int_{z \leq R}
	\frac{\dif^2 \bz}{2\pi }\,
	\frac{r_{\bx\bbx}^2}
	{\big[z^2 
	+\lambda(1-\lambda) r_{\bx\bbx}^2\big]^2}\
	\left[\cdots \right],
\end{align}
with $\br_{\bbx\bx} \equiv \bbx-\bx$ and where in order to arrive at the last expression we have further let $\bz \to \bz + \lambda \br_{\bbx\bx}$. When doing so, in principle one should have changed the integration regime according to $z \leq R - \lambda r_{\bx\bbx} \cos\phi$ (plus corrections of order $r_{\bx\bbx}^2/R$), where $\phi$ is the angle between $\bz$ and $\br_{\bbx\bx}$. However, it is a simple exercise to show that such a correction in the upper limit of the $z$-integration leads to terms which can be neglected to the accuracy of interest, since they are suppressed by a factor $r_{\bx\bbx}^2/R^2$ when compared to the leading one (given in Eq.~\eqref{appeq:int_GA} below). Whenever it is convenient in this Appendix, we shall make use of this property in order to shift the center of the disk by an amount of the order of the desired intercharge distance and keep at the same time $R$ as the upper limit in the $z$-integration.

The combined shift $\bz \to \bz + \bx + \lambda \br_{\bbx\bx}$ for the amplitudes inside the integrand gives
\begin{align}
	\label{appeq:SumT_GBW}
	\mcal{T}_{\bx\bz} + \mcal{T}_{\bbx\bz} -\mcal{T}_{\bx\bbx} 
	\to
	\frac{\qs^2}{4} \left[ 2 z^2 +(2\lambda-1) \bz \cdot \br_{\bbx\bx} - 2 \lambda(1-\lambda)r_{\bx\bbx}^2 \right]. 
\end{align}
The term linear in $\bz$ will vanish upon integration since the integrand in the last line in Eq.~\eqref{appeq:feynman} depends only on $z^2$. The two remaining terms give
\begin{align}
	\label{appeq:int_GA}
	\hspace{-0.6cm}
	\int
	\frac{\dif^2 \bz}{2\pi }\,
	\mcal{M}_{\bx\bbx\bz}\,
	( \mcal{T}_{\bx\bz} + \mcal{T}_{\bbx\bz} -\mcal{T}_{\bx\bbx})
	&\simeq
	\frac{r_{\bx\bbx}^2\qs^2}{4}
	\int_{0}^1
	\dif \lambda 
	\int_0^{R^2}
	\dif z^2 \,\frac{z^2 -\lambda(1-\lambda)r_{\bx\bbx}^2}
	{\big[z^2 
	+\lambda(1-\lambda) r_{\bx\bbx}^2\big]^2}
	\nn[0.1cm] 
	& \simeq \frac{r_{\bx\bbx}^2\qs^2}{4} 
	\int_{0}^1
	\dif \lambda 
	\left[
	\ln \frac{R^2}{\lambda(1-\lambda)r_{\bx\bbx}^2}
	-2
	\right]
	= \frac{r_{\bx\bbx}^2\qs^2}{4} 
	\ln \frac{R^2}{r_{\bx\bbx}^2},
\end{align} 
where we have neglected terms which vanish when $R\to \infty$. Similarly, the terms in Eq.~\eqref{dW_GA_1332_weak_dY} with kernels $\mcal{M}_{\bx\by\bz}$ and $\mcal{M}_{\bbx\by\bz}$ are obtained from the above just by replacing $r_{\bx\bbx}^2$ with $r_{\bx\by}^2$ and $r_{\bbx\by}^2$ respectively. Finally we multiply with the prefactor in Eq.~\eqref{dW_GA_1332_weak_dY} to arrive at Eq.~\eqref{dW_GA_1332_GBW}.

\subparagraph{Dipole term:\!\!} Let us move on to the (weak field) JIMWLK evolution of $\mcal{W}_{\bx\by\by\bbx}$ and in particular to the dipole contribution in Eq.~\eqref{dW_1332_dY_D_T}. Clearly it suffices to consider only one term, say the one involving $\mcal{M}_{\bx\by\bz}$, since the other term can be obtained by the simple exchange $\bx \leftrightarrow \bbx$. In complete analogy to Eq.~\eqref{appeq:feynman} we can write 
\begin{align}
	\label{appeq:M13z}
	\int
	\frac{\dif^2 \bz}{2\pi }\, \mcal{M}_{\bx\by\bz} 
	\left[\cdots \right]
	\simeq 
	\int_{0}^1
	\dif \lambda 
	\int_{z \leq R}
	\frac{\dif^2 \bz}{2\pi }\,
	\frac{r_{\bx\by}^2}
	{\big[z^2 
	+\lambda(1-\lambda) r_{\bx\by}^2\big]^2}
	\left[\cdots \right],
\end{align}
where now the combined shift is $\bz \to \bz + \by + \lambda \br_{\bx\by}$. Such a shift in the combination of the $\bz$-dependent amplitudes gives
\begin{align}
	\label{appeq:TT_D}
	\mcal{T}_{\by\bz} - \mcal{T}_{\bbx\bz}
	\to
	\frac{\qs^2}{4} 
	\left(
	2\bz \cdot \br_{\bbx\by} 
	+2 \lambda \br_{\bx\by} \cdot \br_{\bbx\by} 
	- r_{\bbx\by}^2
	\right) 
\end{align}
and then
\begin{align}
\hspace*{-0.8cm}
	\label{appeq:TT2_D}
	(\mcal{T}_{\by\bz} - \mcal{T}_{\bbx\bz})^2
	\to
	\frac{\qs^4}{16} 
	\left[4(\bz \cdot \br_{\bbx\by})^2
	+4 (2 \lambda \br_{\bx\by} \cdot \br_{\bbx\by}
	  - r_{\bbx\by}^2) (\bz \cdot \br_{\bbx\by}) 
	  +(2 \lambda \br_{\bx\by} \cdot \br_{\bbx\by}
	  - r_{\bbx\by}^2)^2  
	  \right]. 
\end{align}
The linear terms in $\bz$, that is the first term in Eq.~\eqref{appeq:TT_D} and the second term in Eq.~\eqref{appeq:TT2_D}, will vanish upon integration. Making use of these expressions, we find after straightforward algebra that the square bracket in the integrand in Eq.~\eqref{dW_1332_dY_D_T} reduces to
\begin{align}
\label{appeq:allT_D}
\hspace*{-1cm}
	\left[ \cdots \mkern-0.1mu \right]
	\to 
	\frac{\qs^4}{16} 
	\left[
	4(\bz \cdot \br_{\bbx\by})^2
	-4 \lambda (1-\lambda)(\br_{\bx\by} \cdot \br_{\bbx\by})^2
	\right]
	\to
	\frac{\qs^4}{16} 
	\left[
	2 r_{\bbx\by}^2 z^2
	-4 \lambda (1-\lambda)(\br_{\bx\by} \cdot \br_{\bbx\by})^2
	\right],
\end{align}
where we have been allowed to average over the angle between $\bz$ and $\br_{\bbx\by}$ in order to write the last expression. Using \eqref{appeq:M13z} and \eqref{appeq:allT_D}, the term under study in Eq.~\eqref{dW_1332_dY_D_T} becomes
\begin{align}
	\label{appeq:int_D_temp}
	\int
	\frac{\dif^2 \bz}{2\pi }\, \mcal{M}_{\bx\by\bz} 
	\left[\cdots \right]
	=
	\frac{\qs^4}{16}\,r_{\bx\by}^2 
	\int_{0}^1
	\dif \lambda 
	\int_0^{R^2}
	\dif z^2\,
	\frac{r_{\bbx\by}^2 z^2
	-2 \lambda (1-\lambda)(\br_{\bx\by} \cdot \br_{\bbx\by})^2}
	{\big[z^2 
	+\lambda(1-\lambda) r_{\bx\by}^2\big]^2}.
\end{align}
Performing the integrations, neglecting terms which are suppressed for large $R$ and using $2 \br_{\bx\by} \cdot \br_{\bbx\by} =r_{\bx\by}^2+r_{\bbx\by}^2-r_{\bx\bbx}^2$ we find
\begin{align}
	\label{appeq:int_D}
	\int
	\frac{\dif^2 \bz}{2\pi }\, \mcal{M}_{\bx\by\bz} 
	\left[\cdots \right]
	=
	\frac{\qs^4}{16}
	\left[r_{\bx\by}^2 r_{\bbx\by}^2
	\left(\ln \frac{R^2}{r_{\bx\by}^2}+1
	\right)
	-\frac{1}{2}
	\left( r_{\bx\bbx}^2 - r_{\bx\by}^2 - r_{\bbx\by}^2 \right)^2
		\right].
\end{align}
As said, to find the $\mcal{M}_{\bbx\by\bz}$ contribution it suffices to let $\bx\leftrightarrow \bbx$ in the above. Putting together the two pieces we arrive at Eq.~\eqref{dW_1332_GBW_D}.

\subparagraph{Non-dipole term:\!\!} Finally we consider the non-dipole contribution in Eq.~\eqref{dW1332_dY_ND_T}. For the $\mcal{M}_{\bx\bbx\bz}$ term we do the combined shift $\bz \to \bz + \bx + \lambda \br_{\bbx\bx}$, so that we can use Eq.~\eqref{appeq:feynman}. After some tedious algebra we find that the square bracket which contains all the quadratic scattering amplitude terms reduces to
\begin{align}
\label{appeq:allT_ND_temp}
	\hspace{-0.4cm}
	\left[ \cdots \right]
	\to 
	\frac{\qs^4}{16} 
	\left\{
	4(\bz \cdot \br_{\bx\by})(\bz \cdot \br_{\bbx\by}) 
	-4 \br_{\bx\by} \cdot \br_{\bbx\by}\, z^2 
	+4 \lambda (1-\lambda) 
	\left[r_{\bx\by}^2 r_{\bbx\by}^2 -(\br_{\bx\by} 
	\cdot \br_{\bbx\by})^2\right]
	\right\},
\end{align}
where we have neglected the terms linear in $\bz$ since they vanish upon integration. When we average over the angle between $\bz$ and (say) $\br_{\bx\by}$, \eqref{appeq:allT_ND_temp} simplifies to
\begin{align}
\label{appeq:allT_ND}
	\left[ \cdots \right]
	\to 
	\frac{\qs^4}{16} 
	\left[
	-2 r_{\bx\by} r_{\bbx\by} 
	\cos\theta z^2 
	+ 4 \lambda (1-\lambda) r_{\bx\by}^2 r_{\bbx\by}^2 \sin^2\mkern-2mu\theta	
	\right],
\end{align}
where $\theta$ is the angle between $\br_{\bx\by}$ and $\br_{\bbx\by}$ already introduced in Sect.~\ref{sec:break}. Using \eqref{appeq:feynman} and \eqref{appeq:allT_ND}, the term under study in Eq.~\eqref{dW1332_dY_ND_T} becomes
\begin{align}
	\label{appeq:int_ND_temp}
	\hspace*{-0.6cm}
	\int
	\frac{\dif^2 \bz}{2\pi }\, \mcal{M}_{\bx\bbx\bz} 
	\left[\cdots \right]
	=
	\frac{\qs^4}{16}\,r_{\bx\bbx}^2 
	\int_{0}^1
	\dif \lambda 
	\int_0^{R^2}
	\dif z^2\,
	\frac{-r_{\bx\by} r_{\bbx\by} \cos\theta z^2 
	+ 2 \lambda (1-\lambda) 
	r_{\bx\by}^2 r_{\bbx\by}^2 \sin^2\mkern-2mu\theta}
	{\big[z^2 
	+\lambda(1-\lambda) r_{\bx\bbx}^2\big]^2}.
\end{align}
We perform the integrations, we neglect once again terms which are suppressed for large $R$ and we use Eq.~\eqref{theta} to express $\cos\theta$ and $\sin^2\mkern-2mu\theta$ in terms of the three sizes to obtain 
\begin{align}
	\hspace*{-1cm}
	\label{appeq:int_ND}
	\int\!
	\frac{\dif^2 \bz}{2\pi }\, \mcal{M}_{\bx\bbx\bz} 
	\left[\cdots \mkern-0.1mu \right]
	=
	\frac{\qs^4}{32}
	\left[r_{\bx\bbx}^2 
	\big(r_{\bx\bbx}^2 - r_{\bx\by}^2 -r_{\bbx\by}^2\big)
	\ln \frac{R^2}{r_{\bx\bbx}^2}
	+r_{\bx\bbx}^2 \big(r_{\bx\by}^2 + r_{\bbx\by}^2 \big) 
	\!-\! 
	\big(r_{\bx\by}^2 - r_{\bbx\by}^2\big)^2
	\right].
\end{align}
For the term involving $\mcal{M}_{\bx\by\bz}$ in Eq.~\eqref{dW1332_dY_ND_T} we do the combined shift $\bz \to \bz + \by + \lambda \br_{\bx\by}$ and we can use Eq.~\eqref{appeq:M13z}. The sum of the quadratic scattering amplitude terms gives
\begin{align}
\label{appeq:allT_ND_M13z}
	\left[ \cdots \right]
	\to 
	\frac{\qs^4}{16} 
	\left[
	4(\bz \cdot \br_{\bx\by})(\bz \cdot \br_{\bbx\by}) 
	-4 \br_{\bx\by} \cdot \br_{\bbx\by}\, z^2
	\right]
	\to 
	-\frac{\qs^4}{16}\,
	\big( 2 r_{\bx\by} r_{\bbx\by} \cos\theta z^2  \big),
\end{align}
where as usual we neglected the terms linear in $\bz$ when writing the first expression and subsequently we averaged over the angle between $\bz$ and $\br_{\bx\by}$. When we use \eqref{appeq:M13z} and \eqref{appeq:allT_ND_M13z}, we see that the term under study in Eq.~\eqref{dW1332_dY_ND_T} becomes
\begin{align}
	\label{appeq:int_ND_M13z_temp}
	\hspace*{-0.1cm}
	\int
	\frac{\dif^2 \bz}{2\pi }\, \mcal{M}_{\bx\by\bz} 
	\left[\cdots \right]
	=
	\frac{\qs^4}{16}\,r_{\bx\by}^2
	r_{\bx\by} r_{\bbx\by} \cos\theta  
	\int_{0}^1
	\dif \lambda 
	\int_0^{R^2}
	\dif z^2\,
	\frac{z^2}
	{\big[z^2 
	+\lambda(1-\lambda) r_{\bx\by}^2\big]^2},
\end{align}
which, after we neglect the terms that vanish when $R\to \infty$ and express $\cos\theta$ in terms of the sizes, gives
\begin{align}
	\hspace*{-0.3cm}
	\label{appeq:int_ND_M13z}
	-\int
	\frac{\dif^2 \bz}{2\pi }\, \mcal{M}_{\bx\by\bz} 
	\left[\cdots \right]
	=
	- \frac{\qs^4}{32}
	\left[r_{\bx\by}^2 
	\big(r_{\bx\bbx}^2 - r_{\bx\by}^2 -r_{\bbx\by}^2\big)
	\left( 
	\ln \frac{R^2}{r_{\bx\by}^2}
	+1
	\right)
	\right].
\end{align}
In order to find the last, $\mcal{M}_{\bbx\by\bz}$, term it is enough to let $\bx\leftrightarrow \bbx$ in Eq.~\eqref{appeq:int_ND_M13z}. Putting all three pieces together and performing some easy simplifications in the non-divergent term we arrive at the complete non-dipole contribution given in Eq.~\eqref{dW_1332_GBW_ND}.

\bibliographystyle{JHEP-2modlong}
\bibliography{refs}

\end{document}